\documentclass[aps,prb,reprint,
notitlepage]{revtex4-2}
\usepackage{amsmath}
\usepackage{graphicx} 
\usepackage{physics} 
\usepackage[dvipsnames]{xcolor}

\usepackage{minted}

\newcommand{\mctwo}{Department of Microtechnology and Nanoscience -- MC2,
Chalmers University of Technology, SE-41296 Gothenburg, Sweden}

\begin{document}

\title{Interaction energies of H$_2$ and CO on transition-metal surfaces computed by a\\
range-separated hybrid van der Waals density functional}

\author{Per Hyldgaard}
\affiliation{\mctwo}

\date{\today}

\begin{abstract}
Dissociative chemisorption (DC) of H$_2$ on the Cu(111) surface is a prototypical problem for understanding elements of heterogeneous catalysis [Science \textbf{326}, 832  (2009)]. The challenge lies in modeling the reaction dynamics that in turn reflects a classical potential for atomic deformations, friction, and inelastic scattering. Here, I test the use of a set of range-separated hybrid (RSH) van der Waals density functionals (vdW-DFs) [JPCM \textbf{37}, 211501 (2025)] on their ability to describe the classical barrier for dynamics in this H$_2$+Cu(111) DC problem. I furthermore document  
use of a variant for fast accurate predictions of the molecular quasi-particles (QPs), finding excellent performance across a set of small molecules that are often studied in catalysis. Finally, I suggest and implement a way to use that QP focus 
to identify what I consider a best-possible non-empirical
(yet adsorbate specific) RSH vdW-DF version, denoted AHBR($\gamma^*$) for H$_2$ DC modeling, navigating what are partly conflicting requirements on the molecule and metal sides. I find that the AHBR($\gamma^*$) can determine the classical H$_2$+Cu(111) DC barrier height close to chemical accuracy. I suggest that DC modeling can test broader relevance of the physics underpinning these RSH vdW-DFs.

\end{abstract}

\maketitle

\section{Introduction}

Dissociative chemisorption (DC) plays a key role in catalysis \cite{Ertl1976,Ertl1981,Ertl1982,AdsCatBIL1983,EngdahlBILJKN,Diaz2009,JiangDC2019,KroesDC2021}. For example, a Cu surface can cause an incoming hydrogen molecule of sufficiently high initial kinetic energy to break up \cite{BILsurfReact1979,Berger1990,EngdahlBILJKN,Michelsen1993,Rettner1995,McCormackH21998,BranchDCGao2001,KroesScatter2008,Diaz2009,KroesDC2021,AuerbachScatter24,Leiden26}. There is a probability for creating atomic-H adsorption \cite{BILsurfReact1979,ChemSorb1987,Diaz2009,catalysisvdW15}, rather than molecular physisorption \cite{anderssonetal1988,anderssonpeha96,lee11p193408}, and atomic adsorbates are more reactive; They can enable a subsequent formation of a desired chemical product. The product may sometimes also be produced directly in gas reactions and, either way, heat controls the rate of production \cite{Arrhenius1889} as it defines the typical incident energies $E_i$ for the molecules \cite{KroesDC2021}.  However, the presence of the metal surface often lowers heating costs for production. 

Quasi-particles (QPs), such as the lowest unoccupied molecular orbital (LUMO) and the highest 
occupied molecular orbital (HOMO), play a crucial role together with
substrate-induced shifts of their energy-level positions \cite{hoko64,Blyholder,Hedin65,jerry65,lu67,la70,helujpc1971,gulu76,lape77,JKNBILloss1979,lape80,Hedin80,PePaLe82,LePeSa84,HybersenLouieGPP,jogu89,ma,ra,Noblest,HaNo95,Baerends1997,AuJoWi00,Baerends2002,Baerends2003,davo2010,OTRSHalga,KraKro13,nguyen2015,OTRSHadsorb17,WiOhHa21,JPCMreview,ChiDFT23,AHBRmRSH25,hBN2026}. The QPs are defined within many-body perturbation theory (MBPT) \cite{pinesnozieres,Hedin65,FW7,mahansbok} as the formally-exact characterizations of probability amplitudes for adding or extracting an electron in the presence of full electron-electron interaction and subject to the Born-Oppenheimer approximation (BOA); The QP levels are the
associated energy costs \cite{Hedin65,lu67,la70,helujpc1971,Hedin80,HybersenLouieGPP}. The QP picture of such virtual electron excitations (vertical-ionization and electron-attachment processes)
permits a precise definition of molecular orbitals \cite{jogu89,Baerends2003} and practical formulations of exchange-correlation (XC) functionals \cite{helujpc1971,JPCMreview} for density functional theory (DFT) \cite{hoko64,kosh65}.
The QPs and hence DFT are by construction closely related to (in-situ) charge-transfer processes \cite{PePaLe82,LePeSa84,AuJoWi00,ChiDFT23,ChiDFT24,hBN2026}.

For the DC problems I note that 
the inbound dynamics of the molecule shifts the QP levels \cite{BILAdsRev1983} and enhances
charge transfer into the LUMO and out from (primarily) the HOMO \cite{Blyholder}. The
(initial) adsorbate dynamics is represented by a decreasing height $z$ 
which, in turn, further accelerates the rate of tunneling. The consequences are both shorter life times in electron exchange processes \cite{Noblest,HaNo95,kelkkanen11p113401} and enhanced electronic friction \cite{JKNBILloss1979,DampingCTH1983,HellsingPersson84,HeadGordonTully95,Huang2000,SurfChallenge2020,KroesDC2021}. The attraction and proximity also causes acceleration and scattering 
\cite{anderssonetal1988},  converting what initially may be a large center-of-mass kinetic energy $E_i$ connected to the inbound motion.
There are elastic-scattering events that reveal details of the H$_2$ physisorption well on Cu(111) \cite{zarembakohn1976,zarembakohn1977,harrisandliebsch1982b,harrisnordlander1984,andersson1993,anderssonpeha96,lee11p193408,lee12p424213}, at $ z_{\rm phys.} \approx 3.5$ {\AA}. At large incident energies $E_i$ molecules also get closer and experience inelastic-scattering events 
\cite{AdsAspectBIL1991,OsterlundZoricKasemo1997}, some of which can cause an incoming molecule to  dissociate  \cite{Diaz2009}.

Building from an adsorbate-dynamics framework \cite{Berger1990,Michelsen1993,Rettner1995,McCormackH21998,LauhonDyn2000,LauhonDyn2002err,ReppDyn2003,KroesScatter2008,Diaz2009,KroesDC2021,AuerbachScatter24} we can expect the existence of 1) some critical level of deformation (expressed as a state of the atom configuration) that suffices to tricker H$_2$-on-Cu(111) dissociation and 2) some associated (minimum) classical barrier $E_B$ for the incident molecule. The rate for what we below call H$_2$+Cu(111) DC  can be inferred by measurements of the sticking probability `$S_{\rm DC}^{\rm obs}$', at least when maintaining a low coverage
\cite{KroesScatter2008,KroesDC2021}. This follows because physisorption can be ignored at elevated temperatures while 
the probability of two ad-atoms meeting up and approaching
the conditions defining the classical barrier  is 
considered low. However, the challenge for seeking a better insight on these dissociation mechanisms lies in providing theory characterizations. They are needed to detail the atomic and electronic configurations at the critical deformation that constitutes the reaction bottleneck or barrier. Theory must provide good descriptions of the nature of required critical deformation and of the energy height $E_B$ of the classical barrier, measured relative to placing the H$_2$ molecule at rest at $z\to \infty$.

Ground-state (GS) density functional theory (DFT) \cite{hoko64,kosh65,GKSstart,BurkePerspective,beckeperspective,burke} 
is a powerful tool for materials modeling and seeking development in heterogeneous catalysis \cite{Ertl1976,Ertl1982,BILAdsRev1983,Noblest,HaNo95,KroesDC2021}. 
Within GS DFT, accuracy is 
generally defined by the quality of the approximation we must
make for the exchange-correlation (XC) energy density functional $E_{\rm xc}[n]$. However, it comes with an implicit assumption of adiabaticity: DFT use relies on the BOA,  separating
the dynamics of electrons and atoms, while also assuming that the 
electrons reside in the fully-interacting GS as defined by any specific (frozen) atomic geometry \cite{hoko64,Huang2000,SurfChallenge2020}. 

We can use DFT
in an empirical form seeking to bring alignment with observations, for example, the probability  $S^{\rm orb}_{\rm DC}(E_i)$ of the sticking for H$_2$+Cu(111) DC at incident energy $E_i$. The so-called specific-reaction parameter (SRP) DFT \cite{SRPdef1999,Diaz2009}, discussed further within, permits a modeling path with a minimum of wiggle room. It is used to seek a fitted-DFT form via inverse-scatting analysis of
sticking observations  \cite{Diaz2009,JiangDC2019,KroesDC2021}. At any given choice for the adjustable parameter $\lambda$, the corresponding SRP-DFT($\lambda$) provides a total-energy  landscape, i.e., it defines a guess
for the potential that guides the molecule dynamics. This landscape is used for modeling both the friction and the scattering effects to yield a $\lambda$-specific prediction for the sticking. Finally, the parameter $\lambda$ is adjusted iteratively to get alignment with the observed sticking behavior \cite{Diaz2009,KroesDC2021}, and 
in the process identify an optimal SRP-DFT that is specific to the
DC challenge, like H$_2$+Cu(111).
Importantly, by use of the resulting SRP-DFT, we have a relevant characterization of the atomic structure at the barrier, Fig.\ 1, top left panel. We also get a best-guess determination for the barrier height, $E_B^{\rm SRP} = 0.628$ eV, for H$_2$+Cu(111) DC \cite{Diaz2009,SurfChallenge2020}.
 
We often want to instead use DFT as a first-principle tool for predictions and characterizations. Such non-empirical
DFT has inherent mechanisms \cite{BurkePerspective,burke,bearcoleluscthhy14,AHBRlaunch} that give robustness and could
empower it with a general-purpose 
status \cite{BurkePerspective,bearcoleluscthhy14,AHBRlaunch}. We therefore seek to set the XC functional $E_{\rm xc}[n]$ directly and exclusively 
from generic physics concepts and principles, including analysis provided by use of formal MBPT. The benefits are 
that we 1) avoid free parameters in our DFT, allowing comparison of mechanisms between different problems, and 2) can interpret possible successes 
more directly from electronic-structure details \cite{helujpc1971,Hedin80,hBN2026}. Specifically, such DFTs allow us (with extra work) to extract
system-specific QP descriptions, in various approximations  \cite{Baerends1997,Baerends2002,Baerends2003,cococcioni2005,KuismaGB,davo2010,OTRSHalga,Ma2016,nguyen2016,OTRSHadsorb17,WiOhHa21,colonna2022,ChiDFT23,GoGaOh2024,AHBRmRSH25,NitrogenBasesAHBR-mRSH26,hBN2026}. In turn, that work reveals the nature of (virtual and actual) charge transfers in DC and chemisorption. Charge transfer is, for example, understood as key in setting details in adsorption \cite{Noblest,HaNo95,OsterlundZoricKasemo1997} for CO chemisorption on transition metals \cite{Blyholder,olsen2003co,DefineAHCX,AHBRlaunch} 
and O$_2$+Al(111) DC \cite{OsterlundZoricKasemo1997,OadsYourdBIL2002,adsHellmanBIL2005,YinZhang2018}.

We may set the XC energy description via the adiabatic connection formula (ACF) for $E_{\rm xc}$ \cite{gulu76,lape77}. Use of the random phase approximation (RPA) as an implicit density functional is an example,
which, however, ignores all
screening in defining the assumed
local-field response behavior \cite{lape77,rpa:ads,rpa_single,hybesc14} and is also computationally expensive. 
We typically approximate the XC design while making sure that the XC energy can be generally motivated for descriptions of  metals, i.e.,
includes an asymptotic screening in the (exchange) potential \cite{lu67,helujpc1971,lape77}. Then, further
MBPT analysis and insight \cite{mabr,Singwi68,Singwi69,Singwi70,rasolt,lape80,lavo87,thonhauser,JPCMreview} is used as a 
guide, for example, when defining the local density approximation (LDA) \cite{helujpc1971,gulu76,pewa86}, popular versions of the generalized gradient approximation \cite{pebuwa96,pebuer96,PBEsol} (GGA) and of meta-GGAs \cite{SCAN,SCANvdW}, as well as of non-local correlation 
functionals defined within the van der Waals (vdW) density functional 
(vdW-DF) method \cite{anlalu96,Dion,thonhauser,Berland_2015:van_waals,Thonhauser_2015:spin_signature,DFcx02017,DefineAHCX}. It is also possible to port the logic of any of these XCs for traditional Kohn-Sham (KS) DFT 
into versions for generalized KS 
DFT \cite{GKSstart}, namely by replacing part or all
of the exchange description with Fock exchange
\cite{Burke97,PBE0,DFcx02017,DefineAHCX,AHBRlaunch,AHBRmRSH25}, forming hybrid GGAs and vdW-DFs.
Importantly, we still desire chemical accuracy, that is, a maximum deviation of 1 kcal/mol or 0.043 eV relative to the $E_B^{\rm SRP}$ values (when available). Among these parameter-free MPBT-based XCs, it is exclusively two RPA-type studies \cite{Wei2023,Oudot2024}, from 2023 and 2024,  
that deliver chemical accuracy 
for the H$_2$-Cu(111) DC problem and RPA has not been tested against the broader DC benchmark set, called SBH17 \cite{SBH17} that is based on the DFT-SRP modeling approach. Worse, there are DC challenges for which we do not already have a SRP-DFT barrier characterization \cite{Leiden26} as well as some where there are discussions of whether it suffices to use the BOA (as in GS DFT) and implicitly assume adiabaticity in modeling
the sticking \cite{OsterlundZoricKasemo1997,SurfChallenge2020,AuerbachScatter24}.

Here I test a recently introduced
range-separated hybrid (RSH) vdW-DF, denoted vdW-DF2-ahbr \cite{AHBRlaunch} (abbreviated AHBR) by discussing its
description of the classical barrier that guides dynamics for H$_2$+Cu(111) DC.
The naming AHBR is chosen to highlight that it is 
crafted off an analytical hole modeling of the exchange 
component "B86R" of rev-vdW-DF2 \cite{lee10p081101,hamada14}. The latter, regular vdW-DF version, is also called vdW-DF2-b86r because it uses insight 
from analysis that led Becke to his 1986 GGA-type
exchange formulation \cite{becke1986p7184}. I assert the
performance of both AHBR and of the vdW-DF2-b86r, in terms of H$_2$+Cu(111) DC classical-barrier predictions, denoted 
$E_B^{\rm AHBR}$ and $E_B^{\rm br}$;
Comparing with the SRP-DFT result $E_B^{\rm SRP}=0.628$ eV value,  I find
that the former (latter) overestimates (underestimates) the 
$E_B^{\rm SRP}$ target by about 2 kcal/mol. I cross check this AHBR performance (at the barrier energy prediction) by also looking for indicators of quality with regards to predictions of charge transfer in chemisorption. For example, I document that AHBR systematically predicts the 
correct site preferences for CO adsorption on four transition 
metals \cite{Feibelman01p4018,DefineAHCX,AHBRlaunch}, 
within the present focus on
pursuing DFT with high convergence.

I furthermore note that the officially released `AHBR' \cite{AHBRlaunch} is
only the default form of a set of closely related RSH vdW-DFs,  here
denoted AHBR($\gamma$) \cite{AHBRlaunch,AHBRmRSH25}. Much as in HSE \cite{HSE03,HSE06,HJS08}, there is a choice of an inverse length scale (here denoted 
$\gamma$) that controls 
the cross over (with increasing electron-hole separation) 
between an assumed  short-range (SR) Fock-exchange inclusion, $\alpha=0.25$, and fully screened asymptotic exchange 
(set by the B86R form at all ranges in vdW-DF2-b86r). The present paper shows that there exists some AHBR'=AHBR($\gamma'$) for which the barrier 
characterization $E_B^{\rm AHBR'}$ aligns with the $E_B^{\rm SRP}$ value. The AHBR' construction is presented
in the spirit of the SRP-DFT, but stands out by being \textit{a single}
MBPT-guided XC design. That is, it is a member of a closely related set of RSH vdW-DFs \cite{AHBRmRSH25} within which all XC functionals formally have the same MBPT input \cite{becke1986p7184,Dion,HJS08,lee10p081101,hamada14,Thonhauser_2015:spin_signature,DefineAHCX,AHBRlaunch}.

The main suggestion of the paper is the use of a mole\-cular QP focus to define a best-possible non-empirical (AHBR-based)
barrier descriptor, denoted AHBR($\gamma^*$). In practice, I first limit the AHBR($\gamma$) tuning to remains within a maximum value, 
$\gamma'' \equiv 0.5\, a_0^{-1}$, defined by the inverse of the 
size of the H$_2$ molecule. 
This constraint means that one can approximate the AHBR($\gamma$) exchange-energy description by that
of a corresponding molecule-optimized
AHBR-mRSH($\gamma$) form \cite{AHBRmRSH25,NitrogenBasesAHBR-mRSH26}, as argued within. Constrained optimally tuning
\cite{OTRSHalga,OTRSHadsorb17,WiOhHa21,AHBRmRSH25} (OT) within the set of AHBR-mRSH($\gamma$) next identifies a 
$\gamma^* \leq \gamma''$ choice where AHBR-mRSH($\gamma^*)$ provides accuracy 
for the set of molecular QP levels \cite{AHBRmRSH25}. I suggest
use of the corresponding AHBR($\gamma^*)$ form as a DC descriptor because it is motivated on the metallic 
side \cite{DefineAHCX,AHBRmRSH25} and is found robust
on the molecular-energy description \cite{AHBRlaunch,AHBRmRSH25}. Also, the choice ensures consistency between the (molecule) total-energy variation and molecular QP levels
\cite{ChiDFT23,AHBRmRSH25,hBN2026}, well below $z_{\rm phys.}$ and to heights where charge transfer may impact the stated logic of taking guidance from molecule QPs \cite{OTRSHadsorb17}. I motivate use even at the barrier height $z_B^{\rm SRP}=1.16$ {\AA} by documenting that AHBR($\gamma^*$) retains a clear derivative discontinuity (in the energy variation with
partial charging) \cite{PePaLe82,Baerends2003}; This AHBR($\gamma^*$) behavior increases accuracy in descriptions of QPs \cite{Baerends2003,davo2010,nguyen2016,hBN2026} and of molecule-substrate charge transfers \cite{Blyholder}. More generally, this AHBR($\gamma^*$) behavior limits density-driven DFT errors \cite{PePaLe82,BurkeSIE,AHBRlaunch}.

The paper is organized as follows.
The following section presents theory, summarizing both the SRP-DFT modeling approach and the AHBR design, while also giving the rationale for using this RSH vdW-DF to extract molecular QP predictions \cite{OTRSHalga,WiOhHa21,AHBRmRSH25,NitrogenBasesAHBR-mRSH26}. Section III presents computational details
documenting also convergence. Section IV  contains the results and discussion followed by a summary and outlook. There is also an Appendix documenting that
use of non-empirical 
OT permits the here-explored set of vdW-DF2-b86r/AHBR-based XC functionals to accurately predict small-molecule QP  levels, at and below the HOMO.

\section{Theory}

The SRP-DFT is fitted to the specific system, for example the H$_2$+Cu(111)
DC problem. It serves to thus give an impression of the nature of the 
classical barrier for DC of incident H$_2$ molecules and give us a benchmark for succeeding at describing
this barrier via a non-empirical DFT of choice. As summarized above and detailed below, it emerges as  set by an inverse-scattering process, thus seeking
the potential that guides the molecule dynamics in the presence of scattering 
and friction \cite{KroesDC2021,AuerbachScatter24,Leiden26}. However, any such SRP-DFT, including the AHBR-based alternative that we suggest here, should be independently verified. This can be done against experiments on both scattering and sticking.  Testing can also be pursued by an additional focus on prediction of QPs as described within the XC functional in use \cite{kronik2012,OTRSHalga,OTRSHadsorb17,WiOhHa21,NitrogenBasesAHBR-mRSH26,hBN2026}. This is an option for QP-based validation that the AHBR and extensions are 
crafted to also enable \cite{AHBRlaunch,AHBRmRSH25,NitrogenBasesAHBR-mRSH26}.

\begin{figure}

\centering
\includegraphics[width=0.33\linewidth]{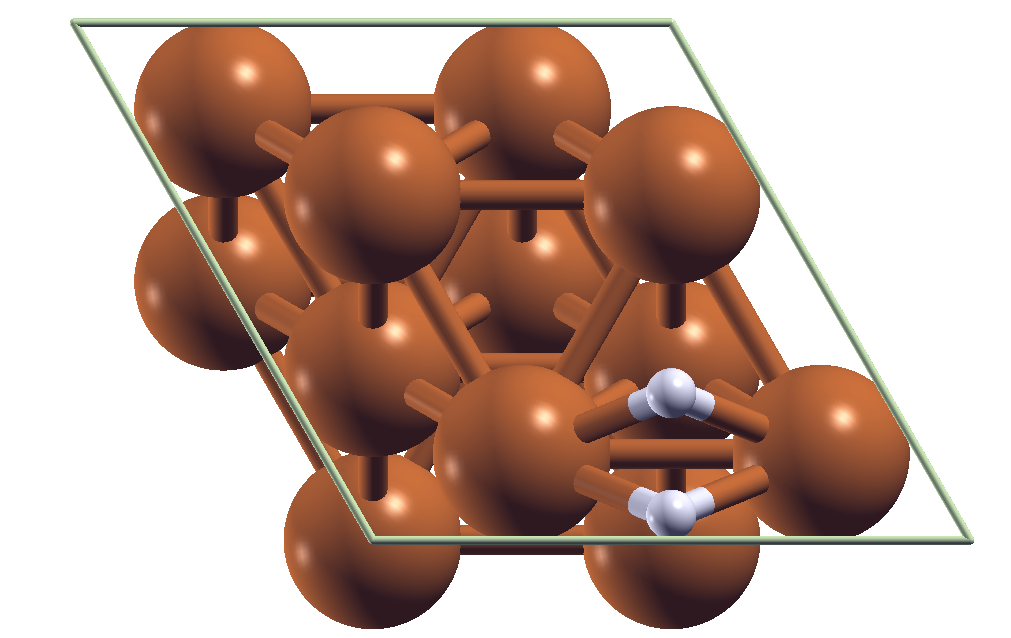}
\includegraphics[width=0.30\linewidth]{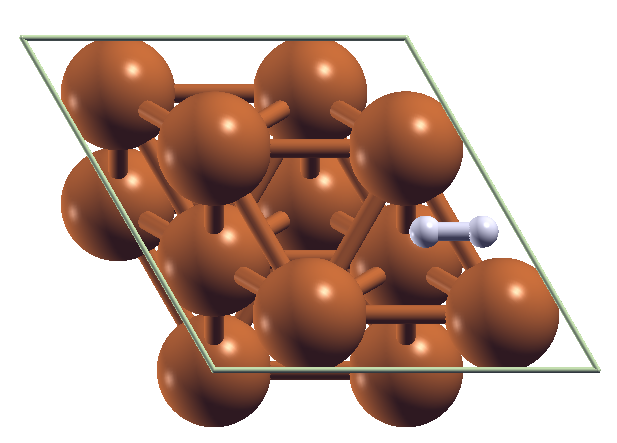} 
\includegraphics[width=0.31\linewidth]{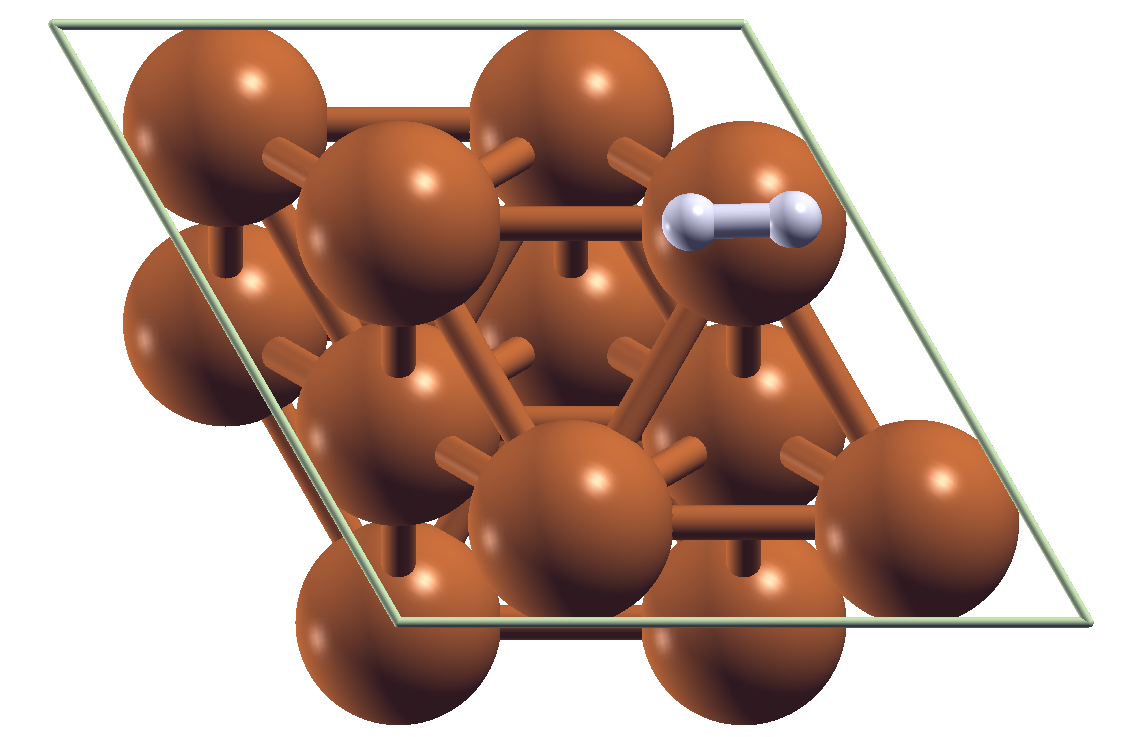} 
\\
\includegraphics[width=0.98\linewidth]{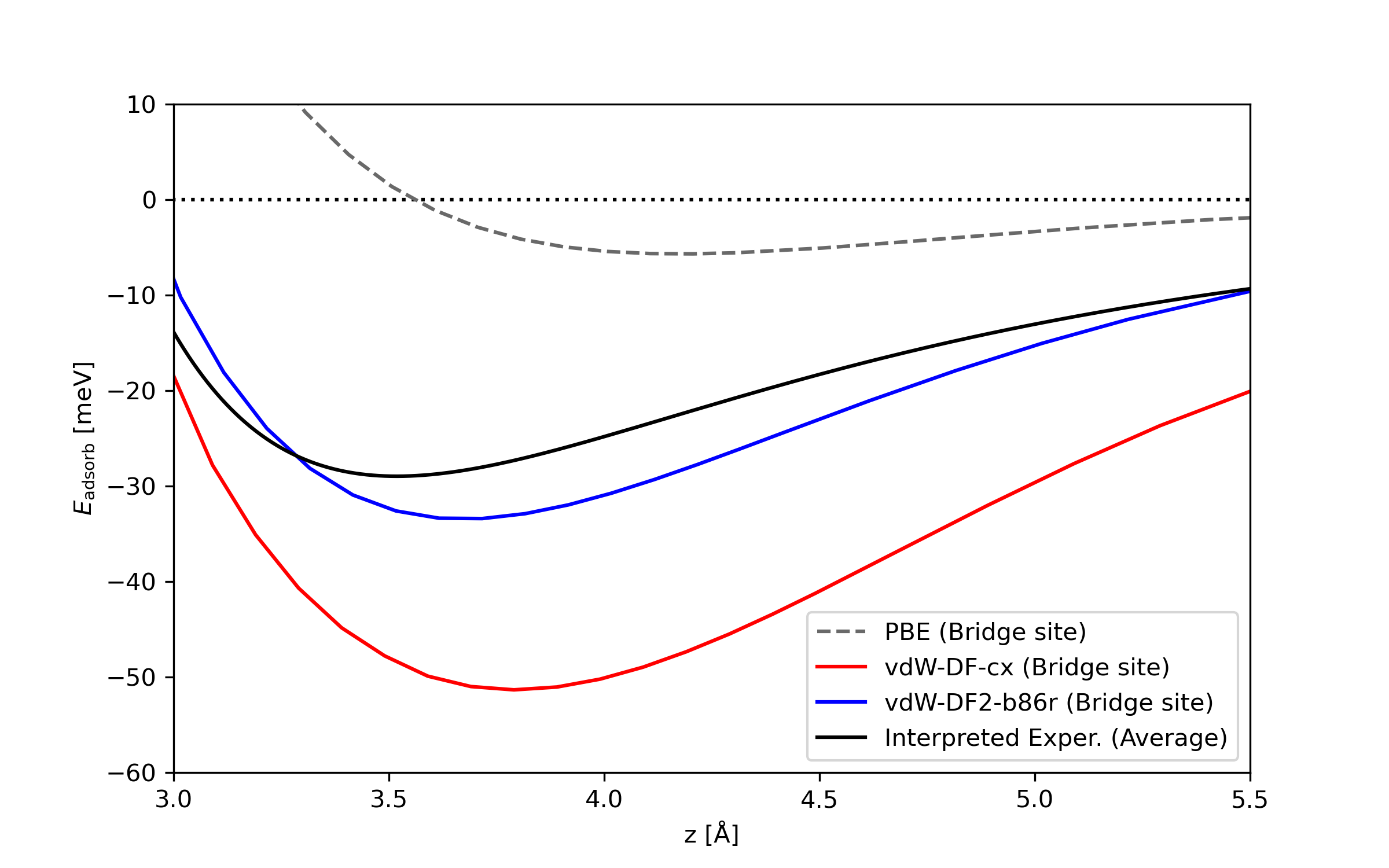}
\caption{\textit{Top row of panels:} Schematics for the barrier geometry that effectively limits dissociative 
adsorption for H$_2$+Cu(111) DC (left) and two possible atomic geometries for non-dissociative H$_2$-adsorption, i.e., H$_2$ physisorption 
on Cu(111) (middle and right). 
\textit{Bottom panel:} Comparison of KS-DFT physisorption-energy predictions $E_{\rm adsorb}$ (as a function of the height $z$ of the H$_2$ 
molecule over the surface) with experimental data, see text. The panel contrasts
results obtained by the semilocal functional PBE and 
by vdW-DF-cx \& vdW-DF2-b86r. The latter two are examples of non-hybrid vdW-DFs, i.e., descriptions with a truly nonlocal-correlation component that also captures dispersion interactions and where the interaction potential is therefore both deeper and having a longer reach. The details of the XC functional nature affect both scattering \cite{lee11p193408,lee12p424213,Berland_2015:van_waals} and 
QP predictions \cite{ChiDFT23,hBN2026}.
}
\label{fig:Physisorption}
\end{figure}

\subsection{Adsorption modeling \& SRP-DFT logic}

Figure \ref{fig:Physisorption} contrasts KS DFT studies of the interaction potential for physisorption of H$_2$ on Cu(111) with a curve representing experimental observations of Fano resonances created by interference in elastic-scattering events \cite{roy,anderssonetal1988}. The curve marked `Exper.' emerges after interpretation of that resonance data
within a model designed to capture the nature of physi\-sorption at  metal surfaces \cite{roy,zarembakohn1976,zarembakohn1977,harrisnordlander1984,anderssonetal1988,andersson1993,anderssonpeha96,lee11p193408,lee12p104102}. Meanwhile, each of three theory curves effectively carves
out the part of the corresponding full-energy variation (sometimes denoted `elbow' plot) with molecular deformations \cite{EngdahlBILJKN,BILAdsRev1983} that is relevant for the physisorption and elastic-scattering events. In practice, this `carving' is done by simply letting the molecule relax under the forces predicted in KS DFT (although keeping the height
of both H atoms fixed at a sequence of $z$ values, measured from the top-most atom position at the Cu(111) surface). These theory studies are
completed in PBE \cite{pebuer96}, vdW-DF-cx \cite{behy14} and vdW-DF2-b86r, for the bridge site, illustrated in the top middle panel; There is no large impact of instead tracking the physisorption at the top site
\cite{lee11p193408,lee12p424213,Berland_2015:van_waals}, i.e., at  the potential adsorption site illustrated in the top-right panel.

Figure \ref{fig:Physisorption} shows that the vdW-DF2-b86r performs well at characterizing 
elastic surface-molecular scattering of physisorption of H$_2$ on Cu(111).
As discussed many instances elsewhere,  use of the popular 
PBE \cite{pebuer96} GGA is not sufficient because GGA lacks a proper account of dispersion, i.e., what is here termed vdW interactions. 
 
Characterization and predictions of H$_2$+Cu(111) DC 
require substantially more modeling work and must be based on much larger
DFT input, namely a full potential-energy landscape reflecting a broad set of potentially possible deformations \cite{BILAdsRev1983,KroesScatter2008,JiangDC2019} (of the molecule and system). For example, H$_2$ molecules with a chance for dissociation
come in with a larger initial kinetic energy and will be further accelerated by the forces reflected in the potential. If the variation in potential energy was always instantaneously (fully) dissipated, there would be no elastic-scattering data to reconstruct a classical-potential representation \cite{roy,anderssonetal1988,andersson1993,anderssonpeha96,Berland_2015:van_waals}, i.e., the curve denoted ``Interpreted Exper." in Fig. \ref{fig:Physisorption}. 

Also, the focus for DC modeling is to a large extent on understanding the inelastic scattering events that emerge with a large kinetic energy available for the atomic dynamics \cite{KroesScatter2008,Diaz2009,KroesDC2021,SBH17}.
Some of that large kinetic energy will, at any given $z$, be directed into more general excitation and deformations. To model sticking
one tracks the molecule dynamics, including acceleration in the classical potential, general scattering, charge transfer, 
and electronic-friction effects. This dynamics modeling defines a 
probability that a given H$_2$ molecule traverses the 
classical DC barrier \cite{Berger1990,Michelsen1993,Rettner1995,SRPdef1999,KroesScatter2008,Diaz2009}.
A broad range of standard GGA or vdW-DF-type XC have been tested as candidates for providing good DFT input on the classical potential that is a core component of the above-summarized dynamics modeling. However,
no single such choice leads to a modeling that 
reproduces the measured probability $S^{\rm obs.}_{\rm DC}(E_i)$ for sticking in  
the H$_2$-Cu(111) DC problem, see discussion in Refs.\ \cite{Michelsen1993,Rettner1995,catalysisvdW15}.

What has instead been successfully done is to leverage empirical
DFT for interpretations of such observations.  The SRP-DFT 
modeling strategy for seeking and using such interpretations 
is outlined in the introduction and further described in 
Refs.\ \cite{KroesScatter2008,Diaz2009,KroesDC2021,SBH17}. It can 
sometimes be called semi-empirical DFT since the focus is on 
minimizing the freedom in fitting details in the XC functional
specifications. However, it remains a strategy to fit DFT outcome 
according to the system-specific measurements of the sticking 
$S_{\rm DC}^{\rm obs}$, along the principles of inverse-scattering modeling. 
The resulting fitted-DFT description, also called SRP-DFT
\cite{Diaz2009}, is designed to be the best-possible simple 
(DFT-based) descriptor of the potential that guides dynamics
so that the associated theory modeling reproduces the measured 
$S_{\rm DC}^{\rm obs}(E_i)$ values. The major advantage is that 
with the SRP-DFT form that is thus optimized, we have 
a plausible descriptor of the nature and details 
of what constitutes the classical  barrier for the DC problem in focus
\cite{BILAdsRev1983,JiangDC2019,KroesDC2021}.

In practice, the SRP-DFT modeling strategy involves working with two 
broadly trusted non-empirical XC functionals, denoted
$E_{\rm xc}^{\rm DF1}$ and $E_{\rm xc}^{\rm DF2}$. These XC functionals should  ideally both permit fast DFT total-energy determinations 
at a general variation in elbow plots \cite{BILAdsRev1983} and each have a fair transferability, 
at least 
within \mbox{(sub-)}{\allowbreak}classes of material problems. 
The picks for DF1 and DF2 are often also complementary: DF1 may have more accuracy and traction in some cases and DF2 in others. Differences in expected strengths may concern the abilities to cover bulk and slightly elongated molecular bonds or the ability to ameliorate the impact of erroneous charge transfer \cite{PeGrRo20,jewahy20,AHBRlaunch,Hard2Soft,AHBRmRSH25}.
With choices for DF1 and DF2, the desired  SRP flexibility 
arises by using GS DFT with the set of following \textit{merged} XC energy functionals [of the electron-density variation n($\mathbf{r})$]:
\begin{equation}
E_{\rm xc}^{\rm SRP}[n](\lambda) 
\equiv \lambda E_{\rm xc}^{\rm DF1}[n] + (1-\lambda)
E_{\rm xc}^{\rm DF2}[n] \, ;
\label{eq:SRPxc}
\end{equation}
There is thus restricted options for fitting or tuning.
The present SRP-XC forms, Eq.\ (\ref{eq:SRPxc}), are not, however, themselves XC designs
for MBPT-guided non-empirical DFT. 

With the more flexible XC functional description for GS DFT, 
$E_{\rm xc}^{\rm SRP}[n](\lambda)$, the SRP-DFT modeling proceeds, for a given DC case, by making a sequence of elbow plots \cite{Diaz2009,JiangDC2019,SBH17} at various $\lambda$, completing the sticking modeling for each of these total-energy landscape descriptions,
and picking the $\lambda'$ that best fits the observations of $S^{\rm obs}_{\rm DC}(E_i)$. The resulting fitted modeling tool, also denoted `SRP-DFT', is simply the (empirical) DFT set by use of $E_{\rm xc}^{\rm SRP}[n](\lambda')$. One arrives at a DC-specific XC functional and SRP-DFT that one can use to understand the barrier nature and barrier height $E_B^{\rm SRP}$ \cite{Diaz2009}. This SRP-DFT process has been 
repeated for several DC challenges so that there exists now a 
surface-barrier height benchmark set, denoted 
SBH17 \cite{SBH17}. 

There are ongoing  discussions, because of the potential importance of non-adiabatic effects in scattering and in friction \cite{JKNBILloss1979,OsterlundZoricKasemo1997,Huang2000}, on whether the SRP-DFT descriptions can systematically be trusted \cite{Diaz2009,SurfChallenge2020,Leiden26}, and therefore whether one would need input beyond GS DFT 
 to understand the sticking and DC in such cases. Direct O$_2$+Al(111) DC is a case
that exemplifies this discussion \cite{SurfChallenge2020}. This DC problem is a long-standing 
challenge for analysis based on DFT, for example, Refs.\ \cite{OsterlundZoricKasemo1997,OadsYourdBIL2002,adsHellmanBIL2005,YinZhang2018}. 

Non-adiabatic effects in scattering can
arise with a coupling of the vibrational effects and the charge transfer that also involves 
the LUMO. The energy of this LUMO should, at 
infinite separation ($z\to\infty$), be set by
the molecule electron affinity (EA), at least assuming we have an XC functional permitting good QP predictions \cite{nguyen2015,AHBRlaunch}.
The level position varies as the molecule approaches
but it is still motivated to expect that the
net charge transfer is proportional to a simple difference
\begin{equation}
    E_{\rm CT} = \Phi_{\rm work}^{\rm M} - \hbox{EA}^{\rm mol.} \, 
    \label{eq:ECT}
\end{equation}
between the metal-surface work function $\Phi_{\rm work}^{\rm M}$ and the molecular EA. Electronic friction arises by electron-hole formations
and it is plausible that a low $E_{\rm CT} < 7$ 
eV value \cite{SurfChallenge2020} is correlated with non-adiabatic couplings 
between the vibrational and atomic dynamics \cite{Huang2000}, thus complicating also
the friction modeling  \cite{SurfChallenge2020}.

The present focus on the H$_2$+Cu(111) DC problem
is characterized by an $E_{\rm CT}\approx 8$ eV
value and hence we are likely set to use associated
SRP-DFT modeling as giving a relevant barrier 
target $E_B^{\rm SRP} = 0.635$ eV. Nevertheless,
the $E_{\rm CT}$ value is so low that it is motivated to check any (AHBR-based) modeling with a discussion of whether it also retains a plausible MBPT and QP description.
  
\subsection{Hybrid vdW-DFs, generalized DFT \& QPs}

As stated in the introduction, the idea of this modeling
paper is to arrive at a QP-guided best-suggestion for a 
non-empirical (yet adsorbate-optimized) AHBR($\gamma^*)$ descriptor of DC and chemisorption problems. To that end it is instructive to consider the link between the exchange energy
$E_{\rm x}$ and the (analytical-) exchange-hole model $n_{\rm x}(\mathbf{r}; |\mathbf{r}-\mathbf{r'}|)$. The definition of AHBR and extensions has at its core a focus on modeling this hole to represent a description of "B86R", i.e., the GGA-type exchange in vdW-DF2-b86r (rev-vdW-DF2) \cite{DefineAHCX,AHBRlaunch}. In general, the exchange hole describes a motivated approximation for how an electron at position $\mathbf{r}$ creates a depletion by same-spin Pauli exclusion in the electron density at position $\mathbf{r'}$, at the indicated distance. This modeling is completed subject to the implicit GGA-exchange assumption of asymptotic screening of exchange contributions  \cite{lape77}, spin scaling \cite{PerZun81,Thonhauser_2015:spin_signature}
and formal MBPT inputs \cite{lavo87,Dion,thonhauser,lee10p081101}. In practice, such GGA-type exchange-hole models  are crafted in a framework that implements a Gaussian-type decay with $|\mathbf{r}-\mathbf{r'}|$. The prefactor for this decay is set by the electron 
concentration $n(\mathbf{r})$, via a so-called scaled gradient $s(\mathbf{r})$, as is done in all aspects of setting GGA exchange details  \cite{lavo87,pebuer96}. 

One can in general express the exchange energy entering in KS DFT and in generalized-KS (of hybrid) DFT as a Coulomb interaction term,
\begin{equation}
   E_{\rm x}^{\rm DF}=\frac{1}{2} \,
   \int_{\mathbf{r}} \, \int_{\mathbf{r'}}
   \, \frac{n(\mathbf{r}) \, 
   n_{\rm x}^{\rm DF}(\mathbf{r};|\mathbf{r}-\mathbf{r'}|)}
   {|\mathbf{r}-\mathbf{r'}|)} \, .
   \label{eq:ExEnergyForm}
\end{equation}
The GGA-type exchange $E_{\rm x}$ functional, 
that is part in vdW-DF2-b86r, emerges when inserting (for $n_{\rm x}^{\rm DF}$) the exchange-hole model form $n_{\rm x}$ that
reflects the "B86R" exchange behavior \cite{becke1986p7184,hamada14,HJS08,DefineAHCX,AHBRlaunch}. Unscreened Fock exchange, denoted $E_{\rm x}^{\rm Fo.}$, arises instead when we first set the single-particle density matrix from (generalized KS) DFT orbitals, define a Fock-exchange hole $n_{\rm x}^{\rm Fo.}$, and insert that hole  in Eq.\ (\ref{eq:ExEnergyForm}).

MBPT-guided functionals with a semilocal, GGA type exchange, like vdW-DF-cx or vdW-DF2-b86r, cannot give good predictions of QP level \cite{AuJoWi00} directly, when used in traditional KS-DFT \cite{kosh65}. This follows because of the KS DFT emphasis on using a \textit{local} effective potential and, in particular, local 
exchange-potential, for fast, in principle exact, determinations of
both the total internal energy (variation) and the electron density, $n(\mathbf{r})$ \cite{kosh65,helujpc1971,GKSstart}. If we rely directly on
KS eigenlevels (nominally an auxiliary construction to keep KS orbitals
normalized), we underestimate, for example, HOMO-LUMO gaps. If, 
however, for small molecules, one adjusts the KS-description of the XC potential by enforcing an asymptotic $-1/r$ variation (corresponding to 
unscreened or Fock exchange) and adjust the raw KS-orbital 
levels (those reported in KS DFT) for the so-called derivative discontinuity, the impression and status changes. One can then get an excellent alignment between what can be called  `true KS levels' and the set of vertical ionization potentials \cite{Baerends1997,Baerends2002,Baerends2003,KuismaGB}, i.e, 
the actual molecular  QP levels \cite{FW7,AuJoWi00,hBN2026}, 
see also appendix A and Ref.\ \cite{ChiDFT23}. 

In fact, MBPT-based XC functionals, like PBE or the vdW-DFs, demonstratively retain the underlying QP (and hence fully interacting Green function) character directly within the XC formulation \cite{FW7,helujpc1971,JPCMreview,ChiDFT23}. This is clear because 
we can use the energy components of a given DFT calculation 
\cite{kosh65,Burke97,JiScHy18a} to provide an explicit
DFT-based determination of the first frequency moment of 
the occupied part of the fully interacting spectral function \cite{ChiDFT23,ChiDFT24}. This holds to the extent a specific MBPT detail (e.g., vdW forces) is retained in the specific XC choice \cite{helujpc1971,lavo87,thonhauser}. 
Again, for small molecules, where exact QP descriptions are available,
and when starting from PBE \cite{pebuer96} suffices, there is excellent alignment on this frequency-average MBPT measure 
and exact configuration-integration (CI) studies \cite{ChiDFT23}. 

More generally, for molecules and for extended systems 
alike, we can extract and compare the 
details of the QP-level description. 
We can therefore also uncover
the MBPT content \cite{helujpc1971,JPCMreview} in, e.g., the consistent-exchange 
vdW-DF-cx version \cite{behy14} and PBE \cite{pebuer96} by slightly
adjusting the orbital-solution strategy
\cite{davo2010,ferretti2014,colonna2022,hBN2026}.
Specifically, we can insert such MBPT-based XC 
functionals (originally designed for KS DFT) into 
the so-called Koopmans-integer DFT, thus explicitly
enforcing piecewise linearity of the XC energy 
with fractional changes in charging \cite{colonna2018,nguyen2018,gennaro2022,linscott2023}.
This approach to fast QP predictions
stands out because the solution (always at 
integer electron occupation per unit cell) keeps
the density (and hence actual structure as we use the BOA) 
and the actual total energy unchanged 
\cite{linscott2023}. However,  
Koopmans-DFT is presently restricted to non-metallic 
systems \cite{nguyen2018,colonna2022}.

Since H$_2$+Cu(111) DC involves a metallic substrate, 
this paper proposes to, instead, rely on the
RSH vdW-DF route \cite{DefineAHCX,AHBRlaunch} to 
simultaneously seek 1) good descriptions of
the barrier energy for H$_2$+Cu(111) DC and 2)
directly-related characterizations of the QP. 
The latter is implicitly a check on whether we land with a 
plausible barrier description because the DFT potential
description in chemisorption and DC (and overall dynamics 
modeling) depends on getting a correct description of 
charge transfer, e.g., avoiding delocalization 
errors \cite{AHBRlaunch}.

\begin{figure}
    \centering
\includegraphics[width=0.95\linewidth]{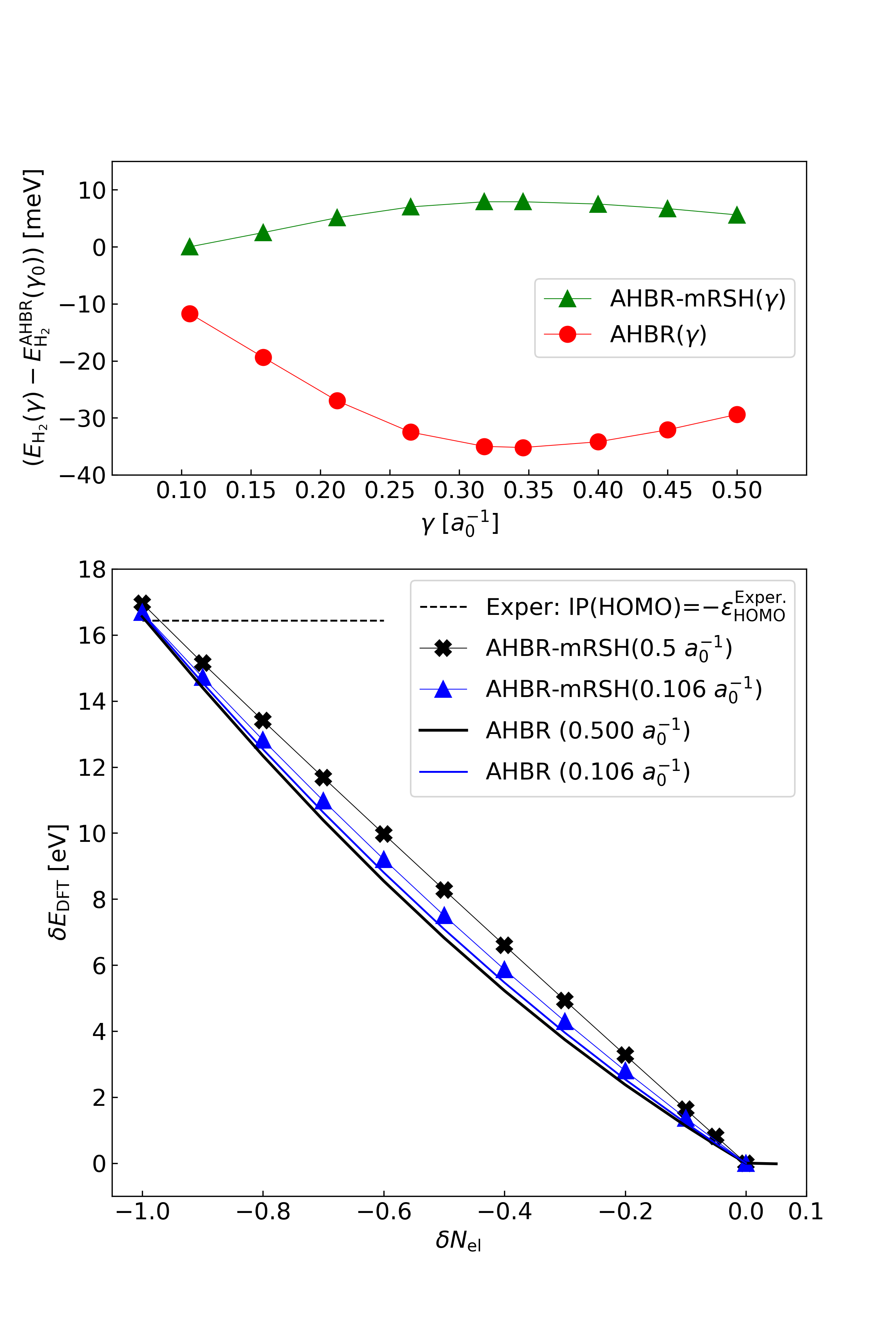} 
\caption{\textit{Top:} Variation in H$_2$ binding energy with the inverse-length scale $\gamma$ tuning
for the metal-compatible AHBR($\gamma$) and the molecule-optimized AHBR-mRSH($\gamma$).
The energy reference is here set by the 
binding-energy description of the default
AHBR=AHBR($\gamma_0 = 0.106\, a_0^{-1}$) \cite{AHBRlaunch}
\textit{Bottom:} Total-internal energy dependence on changes in the  partial electron occupation of the H$_2$ molecule as predicted in 
a set of key AHBR-based RSH vdW-DFs. The value of $\delta N_{\rm el}=0$ (=1) corresponds
to the neutral (singly charged) H$_2$ molecule, as measured relative to the 
GS form at the set of stated generalized KS descriptions. For the OT AHBR-mRSH($\gamma^*=0.5\, a_0^{-1}$) form, the linearity reflects a correct 
prediction of the HOMO energy level, see Appendix A, but it is not possible to directly illustrate the derivative discontinuity in this case. We find that the here-suggested H$_2$+Cu(111) DC modeling tool, AHBR($\gamma^*=0.500\, a_0^{-1}$), with energy variation tracked by the thick black curve, does not retain
full linearity with partial charging. However, the energy variation 
around $\delta N_{\rm el}=0$ shows that it retains a clear derivative discontinuity 
with partial charging \cite{PeGrRo20,KraKro13,Baerends2003,davo2010,nguyen2015}.
}
\label{fig:H2piecewise}
\end{figure}

The specification of hybrid and RSH vdW-DFs also involves use of
a current-conserving nonlocal-correlation term \cite{Dion,dionerratum,hybesc14}, in this case 
the $E_{\rm c}^{\rm nl2}$ from vdW-DF2 \cite{Dion,lee10p081101} and thus from vdW-DF2-b86r. This $E_{\rm c}^{\rm nl2}$ component is kept unchanged when setting up both the original RSH AHBR form
and the wider class `AHBR-gRSH($\alpha, \beta, \gamma$)' \cite{AHBRmRSH25} that also includes simple hybrids (besides both the vdW-DF2-b86r and the 
default AHBR itself). We get to a vdW-DF2-b86r-based simple-hybrid
XC, denoted vdW-DF2-br0 \cite{lee10p081101,hamada14,DFcx02017,AHBRlaunch}, by simply mixing $E_{\rm x}^{\rm br}$ and 
$E_{\rm x}^{\rm Fo.}$ as follows:
\begin{equation}
    E_{\rm xc}^{\rm br0}
    = \alpha E_{\rm x}^{\rm Fo.}
    +(1-\alpha) E_{\rm x}^{\rm br}
    + E_{\rm c}^{\rm nl2} \, .
\end{equation}
Moreover, we get a completely general class, denoted
AHBR-gRSH($\alpha, \beta, \gamma$), of RSH vdW-DF2s by furthermore
using a range separation (defined by the error-function `$\erf$') of the Coulomb-interaction matrix element \cite{OTRSHalga}
\begin{equation}
\frac{1}{y} = \frac{\alpha+\beta \erf(\gamma y)}{y} + \frac{1-[\alpha+\beta \erf(\gamma y)]}{y} \, .
\label{eq:SplitCoul}
\end{equation}
This separation of the Coulomb interaction between an electron and its associated hole is inserted in exchange-hole modeling form, Eq. (\ref{eq:ExEnergyForm}), while weighting the first (second) term by the 
exchange-hole modeling $n_{\rm x}^{\rm Fo.}$
(by the B86R-relevant $n_{\rm x}$ hole modeling). 

Importantly, all such descriptions rely on an identification
of a static dielectric constant, denoted  $\varepsilon$ and characteristic of the system, to set $\beta\equiv\varepsilon^{-1}-\alpha$ \cite{OTRSHalga,OTRSHadsorb17}. This is done  so that we get a system-relevant
screening of exchange contributions at large and asymptotic
electron-hole separations, as explained in Ref.\ \cite{AHBRmRSH25}.
The original AHBR is crafted, much like the PBE-based HSE06 \cite{HSE06} with a set of defaults, $\alpha_0=0.25$, $\beta=-\alpha$, and $\gamma_0=
0.106\, a_0^{-1}$, setting the extent `$\alpha$', 
of the SR mixing of Fock exchange,
and the assumed inverse length scale 
`$\gamma$' for a roll-over in the exchange description with asymptotic electron-hole separations, respectively \cite{EP98,HSE03,HJS08,AHBRlaunch}. More generally
one can consider related descriptors relevant for
metal systems like those of DC challenges, enforcing again $\varepsilon^{-1}=0$.
Because we also then keep the $\alpha_0$ choice deliberately fixed, the range of relevant such RSH vdW-DFs is restricted to the aforementioned subset,
AHBR($\gamma$). Meanwhile, for molecular problems, we ideally should set $\varepsilon^{-1}=1$ \cite{kronik2012,OTRSHalga,AHBRmRSH25},
the relevant descriptors are instead the closely 
related AHBR-mRSH($\gamma$) set \cite{AHBRmRSH25,NitrogenBasesAHBR-mRSH26}.

Here I use the AHBR($\gamma$) to 
discuss the rationale for using generalized KS-DFT descriptions on H$_2$+Cu(111) DC. That is, I use these metal-compatible RSH vdW-DFs to seek both accuracy and QP consistency in predictions of  
the H$_2$+Cu(111) DC barrier, for example, 
comparing with the reference
barrier energy $E_B^{\rm SRP}$.
Reliance on an apparent flexibility in AHBR($\gamma$), with respect to $\gamma$ tuning, to seek SRP-DFT alignment is not the end
purpose here.
The intention is rather to motivate choice of one (adsorbate-specific, yet non-empirical) form in generalized KS DFT \cite{GKSstart}, even if 
that pick may only approach chemical accuracy in the DC-barrier prediction.  I cross 
check the approach by also reporting AHBR predictions for 
the site-preference for
CO chemisorption on transition metals \cite{DefineAHCX,AHBRlaunch}, for reasons explained in Sec.\ IV.

The primary idea 
is to use constrained OT of a $\gamma^*$ setting within molecule-optimized generalizations of the standard RSH designs 
\cite{kronik2012,OTRSHalga,OTRSHadsorb17,AHBRmRSH25} and then import that to an AHBR($\gamma^*$) pick for DC
modeling. The molecule-focused OT step is set by seeking alignment between the molecule HOMO level and its adiabatic ionization potential (IP), see Appendix A.  In effect, I look
at the 
molecular side as in Ref.\ \cite{OTRSHadsorb17},
but I am instead seeking a non-empirical adsorption-energy descriptor within the AHBR set. For isolated molecules, the relevant set of B86R/AHBR-based functionals to pursue this OT strategy is the AHBR-mRSH($\gamma$) set 
(as it ensures a $-1/r$ asymptotic decay in the exchange potential \cite{LePeSa84,OTRSHalga,AHBRmRSH25}). 
Use of this QP focus for crafting 
also a non-empirical DC modeling by use of AHBR($\gamma^*$) is possible when the OT value $\gamma^*$ falls within a molecule-specific limit $\gamma''$, essentially the inverse of the adsorbate size, see below. When that condition holds, one can connect AHBR($\gamma^*$) and 
AHBR-mRSH($\gamma^*$) results of
exchange- and hence total-energy contributions originating from the molecule region. The connection implies consistency between molecular energies and the QPs \cite{ChiDFT23}: I am merely using the
AHBR($\gamma^*$) XC form, not AHBR-mRSH($\gamma^*)$, in the molecular region.

Figure \ref{fig:H2piecewise} illustrates the crux of the argument for use 
of AHBR($\gamma^*$) as a DC descriptor, subject to the $\gamma^* \leq \gamma''$ constraint.  
The top panel contrasts
the AHBR($\gamma$) and AHBR-mRSH($\gamma$) variations in their predictions of binding
in neutral, isolated H$_2$ molecules, relative to the default-AHBR results, corresponding to 
AHBR($\gamma_0=0.106\, a_0^{-1}$). The variation is small, an observation that in fact holds across almost all molecular-energy differences and barrier, as tested within the GMTKN55 benchmark suite \cite{gmtkn55,AHBRlaunch,AHBRmRSH25}. The main
observation of the panel is that there are only small 
\textit{quantitative} differences between the AHBR($\gamma$) 
[in particular AHBR($\gamma^*$)] and AHBR-mRSH($\gamma$) [in particular AHBR-mRSH($\gamma^*$)] results
for small molecules. This observation reflects in turn 
a close similarity
between their \textit{exchange-energy} formulations, Eq.\ (\ref{eq:ExEnergyForm}): They differ only in screening at asymptotic electron-hole separations \cite{AHBRmRSH25}. This is of limited relevance since $1/\gamma''$ is given by the molecule size.

The bottom panel of Fig.\ \ref{fig:H2piecewise}
nevertheless documents that there exists a significant \textit{qualitative} difference between the OT 
AHBR-mRSH($\gamma^* \leq \gamma''$) form and 
the set of AHBR($\gamma$)'s or other AHBR-gRSH($\alpha,\beta,\gamma$)'s. At fixed $\alpha_0$, it is exclusively AHBR-mRSH($\gamma^*$) that succeeds at ensuring a piecewise linearity in the energy variation with partial charging
(ionization). This is a formal thermodynamical criterion on using DFT on molecular components of systems \cite{PePaLe82,davo2010,KraKro13,nguyen2015}. As documented in Appendix A, the OT pick $\gamma^*$ means that the H$_2$ HOMO level sets the QP behavior, with a level which again must (here) align with the adiabatic IP \cite{LePeSa84,Baerends2003}. I add that difference in QP-level predictions results because the QP energy levels are sensitive to the asymptotic form of the \textit{exchange potential}
\cite{PePaLe82,LePeSa84,Baerends2003,AHBRmRSH25}.
The advantage of using AHBR-mRSH($\gamma^*$) would be that we ensure alignment of the energies and the QP levels, i.e., we respect the MBPT input on which the AHBR-type RSH vdW-DFs are crafted \cite{FW7,helujpc1971,ChiDFT23,ChiDFT24,hBN2026}
but it is not directly useful for our metal-containing systems. 

However, for the H$_2$ molecule the bond length is 0.7~{\AA}
while the vdW radius is 1.2 {\AA} and so we can see the
AHBR and AHBR-mRSH descriptions as connected, again as also illustrated in
Fig.\ \ref{fig:H2piecewise}. This connection holds as long as we limit
$\gamma$ adjustments so that $1/\gamma''=2\, a_0$, for reasons detailed in Sec.\ IV.
In the H$_2$ case, we find that the OT value $\gamma^*$, Appendix A, is in fact equal to the H$_2$-specific limiting value $\gamma''= 0.5\, a_0^{-1}$ that we must place on tuning for our modeling purposes.

Accordingly, it is possible for this DC modeling to proceed with a bootstrap inspired by Ref.\ \cite{OTRSHadsorb17}: In the absence of any significant actual charge transfers, we may use AHBR($\gamma^*)$ in lieu of AHBR-mRSH($\gamma^*$) for molecular-region energy descriptions. In effect, we arrive at a QP-guided setting for a non-empirical, yet adsorbate-specific RSH vdW-DF AHBR($\gamma^*)$
that is motivated at both the metal sides and
molecule sides. Full motivation holds at least down to adsorbate heights $z$ 
between those characteristic of the
physi\-sorption well, Fig.\ \ref{fig:Physisorption},
and the barrier location $z_B^{\rm SRP}$.
 
The bottom panel of Fig.\ \ref{fig:H2piecewise} shows that AHBR($\gamma^*=0.500\, a_0^{-1}$) has a key property that allows it to retain fair accuracy even at or inside $z_B^{\rm SRP}$.
There we must expect actual charge transfers \cite{Blyholder,Noblest} and with DFT there is always a risk for spurious enhancements and other density-driven errors \cite{PePaLe82,BurkeSIE,AHBRlaunch}. The panel shows that a switch from AHBR-mRSH($\gamma^*)$ to AHBR($\gamma^*$) 
does imply a partial loss of the ideal linearity in ionization energies \cite{PePaLe82}.
Importantly, however, for AHBR($\gamma^*$), with a variation in charging energies tracked by the thick black curve,  I am 
able to converge also a few descriptions of EA-type excitations (adding fractions of an electron), corresponding to $\delta N_{\rm el} > 0$ results.
I thus illustrate that AHBR($\gamma^*$) retains a fair account of the derivative discontinuity
behavior \cite{PePaLe82,davo2010} that helps suppress \textit{spurious\/} charge transfers. Avoiding these and other density-driven errors \cite{BurkeSIE,AHBRlaunch} means that AHBR($\gamma^*)$ fulfills necessary conditions for keeping accuracy, also when the molecule gets to 
conditions representative of the reaction barrier.

\begin{table}
\caption{\label{tab:H2atCu}
Convergence of the AHBR predictions for 
the H$_2$-Cu(111) dissociation barrier,
$E_B^{\rm AHBR}$, with the extent of the 
in-surface $k\times k$ sampling and with 
the extent of the additional $q\times q$ grid 
(of $k$-point differences) used in
evaluating the Fock-exchange component 
of the RSH vdW-DF. All energies in eVs. The results, subsequently denoted AHBR-12-4
(and AHBR-12-3) for $k=12$ in combination with $q=4$ ($q=3$) is considered fully (fairly) converged and this $k$ point sampling is systematically used in all adsorption 
studies reported here.}
\begin{tabular}{c|ccccc}
\hline
$k\times k$ sampling & $q$ = 2 & $q$ = 3 & $q$ = 4 & $q$ =5 & $q$ =6 \\
\hline
6 & 0.618 & 0.623 & - & - & 0.617 \\
8 & 0.660 & - & 0.670 & - & - \\
10 & - & - & - & 0.760 & -\\
12 & 0.736 & 0.728 & \textbf{0.725} & - & 0.724 \\
\hline
\hline
\end{tabular}
\end{table}

\begin{table}
\caption{\label{tab:COsite}
AHBR characterizations of the CO (top-vs-fcc) adsorption-site preference $\Delta_{\rm site} E_{\rm CO}^{\rm Surf}$ (in eV), for Au, Ag, Cu and Pt surfaces and 25\% coverage as described at increasing accuracy of the choice of norm-conserving PPs and extent in the $k$- and $q$-point sampling, see text. The RSH vdW-DF functionals are here systematically used with the default inverse screening length,  
$\gamma=0.106\, a_0^{-1}$, kept deliberately fixed; For comparison we also report results of the regular vdW-DF2-b86r that corresponds to using AHBR in the $\gamma\to\infty$ limit \cite{AHBRlaunch,AHBRmRSH25}. 
Experimental observations for these surfaces (at this coverage) consistently find $\Delta E_{\rm CO}^{\rm (111)} \ll -0.026$ eV.  Results for vdW-DF2-b86r are also provided, for a $12\times12\times1$ $k$ surface point sampling. We also summarize results by the related RSH vdW-DF-ahcx (abbreviated AHCX) \cite{DefineAHCX} described at a $k=6$ and $q=3$ sampling; Entries marked by an asterisk were obtained at PPs with fewer electrons, see text and Ref.\ \cite{DefineAHCX}. For comparison we observe that the AHBR-6-3 characterization for
$\Delta E^{\rm Pt}_{\rm CO}$ yields a -0.072 value \cite{AHBRlaunch}
when used with the present PPs. Highlighted entries are converged RSH vdW-DF descriptions in qualitative agreement with experimental observations.}
\begin{tabular}{l|c|cc|c}
\hline
& AHCX &  AHBR-12-3 & AHBR-12-4 & vdW-DF2-b86r \\
\hline
$\Delta E_{\rm CO}^{\rm Cu}$ & -0.024$^a$* &  -0.102 & \textbf{-0.101}  &0.128\\
$\Delta
E_{\rm CO}^{\rm Ag}$ & -0.179$^a$* &  -0.255& \textbf{-0.252} & 0.122\\
$\Delta E_{\rm CO}^{\rm Au}$ & -0.156$^a$* &  -0.162 & \textbf{-0.161} & -0.096 \\
$\Delta E_{\rm CO}^{\rm Pt}$ & -0.034 &   -0.089 &  \textbf{-0.085} & 0.085\\
\hline
\hline
\end{tabular}
\end{table}

\section{Computational details}

All calculations are made with the 
\textsc{Quantum-ESPRESSO} (QE) plane-wave DFT code \cite{QE,Giannozzi17,linACE,PaoloElStruct1}, using the electron-rich optimized norm-conserving Vanderbilt \cite{ONCV} (ONCV)
pseudo-potentials (PPs) 
of the so-called SG15 \cite{sg15} set at a large
energy cutoff (160 Ry). This
choice is made to stay consistent with previous studies: 1) Broad types of 
total-energy and structure-optimization
studies included in launching AHBR=AHBR($\gamma_0$) as an accurate general-purpose XC \cite{AHBRlaunch}, and 2) Accurate simultaneous predictions of both binding energies and of frontier-level QPs in larger molecules and systems, upon OT the molecule-optimized AHBR-mRSH($\gamma$) extension form  \cite{AHBRlaunch,AHBRmRSH25,NitrogenBasesAHBR-mRSH26}.

For our set of adsorption studies, for example,
as illustrated in the top panels of Fig.\ \ref{fig:Physisorption}, we use a five-layer slab to model the 111 surfaces of 
the noble and Pt transition metals. The starting point for further characterizations, including the description of the basic in-surface unit cells, is set by first using vdW-DF-cx to optimize the bulk structure and next porting this structural info to the slab geometry. This slab is described in a unit-cell of height $c=30$ {\AA}, so that
our slap modeling leaves more than 20 {\AA} vacuum between repeated images of the metal films. 
The top side of the slab is used to model the (111) themselves, by furthermore implementing atomic relaxations 
as described by vdW-DF-cx, except for the bottom two layers. 
This choice of vdW-DF-cx is made because it generally has a higher accuracy than PBE for bulk structure \cite{Gharaee2017,PeGrRo20,jewahy20,JPCMreview,Hard2Soft}. 
The resulting clean-surface
description is used as the starting point for essentially all subsequent adsorption studies, except as noted. 
This holds whether considering physisorption as described in regular vdW-DFs (vdW-DF-cx or vdW-DF2-b86r), 
Fig.\ \ref{fig:Physisorption}, site preference for CO chemisorption on transition metals \cite{DefineAHCX,AHBRlaunch},
or when asserting the AHBR($\gamma$)
performance on describing 
H$_2$+Cu(111) DC.

We limit the (already large) computational costs of using AHBR($\gamma$) by keeping a strict focus on computing total-energy differences when possible. We do not (directly) track adsorption-induced 
coordinate relaxations in such RSH vdW-DFs 
because they must be run with a high $k$-point sampling to converge total-energy differences, even for any given set of fixed geometries, as 
documented below. Fortunately, to characterize the classical barrier $E_{B}$, as predicted in a AHBR($\gamma$), it is only necessary to  contrast RSH-vdW-DF result at two fixed geometries. These are the SBH17 reference geometry and that which corresponds to removing the molecule to far above the surface; The latter case is one where
the surface atomic configuration is set 
via above-summarized vdW-DF-cx studies of the clean Cu(111) surface. 

As a cross-check, motivated in Section IV,
we also consider CO chemisorption across different transition metals, seeking 
AHBR predictions of the site-preference
energy, denoted $\Delta E_{\rm CO}^{\rm M}$ (M=Cu,Ag,Au,Pt). These $\Delta E_{\rm CO}^{\rm M}$ values are defined as
the difference of the total energy 
for CO adsorption on top of a metal atom
and that for a hollow site at the surfaces \cite{Feibelman01p4018}. Here, adsorption-induced relaxations
are important and I track these using vdW-DF-cx before applying AHBR to compute the resulting total-energy differences 
for the $\Delta E_{\rm CO}^{\rm M}$ set of values. I note that the previous papers launching the two RSH vdW-DFs, denoted AHCX and AHBR \cite{DefineAHCX,AHBRlaunch}, provide estimates,
$\Delta E_{\rm CO}^{\rm M} < 0$, that are
consistent with experimental observations \cite{DefineAHCX,AHBRlaunch}. However, AHCX results
for noble-metal cases do not suffice for the desired discussion of AHBR robustness. 
Also, these previous results are based (in three cases) on the use of less electron-rich PPs 
without the present focus on seeking a high convergence with the $k$-point sampling, see below. 

For all molecular adsorption studies we choose to work with a $2 \times 2$ in-surface repetition of the above-summarized basic slab modeling. This representation implies that we assume a 25\% coverage for actual adsorption
problems, and implicitly assumes that this 
lateral separation between repeated 
images suffice to characterize the
dissociation barriers as concerning
independent-H$_2$ (reaction-or-scattering)
events. Even this modeling choice is  motivated by the
high computation costs of RSH vdW-DF2 
studies performed with the necessary 
$k$-point sampling; For consistency, 
the same choice of unit cells
is also made for characterizations 
of the adsorption potential, as described within 
the non-hybrid vdW-DFs, Fig.\ \ref{fig:Physisorption}.

\begin{figure}
    \centering
\includegraphics[width=0.95\linewidth]{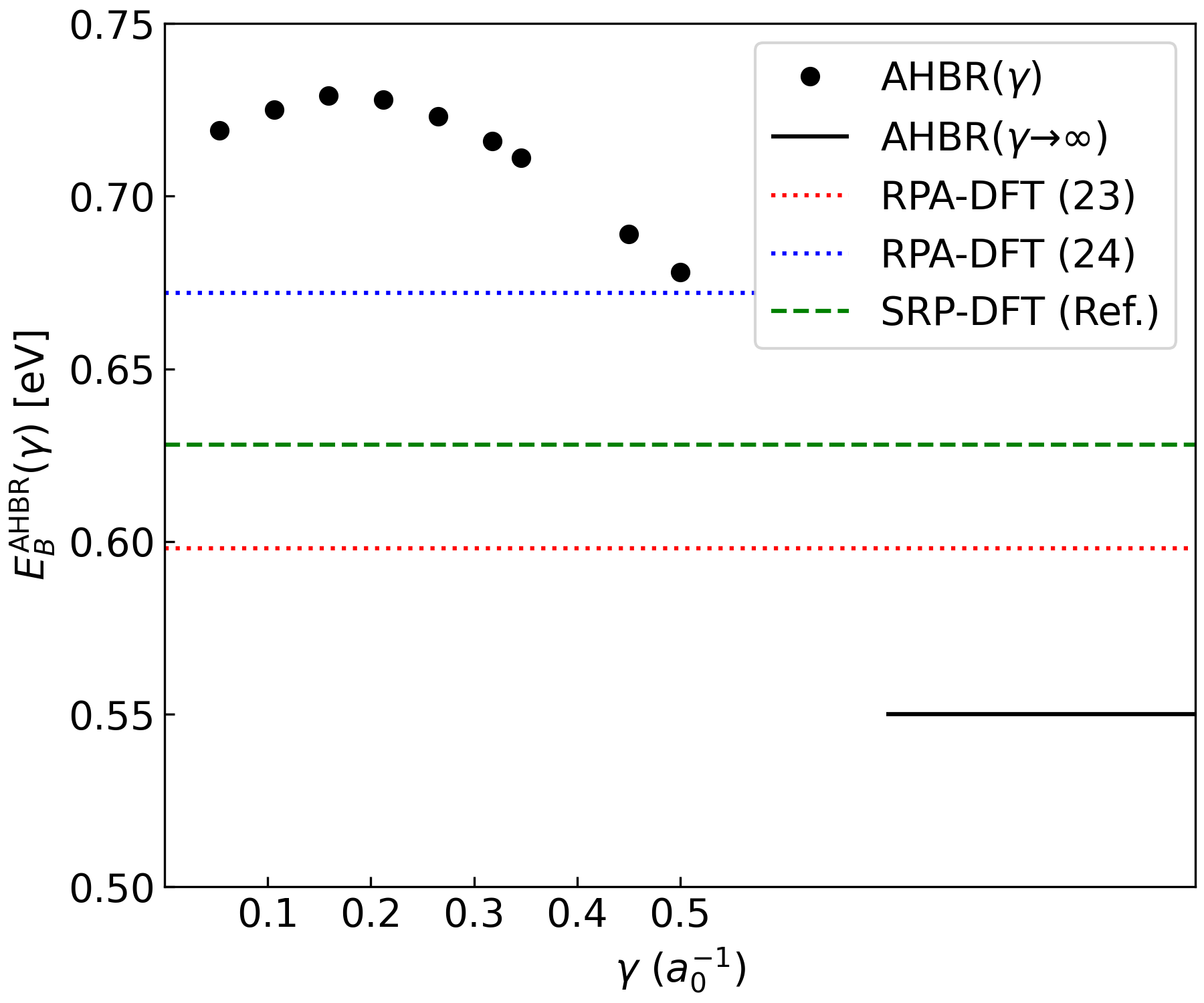}
\caption{Variation in performance of AHBR, as a function of inverse length scale $\gamma$, on description of barrier for 
H$_2$+Cu(111) DC. 
Here I take as a reference (green dashed line) the result $E_B^{\rm SRP} = 0.628$ eV obtained by SRP-DFT \cite{Diaz2009}, 
I also show 
the results obtained by two studies \cite{Wei2023,Oudot2024} based on 
the RPA to the ACF for the exact XC energy \cite{lape77}, i.e.,  
RPA DFT results \cite{Wei2023,Oudot2024}. The here-introduced, non-empirical, QP-guided AHBR($\gamma^*=0.5\, a_0^{-1}$) descriptor
predicts an $E_B^{\rm AHBR}(\gamma^*)=0.678$ eV barrier height.
}
\label{fig:PerformanceBarrier}
\end{figure}

\begin{table}
\caption{\label{tab:BarrierAndQPs}
AHBR-mRSH($\gamma$) and AHBR($\gamma$) characterizations of 
small-molecule fundamental gaps $\Delta_{\rm g}^{\rm mSRH}$ (top section) and AHBR characterizations of the barrier energy $\Delta_{\rm B}E_{{\rm H}_2}^{\rm Cu}$ for H$_2$ dissociation on Cu(111) (bottom section). All energies in eV.  Highlighted entries represent 
molecular descriptions that reflect completion of an OT process
that yields a molecule-specific setting of a $\gamma^*$ value, that
permits the (molecule-optimized) AHBR-mRSH($\gamma^*$) forms to serve as 
accurate predictors of QP levels, Appendix A. For comparison, the reference value for H$_2$-Cu(111) DC is $E_B^{\rm SRP} = 0.628$ eV; Barrier predictions by the regular-vdW-DF-b86r 
[corresponding to AHBR($\gamma\to \infty$)] and by AHBR($\gamma=0.053\, a_0^{-1}$) are instead 0.550 eV and 0.719 eV, respectively.}
\begin{tabular}{l|ccccccc}
\hline
$\gamma$ ($a_0^{-1}$)  
& 0.106 & 0.159 & 0.212 & 0.265 & 0.318 & 0.346 & 0.450 \\
\hline
$\Delta_{\rm B}E_{{\rm H}_2-{\rm Cu}}^{\rm AHBR}$ & 0.725 & 0.729 & 0.728 
       & 0.723 & 0.716 & 0.711 & 0.689  \\
\hline 
N$_2$$\Delta_{\rm g}^{\rm mRSH}$ &  - & 14.23 & 14.79 
       & 15.29 & 15.74 & 15.96 & 16.63 \\
O$\Delta_{\rm g}^{\rm mRSH}$ &  7.51 & 8.58 & 9.56 
       & 10.43 & 11.19 & 11.56 & 12.70 \\      
H$_2$O$\Delta_{\rm g}^{\rm mRSH}$ & - & 10.98 & 11.53 
       & 12.02 & 12.44 & \textbf{12.64} & 13.27 \\
O$_2$-$\Delta_{\rm g}^{\rm mRSH}$ & 8.56 & 9.60 & 10.56 
       & 11.40 & 12.13 & \textbf{12.47} & 13.53 \\
OH$\Delta_{\rm g}^{\rm mRSH}$ & 7.29 & 8.37 & 9.34 
       & 10.21 & 10.97 & 11.33 & 12.46\\
HF$\Delta_{\rm g}^{\rm mRSH}$ & - & 13.75  
       & 14.33 & 14.85 & 15.33 & 15.56 & -16.30\\       
\hline
CO$\Delta_{\rm g}^{\rm mRSH}$ & 12.14 & 12.73 & 13.26 
       & 13.74 & 14.15 & \textbf{14.34} & 14.93 \\
CO$\Delta_{\rm g}^{\rm AHBR}$ & 9.27 & - & 8.74
       & - & 8.18 & 8.09 & 7.81 \\       
H$_2$$\Delta_{\rm g}^{\rm mRSH}$ & 13.35 & 13.93 & 14.46 
       & 14.93 & 15.34 & 15.53 & 16.10 \\
H$_2$$\Delta_{\rm g}^{\rm AHBR}$ & 11.71 & 11.55 & 11.33 
       & 11.18 & 11.04 & 10.98 & 10.79 \\ 
\hline
\hline
\end{tabular}
\end{table}

I note that the SRP-DFT characterization 
that defines the SBH17 H$_2$-Cu(111)
entry is obtained within a 3$\times$3 
surface representation, i.e., having 
a larger separation between the 
repeated images of the H$_2$ molecules.
However, even though the relevant H$_2$ geometry has the molecular axis parallel to the surface (with also a slightly stretched molecular bond length), the lateral molecular extension remains limited, see top right panel of Fig.\ \ref{fig:Physisorption}. The present use of 
a smaller in-surface extension is not 
expected to impact the comparison: I find 
no difference between dissociation-barrier predictions obtained for a 3$\times$3 and 2$\times$2 surface modeling, when the assessment is instead pursued using the (closely related) vdW-DF2-b86r XC functional. 

Similarly, use of a 2$\times$2 surface unit cell is also motivated for a set of CO chemisorption studies. Here the molecular axis is instead aligned with the surface normal and there is again a good lateral separation between repeated images of the CO molecules in our plane-wave-DFT descriptions. Also, these studies
are included to assert the AHBR ability to capture details of the charge transfer processes. The net rate is  expected to be small given
the low $E_{\rm CT} \approx 4$ eV value for CO-Cu(111) \cite{SurfChallenge2020}. However, the small rate of net charge transfer should be understood
as arising from a partial cancellation of forward- and back-donation processes \cite{Blyholder}. At an assumed 25\% coverage, the experimental observations are consistent findings of a room-temperature preference for top-site adsorption across the set of Cu(111), Ag(111), Au(111), and Pt(111) surfaces. This is the \textit{qualitative gauge} by which I also seek to assert the AHBR performance, i.e., my generic choice of using a 2$\times$2 cell for surface modeling suffices also for my
CO-adsorption tests.

Tables \ref{tab:H2atCu} and \ref{tab:COsite} summarize the work to here converge
AHBR [and by extension AHBR($\gamma$)]
performance on adsorption challenges, as described within the $2\times 2$ surface-cell modeling. The tables concern the core challenge, namely converging these RSH vdW-DF studies for metal
surfaces with respect to 1) the scope 
(denoted `$k$') of the overall $k\times k\times 1$ sampling across the Brillouin 
zone, and with respect to 2) the scope  
(denoted $q$) of an additional
$q\times q\times 1$ grid (of $k$-point differences) that permits completion of the
Fock-exchange descriptions in the generalized KS-DFT studies \cite{GKSstart,linACE,PaoloElStruct1,AHBRmRSH25}.
Tables \ref{tab:H2atCu} and \ref{tab:COsite} document that a choice of $k=12$ and $q=4$
is sufficient to converge AHBR predictions of the H$_2$-Cu(111) dissociation  barrier and of the CO site preference descriptions,
respectively. These converged descriptions
are referred to as AHBR-12-4, when relevant;  Below a $k=12, q=4$ scope of
convergence for RSH vdW-DFs ($k=12$ scope
for regular vdW-DFs and PBE) is systematically 
implied for metal-surface
studies, except where noted.

The molecule-optimized AHBR-mRSH($\gamma$) studies 
are instead pursued with a single $k$- \mbox{(and therefore also $q$-)}{\allowbreak}point sampling,
in a cubic unit cell of 30 {\AA} side 
length. We furthermore use a 
Makov-Payne electrostatic 
decoupling between repeated images
for the descriptions of the
of set of molecules that are (except
for H$_2$) included in the G21IP benchmark set, described in Ref.\ \cite{gmtkn55}. 
This electrostatic decoupling is 
motivated not only because some of 
molecules have dipole moments, but 
especially because we have to compute 
total energies of charge systems
as an inherent part of the OT process, 
Appendix A. The choice of unit-cell size is 
larger than that used in an earlier study, 
where we documented that  we have converged 
errors (introduced by spurious couplings
among repeated images) to less than 
0.01 kcal/mol, for GMPTKN55 benchmark sets 
\cite{AHBRlaunch}.

\section{Results and discussion}

Figure \ref{fig:PerformanceBarrier} summarizes the variation in the AHBR($\gamma$) predictions for the H$_2$ dissociation barrier with $\gamma$ (black dots), comparing with the SRP-DFT reference value (green dashed line). Figure \ref{fig:PerformanceBarrier} also reports the corresponding AHBR($\gamma\to \infty$) prediction, short solid black line at right-hand side. This value equals the result of the vdW-DF2-b86r barrier characterization \cite{AHBRmRSH25}.
The data for these new AHBR($\gamma$) characterizations 
are collected in Table \ref{tab:BarrierAndQPs}.
I note that the set of here-employed AHBR($\gamma \leq \gamma''$) 
forms at best approach chemical accuracy with respect to the barrier-energy target defined by the SRP-DFT modeling for 
H$_2$+Cu(111) DC. The default AHBR 
and vdW-DF2-b86r XC functionals are 
about 2 kcal/mol too large and too small,
respectively. In contrast, as indicated 
in Fig.\ \ref{fig:PerformanceBarrier}, 
there exist two literature RPA-DFT results that characterize the
H$_2$+Cu(111) DC barrier energy, $E_B^{\rm SRP}$ to within chemical accuracy \cite{Wei2023,Oudot2024}. RPA-DFT studies do, however, require substantially more computationally resources than regular and
hybrid vdW-DFs like the AHBR($\gamma$)'s.

Figure \ref{fig:PerformanceBarrier} and Table \ref{tab:BarrierAndQPs} suggest potential use of a new merge-type approach that adapts the SRP-DFT logic. 
That is, one could model the H$_2$+Cu(111) DC problem based on taking a 45\% inclusion of AHBR and taking the rest of the merged-XC form from vdW-DF2-b86r; This mixing brings alignment with 
the established $E_B^{\rm SRP}$ value for H$_2$+Cu(111) DC and it is ready for practical use by simply employing the mixed-XC call
DFT-code adjustment that allows practical SRP-DFT studies \cite{Diaz2009,catalysisvdW15,KroesDC2021,SBH17}. 
Use of this new-SRP-DFT candidate would raise the
computational costs over the existing SRP-DFT
approach. Also, it should still be seen as an empirical, system-specific (energy-landscape) descriptor. This follows because it uses the goal, the $E_B^{\rm SRP}$ value, to pick the extent of mixing in the merger-XC form.

Moving towards the above-defined new-SRP-DFT approach 
presents an opportunity: The move makes it meaningful to 
extract and use predictions of corresponding QPs
\cite{GKSstart,Baerends2003,kronik2012,OTRSHadsorb17,WiOhHa21,ChiDFT23,GoGaOh2024,AHBRmRSH25,NitrogenBasesAHBR-mRSH26,hBN2026}. 
The key observation is that identified "mixing-XC" form can still
be seen as  \textit{one MBPT-guided XC functional\/} within the vdW-DF family: It remains just a regular hybrid vdW-DF 
member of the wider AHBR-gRSH($\alpha,\beta,\gamma$). 
Specifically, it emerges when setting $\alpha\approx 0.1$, 
$\gamma=\gamma_0$ (as in the default AHBR) and 
$\beta=-\alpha$ (as relevant for the present 
focus on a DC problem); This identification
follows directly from the discussion of how
variations in the `($\alpha,\beta,\gamma$)' 
parameter set merely reflect assumptions of the 
extent of which we choose to screen Fock exchange by mixing in the
B86R exchange-hole model, at various length 
scales \cite{kronik2012,AHBRlaunch,AHBRmRSH25}. 
Of course, a RSH (vdW-DF) form  set by a low (SR) Fock mixing (0.1) is 
unusual \cite{Gorling93,Burke97,JiScHy18b,WiOhHa21,GoGaOh2024}; 
This is true even if the new-SRP-DFT form is compatible with the presence of the metal substrate. 
However, there is a clarification about the 
MBPT foundation for such a new SRP DC modeling: 
Wherever the mixing lands, we are working with just one framework. 
We stay with one correlation form, defined by a Quantum-Monte Carlo \cite{Perdew_1992:accurate_simple} and by the 
current-serving $E_c^{\rm nl2}$ form \cite{Dion,lee10p081101,hybesc14,JPCMreview}, and by 
exchanges that all originate from an
analytical-hole model of the B86R-exchange nature \cite{hamada14,DefineAHCX,AHBRlaunch,AHBRmRSH25}, and we can view 
the XC specification in terms of QP modeling \cite{FW7,helujpc1971,JPCMreview,ChiDFT23,hBN2026}. 
To employ such QP insight in this modeling, I extract and use (adsorbate) QP predictions by AHBR-mRSH($\gamma$), 
upon a non-empirical OT process
that compares HOMO-level and adiabatic-IP predictions, Appendix A and Refs.\ \cite{OTRSHalga,OTRSHadsorb17}.

The implied idea is that we may get coincidental
QP info out from what is a fair description of the
H$_2$+Cu(111) DC barrier-height description. To make
the idea practical, I choose to bypass an actual
use of the above-summarized simple-merger idea. Instead I
extract a closely-related observation from
the data in Table \ref{tab:BarrierAndQPs} and from Fig.\ \ref{fig:PerformanceBarrier}. Specifically, since AHBR [corresponding to AHBR($\gamma_0$)]
and vdW-DF2-b86r [corresponding to AHBR($\gamma \to \infty$)]
sit on different sides of the SRP-DFT values, there exists
some $\gamma'$ value for which AHBR($\gamma'$) gives an
$E_B^{\rm AHBR'}$ prediction that aligns with $E_B^{\rm SRP}$.
More studies would be required to find that $\gamma'$ value
but by the original SRP-DFT modeling logic \cite{Diaz2009},
we can expect that the resulting AHBR' form would be fairly
represented by the above-stated AHBR/vdW-DF2-b86r merger.
That said, the resulting AHBR' form is still an empirical
construction (set by the modeling goal), at this stage of discussions. Also, since I must restrict the
OT process to $\gamma \leq \gamma'' =
0.5\, a_0^{-1}$, as motivated in Sec.\ II.B,
Fig.\ \ref{fig:PerformanceBarrier} makes it clear 
that the corresponding $E_B^{\rm AHBR}(\gamma^*$)
prediction only approaches $E_B^{\rm SRP}$ with
near-chemical accuracy.

Nevertheless, the above-stated analysis provides contact with
the OT-based description of QPs (for molecules and 
adsorption \cite{OTRSHadsorb17}): It allows identification 
of a plausible best-nonempirical AHBR($\gamma^*$) descriptor
for H$_2$+Cu(111) DC, within the present scope defined by
Refs.\ \cite{lee10p081101,hamada14,AHBRlaunch,AHBRmRSH25}.
Appendix A shows that a choice of 
$\gamma^*=0.346\, a_0^{-1}$ represents a good OT value 
for molecular-QP determinations across small 
molecules CO, O$_2$, P$_2$, and H$_2$O as long as we
work with the corresponding AHBR-mRSH($\gamma$) forms 
\cite{AHBRmRSH25}. It also shows
that a somewhat larger $\gamma^*$ choice is motivated in the OT process for N$_2$ and especially for the tiny H$_2$ molecule, where $\gamma^*\approx \gamma'' = 0.5\, a_0^{-1}$, i.e., right at the edge of what can
be motivated when seeking a connection of 
AHBR and AHBR-mRSH characterizations, Sec.\ II. B.

More broadly, Appendix A details the OT process for smaller molecules and shows
that the OT-AHBR-mRSH($\gamma^*$) success (for the larger nitrogen-base systems \cite{NitrogenBasesAHBR-mRSH26})
is repeated for the here-relevant small molecule cases. 
There is generally strong
agreement, at and below the HOMO level, of the here-predicted QP orbital levels with the set of True-KS levels previously
obtained by Baerends and co-workers \cite{Baerends2002,Baerends2003}
and also with experimental observations collected in Ref.\ \cite{Baerends2003}. 
I note that the set of True-KS 
results started with molecular densities computed in CI
but one must adjust the exchange potential 
to get an asymptotic $-1/r$ form that reflects the
unscreened Fock-exchange behavior \cite{AHBRmRSH25}. This cross-over is, as noted
in Sec.\ II.B, what a switch from AHBR($\gamma$) to
AHBR-mRSH($\gamma$) automatically accomplishes within
the presently explored AHBR-based RSH vdW-DF set. 

I add that the alignment of the here-reported AHBR-mRSH($\gamma^*$)
small-molecule QP predictions and of the Baerends earlier true-KS
characterizations \cite{Baerends2003} is expected; The 
close alignments that is documented in Appendix can therefore also 
be seen as a validation of using the AHBR-mRSH($\gamma^*$) forms for 
fast QP predictions in isolated molecules, upon OT. There should be a
good or even excellent agreement on the resulting QP predictions
because there is no large difference between  GGA- 
and non-hybrid vdW-DF-based descriptions of small 
molecules and there exists a demonstration
that PBE contains an excellent QP content for
small molecules \cite{ChiDFT23}. Refs.\ \cite{Baerends2002,Baerends2003,AHBRmRSH25,NitrogenBasesAHBR-mRSH26},
and now also this paper, are helping to make the observation
that KS-DFT has an underlying QP content \cite{helujpc1971,ChiDFT23}
that we can try to use.

The present suggestion is to rely on an OT process for setting isolated-molecule QPs, Appendix A and Ref.\ \cite{OTRSHadsorb17}, to permit the identification of a best-possible nonempirical AHBR($\gamma^* \approx\gamma''$) for H$_2$+Cu(111) DC. The suggestion needs arguments and an independent
testing, as summarized, in part, in Tables \ref{tab:BarrierAndQPs} 
and \ref{tab:COsite}. 

First,  I observe that with the implied approach (that requires
enforcement of the $\gamma''$ limit) one retains the standard $\alpha_0=0.25$ value, with which
AHBR-type descriptions are known to often work for bulk structure \cite{AHBRlaunch}. Meanwhile, Ref.\ \cite{AHBRmRSH25} shows that the performance of AHBR-mRSH($\gamma$), as a molecular-energy descriptor, has only a 
soft $\gamma$ dependence for broad types of molecular problems
\cite{AHBRmRSH25}, as tested on the GMTKN55 benchmark suite \cite{gmtkn55}. Also, on the small-molecule side of the problem, any  $\gamma \leq \gamma'' = 0.5\, a_0^{-1}$ choice means that whether the molecule is described with
AHBR($\gamma$) or AHBR-mRSH($\gamma$), its energy behavior
is still effectively described as in the vdW-DF2-br0 simple 
hybrid \cite{lee10p081101,hamada14,DFcx02017} (that is again set by a fixed $\alpha_0=0.25$ mixing).
This simple hybrid is documented to have a strong performance on describing general molecular energy differences and on molecular transition states \cite{AHBRlaunch}. 

The key observations for allowing the suggested 
use of AHBR-mRSH OT to set the AHBR($\gamma^*$) pick for DC studies [subject to respecting a 
$\gamma'' \leq 0.5\, a_0^{-1}$ limit for H$_2$]
are as follows. First the H$_2$ has a highly concentrated electron distribution. Second,
the exchange energy is dominated 
by the SR-exchange component of the general 
RSH vdW-DF design, whenever we can ignore 
the error-function weight on the Fock-exchange
part, i.e., when `$\gamma y$' is small in Eqs.\ (\ref{eq:ExEnergyForm}) and (\ref{eq:SplitCoul}). 
The connection of AHBR/AHBR-mRSH exchange-energy descriptions is possible if almost all electrons and almost all holes reside within
some maximum separation, $y_{\rm max} \sim 1/\gamma$.
This is a criterion on the OT process for H$_2$-adsorbate problems that we respect in this paper.

The middle section of Table \ref{tab:BarrierAndQPs} summarizes
the nature of the OT tuning for AHBR-mRSH (appendix A) in terms of
the descriptions of the HOMO-LUMO gaps. These characterizations
are provided as a function of $\gamma$; Bold-faced entries 
identify small-molecule cases where the Appendix-A generic setting (of one overall or average $\gamma^*$ pick) also corresponds to a complete OT specification in the specific case. The fact that there is no large scatter for the set of molecule-specific OT $\gamma^*$ results suggests that there is an overall
robustness, at least as applied to these small molecules in isolation. I note that
given the use of plane-wave DFT, the description of LUMO levels
with a negative EA is inherently challenging. For the larger
nitrogen bases, the OT-AHBR-mRSH form does provide a good description
of even weakly bound (and weakly-unbound) LUMO states and resonances \cite{AHBRmRSH25,NitrogenBasesAHBR-mRSH26}. That is not possible for present cases, especially for the tiny H$_2$ molecule
(that has a large negative EA value). 

The bottom part of Table \ref{tab:BarrierAndQPs} contrasts
the AHBR($\gamma$) and AHBR-mRSH($\gamma$) predictions of the
HOMO-LUMO (or fundamental) gap in isolated CO and in isolated
H$_2$ molecules. The table shows that while descriptions 
of these gaps differ, the AHBR($\gamma$) nevertheless 
maintains a mechanism (a large HOMO-LUMO gap)  to suppress potentially spurious charge transfer 
onto these molecules, even at the $\gamma^*=\gamma''=0.5\, a_0^{-1}$ value 
that is identified as the OT-set descriptor
for H$_2$ QPs, Appendix A. 
The data shows that 
AHBR($\gamma$) predictions of the HOMO energy 
position is not reflecting the adiabatic IP 
for neither H$_2$ nor CO. However, 
use of AHBR($\gamma$) still yields a set of QP-level predictions that are fairly insensitive to 
the $\gamma$ tuning; The use of AHBR($\gamma$) will still, generally, suppress most spurious charge transfers.

Table \ref{tab:COsite} represents the results of a different type
of more \textit{in situ}, chemisorption-type testing of the suggested
non-empirical modeling. 
Specifically, Table \ref{tab:COsite} reports AHBR-based predictions
of the so-called site-energy preference \cite{DefineAHCX}
for CO adsorption on a set of four metal M(111) surfaces, namely
given by M=Cu, Ag, Au, and Pt. The individual site-preference energy, denoted $\Delta E_{\rm CO}^M$, 
is defined as the total energy difference between having the CO sitting at the top and at the hollow site of the indicated 
transition metal surface. The $\Delta E_{\rm CO}^M$ values
should all be negative, in fact, clearly smaller that minus 
the energy (0.026 eV) corresponding to room temperature. This 
is the primary \textit{qualitative\/} observation
that can be extracted from observations and it holds, 
when (as done in the present modeling) there is at 
most 25\% coverage. This target of chemisorption
modeling is discussed in Ref.\ \cite{DefineAHCX} as part of the introduction of an earlier RSH vdW-DF abbreviated AHCX (and instead crafted off an analytical-hole modeling of the exchange description in 
the vdW-DF-cx version \cite{behy14,Thonhauser_2015:spin_signature}).
I also note in passing that for the case of CO on Pt(111), the present site-preference energy-result supersedes 
that reported in the original AHBR paper \cite{AHBRlaunch} but as the Table shows, the impact is just quantitative.

Importantly, Table \ref{tab:COsite}
shows that use of AHBR permits characterizations
that are systematically in agreement with the 
above-discussed qualitative observations for CO
adsorption. This is indicated by the set of 
bold-face entries. I observe that this characterization is provided with just the
default AHBR implementation in QE \cite{AHBRlaunch}.
However, Table \ref{tab:BarrierAndQPs} and Fig.\ 
\ref{fig:PerformanceBarrier} suggest 
that there should be no dramatic impacts on 
the energy descriptions with moderate changes in the  setting of $\gamma$. 

I consider the successful passing of also 
this \textit{in-situ} chemisorption
testing as important for motivating use of 
AHBR($\gamma^*$) as non-empirical DC descriptor.
The idea for using such a QP-guided 
functional design would in the CO case
amount to setting $\gamma=0.346\, a_0^{-1}$,
Appendix A. This value is also a good average 
$\bar{\gamma^*}$ of the OT $\gamma^*$ settings 
for small molecules. Using that average 
$\bar{\gamma^*}=0.346\, a_0^{-1}$ 
(the $\gamma^*=\gamma''$ value relevant for H$_2$ OT)
in AHBR($\gamma$) yields a good (very good) DC prediction,
even if we may then only approach chemical accuracy, Fig.\ \ref{fig:PerformanceBarrier}. Equally important, however, is 
the expectation, from having the QP and MBPT connection, 
that use of such non-empirical (yet adsorbate specific) 
AHBR($\gamma^*$) may deliver robust descriptions of 
adsorbate-surface charge transfer in chemisorption. 
In fact, a key aspect of the Blyholder model for CO adsorption \cite{Blyholder} is the emphasis on getting both outbound and inbound
charge transfers right. Predicting the CO-chemisorption site presence 
has been a long-standing challenge for DFT, and so it is encouraging 
that AHBR has enough transferability to be right across the set of four here-investigated metal surfaces for CO adsorption.

I finally note that Table \ref{tab:COsite} reports results that are completed in the presence of actual charge
transfer instead of merely tracking of the potential for charge transfer. The latter concerns the here-explored idea 
to use (adsorbate-)QP characteristics \cite{OTRSHadsorb17} to finalize the setting of a non-empirical XC 
AHBR($\gamma^*)$ for DC modeling, in the H$_2$+Cu(111) case.  
Success at the former (as in the CO-adsorption cases) rests instead on
being systematically accurate at the cases where the
electron distributions reflect actual adsorption-induced changes. As such,
the CO-adsorption problems represent what I consider a cross-testing of the consistency the present 
overall paper suggestion for QP-guided XC setting and DC modeling.

\begin{table}
\caption{\label{tab:QPtuning}
Optimal tuning (OT) details for a class of relevant small-molecule adsorbates that are also in G21IP. 
The table tracks and compares AHBR-mRSH($\gamma$) characterizations of both the adiabatic IP and the HOMO energy level
positions $\epsilon_H$, seeking a $\gamma^*$ value where we have alignment, i.e., have completed the OT process \cite{OTRSHalga}. 
This tuning to the adiabatic IP is here done at a molecule-per-molecule basis. 
Italicized entries indicate cases where the OT process completed within the listed scope giving OT specifications $\gamma^*$; For HF the OT value is instead even larger, $\gamma^*=0.450\, a_0^{-1}$. 
The boldface label on the varying $\gamma$ values highlights that there exists a
workable average $\bar{\gamma^*}=0.346\, a_0^{-1}$ for approximating OT on the indicated set of molecules. By doing so, however, one has to tolerate incomplete alignment, for  CH$_4$, NH$_3$, P$_2$, N$_2$, OH, and most clearly for HF.  All energies in eV.
}
\begin{tabular}{l|ccccccc}
\hline
$\gamma$ ($a_0^{-1}$)  
& 0.106 & 0.166 & 0.226 & 0.286 & 0.316 & \textbf{0.346} & 0.406 \\
\hline 
CH$_4$ IP &  12.72 & \textit{12.77} & 12.83 
       & - & - & - & - \\
CH$_4$ $-\epsilon_H$ &  12.32 & \textit{12.95} &  13.51
       & 13.98 & 14.18  & 14.37 & 14.67 \\
NH$_3$ IP &  10.17 & - & \textit{10.24} 
       & - & 10.27 & 10.27 & - \\
NH$_3$ $-\epsilon_H$ &  9.05 & 9.61 &  \textit{10.24}
       & 10.71 & 10.91  & 11.10 & - \\ 
N$_2$ IP &  15.93 & 16.01 & 16.01 
       & 16.13 & 16.27 & 16.34 & \textit{16.53} \\
N$_2$ $-\epsilon_H$ & 13.63 & 14.29 & 14.92 
       & 15.47 & 15.72 & 15.95 & \textit{16.36}
       \\      
P$_2$ IP &  - & - & - 
       & \textit{10.47} & 10.47 & 10.46 & - \\
P$_2$ $-\epsilon_H$ & 9.27 & 9.80 &  10.19
       & \textit{10.47} & 10.57 & 10.66 & - \\  
CO IP &  14.16 & 14.20 & 14.29 
       & 14.36 & 14.39 & \textit{14.42} & 14.47 \\
CO $-\epsilon_H$ &  12.14 & 12.79 & 13.38 
       & 13.90 & 14.12 & \textit{14.33} & 14.69 \\  
O$_2$ IP &  12.63 & 12.67 & 12.72 
       & 12.78 & 12.81 & \textit{12.83} & 12.89 \\
O$_2$ $-\epsilon_H$ &  10.47 & 11.14 & 11.76 
       & 12.32 & 12.58 & \textit{12.82} & 13.27 \\   
H$_2$O IP &  - & - & - 
       & - & 12.76 & \textit{12.77} & - \\
H$_2$O $-\epsilon_H$ & 9.27 & 9.80 &  10.19
       & 10.47 & 12.57  & \textit{12.66} & 13.04 \\ 
OH IP & 13.18 & - & - 
       & - & 13.30 & 13.31 & \textit{13.33} \\
OH $-\epsilon_H$ & 10.55 & 11.21 &  11.81
       & 12.35 & 12.59  & 12.81 & \textit{13.21} \\ 
PH IP &  10.33 & - & - 
       & - & 10.46 & 10.47 & \textit{10.49} \\
PH $-\epsilon_H$ &  8.62 & 9.20 &  9.68
       & 10.05 & 10.21  & 10.35 & \textit{10.57} \\ 
HF IP &  16.17 & - & - 
       & - & 16.27 & 16.31 & 16.34 \\
HF $-\epsilon_H$ &  13.19 & 13.86 &  14.49
       & 15.06 & 15.33  & 15.57 & 16.02 \\ 
\hline
\hline
\end{tabular}
\end{table}

\begin{figure}
    \centering
\includegraphics[width=0.95\linewidth]{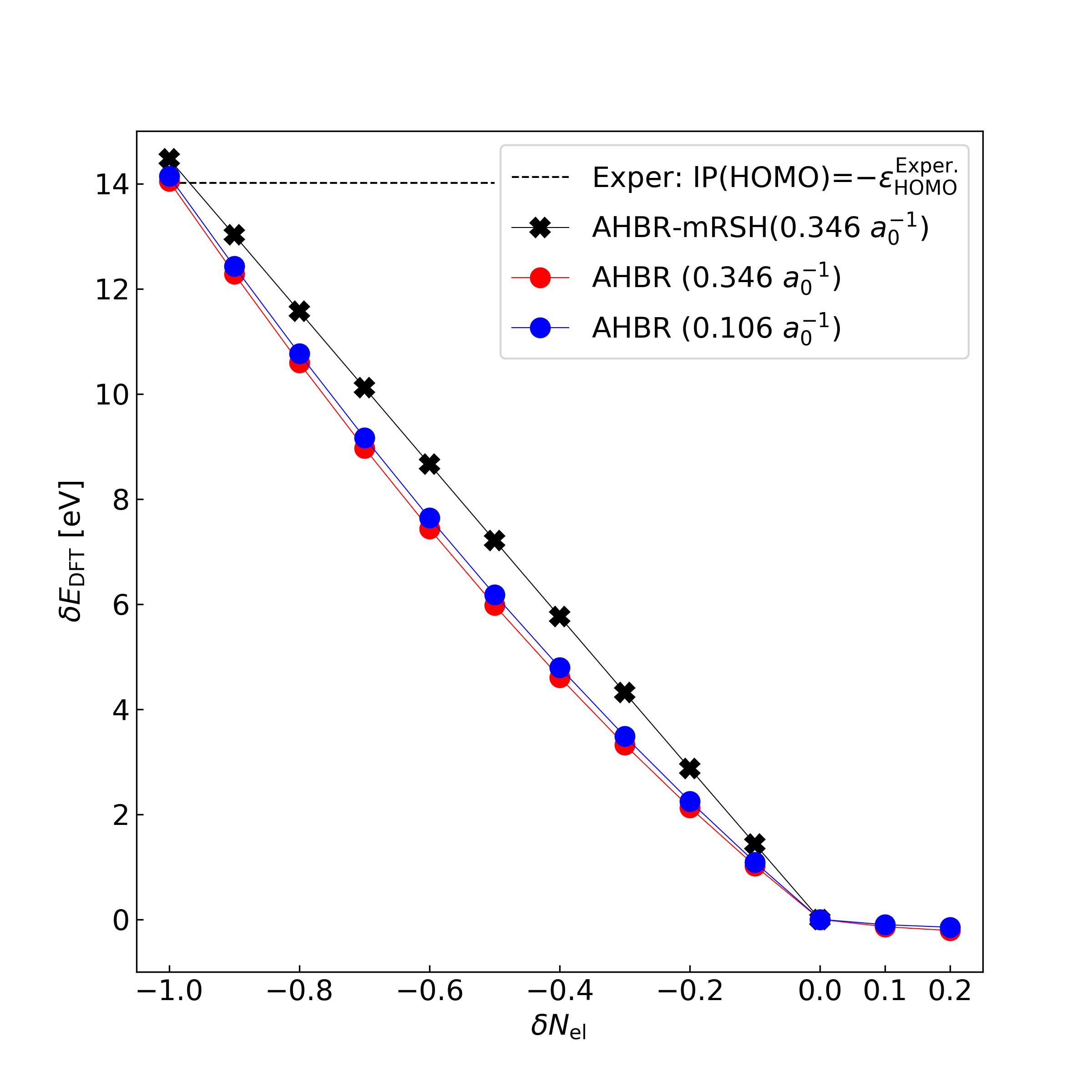} 
\caption{Total-internal energy variation with increasing partial occupation of the CO molecule, at described in the OT AHBR-mRSH($\gamma^*=0.346\, a_0^{-1}$) form
and in AHBR($\gamma^*$). The value of $\delta N_{\rm el}=0$ (=1) corresponds
to the neutral (singly charged) CO molecule, as measured relative to the 
GS as described in these generalized KS DFT forms.
}
\label{fig:COpiecewise}
\end{figure}
\begin{table}
\caption{\label{tab:QPperform}
QP level descriptions by AHBR-mRSH($\bar{\gamma^*}$) 
that is defined by a good average value, $\bar{\gamma^*}=0.346\, a_0^{-1}$, of  individual-small-molecule OT processes reported in Table \ref{tab:QPtuning}; Note that the average does not reflect the OT process for the very small molecule H$_2$ molecule. All energies are in eV.
The AHBR-mRSH($\bar{\gamma^*}$) performance is compared with the CI-based analysis 
of so-called `True KS level' for molecules as extracted upon adjusting a GGA-exchange potential for both derivative-discontinuity effects
and to ensure an unscreened asymptotic form, $-1/r$ \cite{Baerends2003}
and to a set of experimental results collected in that CI-based, asymptotically-corrected analysis  \cite{Baerends2003}. 
I note that roll-over to the implied Fock-type exchange behavior at asymptotic distances between electrons and associated holes is automatic in the here explored AHBR-mRSH($\gamma$) set of RSH vdW-DFs. The alignment
is good except for the case of the tiny H$_2$ molecules, which needs a separate OT analysis, see Table \ref{tab:H2tune}.
}
\begin{tabular}{l|ccc|c}
\hline
 
System & AHBR-mRSH($\bar{\gamma^*}$) & Exper. & True-GGA & Orbital\\
\hline 
CO &  0.007 & - & - 
       & LUMO \\
 & -14.33 & -14.01 & -14.01 & HOMO
       \\   
 & -16.97 & -16.91 & -16.80 & -1, -2   \\   & -20.02 & -19.72 & -19.37 & -3   \\
 & -39.96 & -38.3 & -34.70 & -4   \\
\hline       
N$_2$ &  0.007 & - & - 
       & LUMO \\
 & -15.95 & -15.58 & -15.57 & HOMO
       \\   
 & -16.67 & -16.93 & -16.38 & -1, -2   \\   & -19.34 & -18.75 & -18.77 & -3   \\
 & -35.94 & -37.3 & -33.69 & -4   \\
 \hline       
P$_2$ &  -0.7819 & - & - 
       & LUMO \\
 & -10.66 & -10.65 & -10.65 & HOMO, -1
       \\   
 & -11.27 & -10.84 & -10.91 &  -2  
 \\   & -16.36 & - & -14.95 & -3   \\
 & -23.03 & - & -20.53 & -4   \\
\hline       
H$_2$O &  -0.015 & - & - 
       & LUMO \\
 & -12.66 & -12.62 & -12.62 & HOMO
       \\   
 & -14.67 & -14.74 & -14.73 & -1   \\   
 & -18.55 & -18.55 & -18.33 & -2   \\
 & -32.66 & -32.2 & -30.72 & -3   \\
 \hline       
HF &  0.002 & - & - 
       & LUMO \\
 & -15.57 & -16.19 & -16.18 & HOMO, -1
       \\      
 & -19.38 & -19.9 & -19.90 & -2   \\
 & -38.03 & -39.7 & -36.77 & -3   \\
\hline
H$_2$ &  -0.014 & - & - 
       & LUMO \\
        & -15.53 & -16.44 & -16.44 & HOMO
       \\      
\hline
\hline
\end{tabular}
\end{table}

\begin{table}
\caption{\label{tab:QP-OTperform}
QP level descriptions by the system-specific OT AHBR-mRSH($\gamma^*$) forms, where relevant because $\gamma^*\neq \bar{\gamma^*}$;
All energies are in eV and again the AHBR-mRSH($\gamma^*$) performance is compared with `True KS level' and 
experimental results  \cite{Baerends2003}.  Comparison with  Table \ref{tab:QPperform} shows that alignment 
with reference values \cite{Baerends2003} tends to improve for deep levels.}
\begin{tabular}{l|ccc|c}
\hline
 
System & AHBR-mRSH($\gamma^*$) & Exper. & True-GGA & Orbital\\
\hline 
N$_2$ & -16.36 & -15.58 & -15.57 & HOMO
       \\   
 & -16.99 & -16.93 & -16.38 & -1, -2   \\   & -19.78 & -18.75 & -18.77 & -3   \\
 & -36.58 & -37.3 & -33.69 & -4   \\
 \hline       
P$_2$ & -10.47 & -10.65 & -10.65 & HOMO, -1
       \\   
 & -10.99 & -10.84 & -10.91 &  -2  
 \\   & -16.00 & - & -14.95 & -3   \\
 & -22.45 & - & -20.53 & -4   \\
\hline           
HF & -16.31 & -16.19 & -16.18 & HOMO, -1
       \\      
 & -19.98 & -19.9 & -19.90 & -2   \\
 & -39.14 & -39.7 & -36.77 & -3   \\
\hline
H$_2$        & -16.30 & -16.44 & -16.44 & HOMO
       \\      
\hline
\hline
\end{tabular}
\end{table}

\section{Summary and outlook}

The paper reports AHBR-based generalized KS DFT predictions of the barrier for H$_2$+Cu(111) DC and for CO chemisorption on metal surfaces seeking also a non-empirical (yet adsorbate-specific) RSH vdW-DF form, denoted AHBR($\gamma^*)$,
for possible future exploration.

On a broad note, this paper suggests use the AHBR and its extensions for a description and discussion of H$_2$+Cu(111) DC for several reasons. First, the underlying vdW-DF2-b86r XC functional has a good performance at accurately describing the (non-dissociate) physi\-sorp\-tion at the Cu(111) surface.
Second, the AHBR is expected
to repair occasional delocalization errors \cite{cococcioni2005,BurkeSIE,davo2010,nguyen2015,colonna2022,AHBRlaunch}, for example, those reflecting incorrect predictions of charge transfers among the components \cite{PePaLe82} in adsorption problems \cite{OTRSHadsorb17,MOFdobpdc,AHBRlaunch}. Importantly, use of the underlying RSH-vdW-DF framework permits some checks on whether we also have a reasonable account of the charge transfer onto the molecule, as revealed in QP characterizations.

The main idea of crafting
and using the here-introduced AHBR($\gamma^*)$ adsorption and DC descriptor follows from the observation: One can
link use of the AHBR($\gamma$) for small-molecule energy descriptions, e.g., in DC and chemisorption predictions with AHBR-mRSH($\gamma$)-based QP predictions. One can therefore also use an QP-motivated OT 
process to set $\gamma^*$ for use in a non-empirical AHBR($\gamma^*$)
DC description. This setting proceeds by using part of the logic presented in Ref.\ \cite{OTRSHadsorb17}, but it is here focused on adsorption-energy
predictions. The resulting AHBR-based
modeling strategy can be motivated as long as the effective extent of molecules remains smaller than a molecule-specific tuning constraint 
$1/\gamma''$, as it can for the case of H$_2$ adsorbates.

In the case of larger adsorbates, it is likely that the basic ideas must instead be adopted for
the more general OT process, perhaps along the lines discussed in Refs.\ \cite{OTRSHadsorb17,WiOhHa21,GoGaOh2024}.
However, it is plausible that the here-introduced approach can work for DC modeling and chemisorption predictions also for some other small-adsorbate cases. This is because I generally find that the values obtained for
$\gamma^*$ via the OT process decrease with increasing adsorbate size.

A key step in any such 
attempt at generalization
to other adsorbates involves completing a $\gamma$-OT process, while focusing
on the molecule. This can be done once and for all, as 
completed in the appendix for typical small molecule adsorbates. 
For actual metal adsorption studies one can then use the set of adsorbate-specific OT settings, $\{\gamma^*\}$, as input
for subsequent regular
AHBR-in-QE studies 
when studying adsorption on metal surfaces. This is possible because an option for practitioners to implement a $\gamma$ resetting is already build into the official AHBR 
release as an XC functional in QE \cite{DefineAHCX,AHBRlaunch}.

Finally, as part of that work, the appendix also documents that upon using this OT process we arrive
at accurate characterizations of
molecular QP levels.
This holds for the frontier HOMO level but also  when it comes to description of deeper occupied levels.
I add that as we complete this OT process we seem to recover piecewise
linearity in the change of total energy with partial charging. This is now documented for both the larger nitrogen bases, 
and here for both the small CO and the tiny H$_2$ molecules. 
This finding suggests a complementarity between the present approach (based on OT of RSH vdW-DF) and the  
use of a corresponding vdW-DF based Koopmans-DFT approach \cite{hBN2026}. 
Specifically, in the latter case one explicitly enforces the linearity and out comes excellent QP predictions. 
In the here explored case, instead, one seeks to  enforce a different IP criterion \cite{LePeSa84,PePaLe82,OTRSHalga,OTRSHadsorb17} that leads to an excellent QP description. 
It is noteworthy that this OT process itself implies restoration of piecewise linearity.

\section*{Acknowledgment}

The author thanks R. Quintero-Monsebaiz and M. Rahm for
discussions of the link between QPs, charge-transfer, electro-negativity, and DFT. 
The author also thanks B. I. Lundqvist, 
E. Schr{\"o}der, and G. Wahnstr{\"o}m for discussions on the inter-connection of vdW forces, surface processes, and modeling of adsorbate dynamics, over  many years. Work supported 
by Swedish Research Council (VR) through Grant No.\ 2022-03277, 
and by the Chalmers Area of Advance (AoA) Nano
and Chalmers AoA Production.
The author furthermore acknowledges support in the form of computational and storage resources at 
Chalmers Centre for Computational Science and Engineering (C3SE),
and from the 
National Academic Infrastructure for Supercomputing in Sweden (NAISS), under contracts
NAISS2023/3-22, 
NAISS2023/6-306, 
NAISS2024/3-16,   NAISS2024/6-432,
and NAISS2025-3-25.

\appendix

\section{Computing molecular QPs}

I first observe that MBPT-guided
XC functionals are designed to characterize virtual excitations that 
reflect the zero-point electron dynamics of the interacting GS. 
This XC content is perhaps most conveniently cast as a description of virtual collective excitations, i.e., plasmons that correspond to the resonances in the density and therefore also the dielectric or 
screening response
\cite{mabr,rasolt,lape75,gulu76,lape77,lape77,lape80,lameprl1981,ma,ra,lavo87,anlalu96,rydberg03p126402,Dion,thonhauser,Berland_2015:van_waals,JPCMreview}. However, one can also craft MBPT-guided XC functionals by instead consider the virtual dynamics in terms of QPs \cite{FW7}, seeking to model them through a set of (generalized) KS orbitals, as best one can. This strategy was used in the first practical formulation of the LDA \cite{helujpc1971}; The two approaches to XC designs are formally equivalent, linked by the Hedin equation \cite{Hedin65,JPCMreview}. 

Importantly, one can at any given density variation (obtained with such KS-DFTs) extract and analyze details on the underlying QP information that enters in the XC design, either explicitly \cite{helujpc1971} or implicitly \cite{JPCMreview}.
This can be done directly from the DFT-energy components, 
when simply seeking a measure of the first frequency moment of
the occupied part of the fully interacting spectral function \cite{FW7,ChiDFT23,ChiDFT24}. It can also be done, with greater efforts, at the level of individual QP orbitals. A key observation is here that the occupied QP levels  must align with vertical IPs \cite{AuJoWi00,ChiDFT23}. Also, one can sometimes extract predictions of even unoccupied QPs, for example, Refs.\ 
\cite{FW7,helujpc1971,Baerends2003,cococcioni2005,davo2010,kronik2012,OTRSHalga,KraKro13,ferretti2014,nguyen2015,Ma2016,nguyen2016,OTRSHadsorb17,nguyen2018,JPCMreview,gennaro2022,colonna2022,GoGaOh2024,AHBRmRSH25,NitrogenBasesAHBR-mRSH26,hBN2026}.

An extraction of inherent QP information \cite{ChiDFT23} 
contained in the PBE or in either the consistent-exchange vdW-DF-cx 
or rev-vdW-DF2 (also called vdW-DF2-b86r) version can, 
for example, be done by constructing corresponding 
RSH forms  \cite{EP98,HSE03,HJS08,OTRSHalga,DefineAHCX,AHBRmRSH25}. This adapts them for generalized KS DFT and one then uses them subject to an OT
criterion: Alignment of minus the HOMO level and the adiabatic IP, both as described with the same adjustable generalized KS DFT \cite{OTRSHalga,WiOhHa21,AHBRmRSH25}.

Table \ref{tab:QPtuning} illustrates how the steps, gradually
adjusting the choice of $\gamma$, at some $\gamma^*$ value makes the HOMO level consistent with the adiabatic IP. This $\gamma$ tuning proceeds within the AHBR-mRSH($\gamma$) set of RSH vdW-DFs. 
The table considers a set of molecules
that are both part of the G21IP benchmark set \cite{gmtkn55} and 
previously analysis to yield so-called true KS (or MBPT-corrected KS)
orbital levels \cite{Baerends2003}. 
The optimal $\gamma^*$ value is, in principle, specific to the specific molecule. However, the Table also suggests
that an average choice $\bar{\gamma^*} \approx 0.346 \, a_0^{-1}$ is often a good  average estimate for setting the OT AHBR-mRSH($\gamma^*$) descriptor of 
QPs. This holds here for the class of small molecules that are also often involved in important catalytic processes, but it is noted that H$_2$ is
even smaller and requires wider tuning.

Figure \ref{fig:COpiecewise} shows the
energy variation for a CO molecule with partial charging for AHBR-mRSH($\gamma^*=0.346\, a_0^{-1}$), 
i.e., the descriptor that results upon the CO-molecule specific OT process. The partial
charging is reflected in changes in the
electron count, as represented in an ensemble-DFT framework \cite{PePaLe82}. The CO-specific 
$\gamma^*$ setting happens to coincide with $\bar{\gamma^*}$, so Table \ref{tab:QPperform} 
shows that AHBR-mRSH($\gamma^*$) here delivers a excellent QP characterization. The black curve in this figure shows that this performance is reflected in the restoration of a near-perfect
piecewise linearity, much like what we previously found for the larger nitrogen bases \cite{AHBRmRSH25}.
In contrast, use of the corresponding AHBR($\gamma^*)$
form (with its perfect asymptotic screening in its exchange functional form \cite{DefineAHCX,AHBRlaunch,AHBRmRSH25}) does not 
have this linearity. It does, like AHBR($\gamma_0=0.106\, a_0^{-1}$), however, 
exhibit a derivative discontinuity in its exchange- and therefore also total-energy variation as we go between partial-IP to partial EA descriptions, i.e., around $\delta N_{\rm el}\approx 0$.

Table \ref{tab:QPperform} assumes a generic (if generally approximate) use of the average 
$\bar{\gamma^*}=0.346\, a_0^{-1}$ setting and 
tests the performance of AHBR-mRSH($\bar{\gamma^*}$) as QP descriptor across the set of small molecules. The comparison is made against both the true-KS description obtained in Ref.\ \cite{Baerends2003} and a set of photoemission measurements summarized in that work. There is overall a very good agreement. This holds to some extent also in cases where the average $\bar{\gamma^*}$ setting is not aligned with the molecule-specific OT process, e.g., for N$_2$. 

\begin{table}
\caption{\label{tab:H2tune}
Tuning of 
AHBR-mRSH($\gamma$) on description of the tiny H$_2$ molecule. All energies in eV, length scales $1/\gamma$ in Bohrs. I note that at the upper limit of tuning, $\gamma''=0.5\, a_0^{-1}$, set by the overall modeling needs, the roll over to a full Fock-exchange description happens at electron-hole separations that is roughly set by $1/\gamma''=2 \, a_0\approx 1$
{\AA}. For comparison the bond length and the vdW-radius of H$_2$ molecules are 0.74 {\AA} and 1.2 {\AA}, respectively. 
}
\begin{tabular}{l|cc|c|c}
\hline
$\gamma$ & \multicolumn{2}c{AHBR-mRSH($\gamma$)} & AHBR($\gamma$) & $1/\gamma$\\
$a_0^{-1}$ &  $-\epsilon_H(\gamma)$ & IP($\gamma$) & $-\epsilon_H(\gamma)$ & $a_0$\\
\hline
0.106  & 13.3491 & 16.68 & 11.73 &  9.4 
       \\   
0.159  & 13.9295 & 16.72 & 11.53 & 6.3 
       \\  
0.212  & 14.4619 & 16.77 & 11.35 & 4.7 
       \\   
0.265  & 14.9327 & 16.81 & 11.20 & 3.8 
       \\   
0.318  & 15.3395 & 16.86 & 11.06 & 3.1 
       \\   
0.346  & 15.5288 & 16.88 & 11.00 & 2.9 
       \\   
0.400  & 15.8475 & 16.91 & 10.89 & 2.5 
       \\   
0.450  & 16.0946 & 16.94 & 10.81 & 2.2 
       \\        
0.500  & \textit{16.2997} & \textit{16.96} & 10.75 & 2.0 
       \\        
\hline
Ref.\ \cite{Baerends2003} & 16.44 & - & - & - \\
\hline
\hline
\end{tabular}
\end{table}

Table \ref{tab:QP-OTperform} details the sensitivity on molecular
QP predictions at precise OT setting of AHBR-mRSH($\gamma^*$) for small 
molecules, that is, for every case in Table \ref{tab:QPperform}, like N$_2$
and H$_2$, where the systems-specific OT setting $\gamma^*$ differs from the
small-molecule average, $\bar{\gamma^*}$. I here repeat the 
performance assessment, comparing again with the data in Ref.\ \cite{Baerends2003}.
There are generally improvements in the accuracy of QP-level descriptions at deeper levels and especially for the HF and H$_2$ cases.

Finally, Table \ref{tab:H2tune} considers the OT process as it applies to the case of the very small H$_2$ molecule.
Here, alignment, i.e., OT success, does require one to go to a larger $\gamma''$ value. 
However, the value is still limited, and small enough that $1/\gamma''$ is still larger than the physical extension of the H$_2$ molecule. 
This is relevant, as noted in Sec.\ II.B, for seeing AHBR($\gamma$) and AHBR-mRSH($\gamma$)
as roughly equivalent when it comes to the description of H$_2$ 
exchange energies,  when considered at a given density variation $n(\mathbf{r})$.


\begin{thebibliography}{164}%
\makeatletter
\providecommand \@ifxundefined [1]{%
 \@ifx{#1\undefined}
}%
\providecommand \@ifnum [1]{%
 \ifnum #1\expandafter \@firstoftwo
 \else \expandafter \@secondoftwo
 \fi
}%
\providecommand \@ifx [1]{%
 \ifx #1\expandafter \@firstoftwo
 \else \expandafter \@secondoftwo
 \fi
}%
\providecommand \natexlab [1]{#1}%
\providecommand \enquote  [1]{``#1''}%
\providecommand \bibnamefont  [1]{#1}%
\providecommand \bibfnamefont [1]{#1}%
\providecommand \citenamefont [1]{#1}%
\providecommand \href@noop [0]{\@secondoftwo}%
\providecommand \href [0]{\begingroup \@sanitize@url \@href}%
\providecommand \@href[1]{\@@startlink{#1}\@@href}%
\providecommand \@@href[1]{\endgroup#1\@@endlink}%
\providecommand \@sanitize@url [0]{\catcode `\\12\catcode `\$12\catcode `\&12\catcode `\#12\catcode `\^12\catcode `\_12\catcode `\%12\relax}%
\providecommand \@@startlink[1]{}%
\providecommand \@@endlink[0]{}%
\providecommand \url  [0]{\begingroup\@sanitize@url \@url }%
\providecommand \@url [1]{\endgroup\@href {#1}{\urlprefix }}%
\providecommand \urlprefix  [0]{URL }%
\providecommand \Eprint [0]{\href }%
\providecommand \doibase [0]{https://doi.org/}%
\providecommand \selectlanguage [0]{\@gobble}%
\providecommand \bibinfo  [0]{\@secondoftwo}%
\providecommand \bibfield  [0]{\@secondoftwo}%
\providecommand \translation [1]{[#1]}%
\providecommand \BibitemOpen [0]{}%
\providecommand \bibitemStop [0]{}%
\providecommand \bibitemNoStop [0]{.\EOS\space}%
\providecommand \EOS [0]{\spacefactor3000\relax}%
\providecommand \BibitemShut  [1]{\csname bibitem#1\endcsname}%
\let\auto@bib@innerbib\@empty
\bibitem [{\citenamefont {Ertl}\ \emph {et~al.}(1976)\citenamefont {Ertl}, \citenamefont {Grunze},\ and\ \citenamefont {Weiss}}]{Ertl1976}%
  \BibitemOpen
  \bibfield  {author} {\bibinfo {author} {\bibfnamefont {G.}~\bibnamefont {Ertl}}, \bibinfo {author} {\bibfnamefont {M.}~\bibnamefont {Grunze}},\ and\ \bibinfo {author} {\bibfnamefont {M.}~\bibnamefont {Weiss}},\ }\bibfield  {title} {\bibinfo {title} {{Chemisorption of N$_2$ on an Fe(100) surface}},\ }\href@noop {} {\bibfield  {journal} {\bibinfo  {journal} {J. Vac. Sci. Technol.}\ }\textbf {\bibinfo {volume} {13}},\ \bibinfo {pages} {314} (\bibinfo {year} {1976})}\BibitemShut {NoStop}%
\bibitem [{\citenamefont {Ertl}\ \emph {et~al.}(1981)\citenamefont {Ertl}, \citenamefont {Huber}, \citenamefont {Lee}, \citenamefont {Pa{\'a}l},\ and\ \citenamefont {Weiss}}]{Ertl1981}%
  \BibitemOpen
  \bibfield  {author} {\bibinfo {author} {\bibfnamefont {G.}~\bibnamefont {Ertl}}, \bibinfo {author} {\bibfnamefont {M.}~\bibnamefont {Huber}}, \bibinfo {author} {\bibfnamefont {S.~B.}\ \bibnamefont {Lee}}, \bibinfo {author} {\bibfnamefont {Z.}~\bibnamefont {Pa{\'a}l}},\ and\ \bibinfo {author} {\bibfnamefont {M.}~\bibnamefont {Weiss}},\ }\bibfield  {title} {\bibinfo {title} {{Interactions of nitrogen and hydrogen on iron surfaces}},\ }\href@noop {} {\bibfield  {journal} {\bibinfo  {journal} {Applic. Surf. Sci.}\ }\textbf {\bibinfo {volume} {8}},\ \bibinfo {pages} {373} (\bibinfo {year} {1981})}\BibitemShut {NoStop}%
\bibitem [{\citenamefont {Ertl}(1982)}]{Ertl1982}%
  \BibitemOpen
  \bibfield  {author} {\bibinfo {author} {\bibfnamefont {G.}~\bibnamefont {Ertl}},\ }\bibfield  {title} {\bibinfo {title} {{Reaction mechanisms in catalysis by metals}},\ }\href {https://doi.org/10.1080/10408438208243640} {\bibfield  {journal} {\bibinfo  {journal} {Crit. Rev. Solid State Mater. Sci.}\ }\textbf {\bibinfo {volume} {10}},\ \bibinfo {pages} {349} (\bibinfo {year} {1982})}\BibitemShut {NoStop}%
\bibitem [{\citenamefont {Lundqvist}(1983)}]{AdsCatBIL1983}%
  \BibitemOpen
  \bibfield  {author} {\bibinfo {author} {\bibfnamefont {B.~I.}\ \bibnamefont {Lundqvist}},\ }\bibfield  {title} {\bibinfo {title} {{Theoretical aspects of adsorption and heterogeneous catalysis}},\ }\href {https://doi.org/10.1016/0042-207X(83)90587-0} {\bibfield  {journal} {\bibinfo  {journal} {Vacuum}\ }\textbf {\bibinfo {volume} {33}},\ \bibinfo {pages} {639} (\bibinfo {year} {1983})}\BibitemShut {NoStop}%
\bibitem [{\citenamefont {Engdahl}\ \emph {et~al.}(1992)\citenamefont {Engdahl}, \citenamefont {Lundqvist}, \citenamefont {Nielsen},\ and\ \citenamefont {N{\o}rskov}}]{EngdahlBILJKN}%
  \BibitemOpen
  \bibfield  {author} {\bibinfo {author} {\bibfnamefont {C.}~\bibnamefont {Engdahl}}, \bibinfo {author} {\bibfnamefont {B.~I.}\ \bibnamefont {Lundqvist}}, \bibinfo {author} {\bibfnamefont {U.}~\bibnamefont {Nielsen}},\ and\ \bibinfo {author} {\bibfnamefont {J.~K.}\ \bibnamefont {N{\o}rskov}},\ }\bibfield  {title} {\bibinfo {title} {{Multidimensional effects in dissociative chemisorption: H$_2$ on Cu and Ni surfaces}},\ }\href {https://doi.org/10.1103/PhysRevB.45.11362} {\bibfield  {journal} {\bibinfo  {journal} {Phys. Rev. B}\ }\textbf {\bibinfo {volume} {45}},\ \bibinfo {pages} {11362} (\bibinfo {year} {1992})}\BibitemShut {NoStop}%
\bibitem [{\citenamefont {Diaz}\ \emph {et~al.}(2009)\citenamefont {Diaz}, \citenamefont {Pijper}, \citenamefont {Olsen}, \citenamefont {Busnengo}, \citenamefont {Auerbach},\ and\ \citenamefont {Kroes}}]{Diaz2009}%
  \BibitemOpen
  \bibfield  {author} {\bibinfo {author} {\bibfnamefont {C.}~\bibnamefont {Diaz}}, \bibinfo {author} {\bibfnamefont {E.}~\bibnamefont {Pijper}}, \bibinfo {author} {\bibfnamefont {R.~A.}\ \bibnamefont {Olsen}}, \bibinfo {author} {\bibfnamefont {H.~F.}\ \bibnamefont {Busnengo}}, \bibinfo {author} {\bibfnamefont {D.~J.}\ \bibnamefont {Auerbach}},\ and\ \bibinfo {author} {\bibfnamefont {G.-J.}\ \bibnamefont {Kroes}},\ }\bibfield  {title} {\bibinfo {title} {{Chemically Accurate Simulation of a Prototypical Surface Reaction: H$_2$ Dissociation on Cu(111)}},\ }\href {https://doi.org/10.1126/science.1178722} {\bibfield  {journal} {\bibinfo  {journal} {Science}\ }\textbf {\bibinfo {volume} {326}},\ \bibinfo {pages} {832} (\bibinfo {year} {2009})}\BibitemShut {NoStop}%
\bibitem [{\citenamefont {Jiang}\ and\ \citenamefont {Guo}(2019)}]{JiangDC2019}%
  \BibitemOpen
  \bibfield  {author} {\bibinfo {author} {\bibfnamefont {B.}~\bibnamefont {Jiang}}\ and\ \bibinfo {author} {\bibfnamefont {H.}~\bibnamefont {Guo}},\ }\bibfield  {title} {\bibinfo {title} {{Dynamics in reactions on metallic surfaces: A theoretical perspective}},\ }\href {https://doi.org/10.1039/d1cp00044f} {\bibfield  {journal} {\bibinfo  {journal} {J. Chem. Phys.}\ }\textbf {\bibinfo {volume} {150}},\ \bibinfo {pages} {180901} (\bibinfo {year} {2019})}\BibitemShut {NoStop}%
\bibitem [{\citenamefont {Kroes}(2021)}]{KroesDC2021}%
  \BibitemOpen
  \bibfield  {author} {\bibinfo {author} {\bibfnamefont {G.-J.}\ \bibnamefont {Kroes}},\ }\bibfield  {title} {\bibinfo {title} {Computational approaches to dissociative chemisorption on metals: Towards chemical accuracy},\ }\href {https://doi.org/10.1039/d1cp00044f} {\bibfield  {journal} {\bibinfo  {journal} {Chem. Phys. Phys. Chem.}\ }\textbf {\bibinfo {volume} {23}},\ \bibinfo {pages} {8962} (\bibinfo {year} {2021})}\BibitemShut {NoStop}%
\bibitem [{\citenamefont {Lundqvist}\ \emph {et~al.}(1979)\citenamefont {Lundqvist}, \citenamefont {Gunnarsson}, \citenamefont {Hjelmberg},\ and\ \citenamefont {N{\o}rskov}}]{BILsurfReact1979}%
  \BibitemOpen
  \bibfield  {author} {\bibinfo {author} {\bibfnamefont {B.~I.}\ \bibnamefont {Lundqvist}}, \bibinfo {author} {\bibfnamefont {O.}~\bibnamefont {Gunnarsson}}, \bibinfo {author} {\bibfnamefont {H.}~\bibnamefont {Hjelmberg}},\ and\ \bibinfo {author} {\bibfnamefont {J.~K.}\ \bibnamefont {N{\o}rskov}},\ }\bibfield  {title} {\bibinfo {title} {{Theoretical description of molecule-metal interaction and surface reactions}},\ }\href@noop {} {\bibfield  {journal} {\bibinfo  {journal} {Surf. Sci.}\ }\textbf {\bibinfo {volume} {89}},\ \bibinfo {pages} {196} (\bibinfo {year} {1979})}\BibitemShut {NoStop}%
\bibitem [{\citenamefont {Berger}\ \emph {et~al.}(1990)\citenamefont {Berger}, \citenamefont {Leisch}, \citenamefont {Winkler},\ and\ \citenamefont {Rendulic}}]{Berger1990}%
  \BibitemOpen
  \bibfield  {author} {\bibinfo {author} {\bibfnamefont {H.~F.}\ \bibnamefont {Berger}}, \bibinfo {author} {\bibfnamefont {M.}~\bibnamefont {Leisch}}, \bibinfo {author} {\bibfnamefont {A.}~\bibnamefont {Winkler}},\ and\ \bibinfo {author} {\bibfnamefont {K.~D.}\ \bibnamefont {Rendulic}},\ }\bibfield  {title} {\bibinfo {title} {{A search for vibrational contributions to the activated adsorption of H$_2$ on copper}},\ }\href {https://doi.org/10.1063/0009-2614(90)85558-T} {\bibfield  {journal} {\bibinfo  {journal} {Chem. Phys. Lett.}\ }\textbf {\bibinfo {volume} {175}},\ \bibinfo {pages} {425} (\bibinfo {year} {1990})}\BibitemShut {NoStop}%
\bibitem [{\citenamefont {Michelsen}\ \emph {et~al.}(1993)\citenamefont {Michelsen}, \citenamefont {Rettner}, \citenamefont {Auerbach},\ and\ \citenamefont {Zare}}]{Michelsen1993}%
  \BibitemOpen
  \bibfield  {author} {\bibinfo {author} {\bibfnamefont {H.~A.}\ \bibnamefont {Michelsen}}, \bibinfo {author} {\bibfnamefont {C.~T.}\ \bibnamefont {Rettner}}, \bibinfo {author} {\bibfnamefont {D.~J.}\ \bibnamefont {Auerbach}},\ and\ \bibinfo {author} {\bibfnamefont {R.~N.}\ \bibnamefont {Zare}},\ }\bibfield  {title} {\bibinfo {title} {{Effect of rotation on the translational and vibrational energy dependence of the dissociative adsorption of D$_2$ on Cu(111)}},\ }\href {https://doi.org/10.1063/1.464535} {\bibfield  {journal} {\bibinfo  {journal} {J. Chem. Phys.}\ }\textbf {\bibinfo {volume} {98}},\ \bibinfo {pages} {8294} (\bibinfo {year} {1993})}\BibitemShut {NoStop}%
\bibitem [{\citenamefont {Rettner}\ \emph {et~al.}(1995)\citenamefont {Rettner}, \citenamefont {Michelsen},\ and\ \citenamefont {Auerbach}}]{Rettner1995}%
  \BibitemOpen
  \bibfield  {author} {\bibinfo {author} {\bibfnamefont {C.~T.}\ \bibnamefont {Rettner}}, \bibinfo {author} {\bibfnamefont {H.~A.}\ \bibnamefont {Michelsen}},\ and\ \bibinfo {author} {\bibfnamefont {D.~J.}\ \bibnamefont {Auerbach}},\ }\bibfield  {title} {\bibinfo {title} {{Quantum‐state‐specific dynamics of the dissociative adsorption and associative desorption of H$_2$ at a Cu(111) surface}},\ }\href {https://doi.org/10.1063/1.469511} {\bibfield  {journal} {\bibinfo  {journal} {J. Chem. Phys.}\ }\textbf {\bibinfo {volume} {102}},\ \bibinfo {pages} {4625} (\bibinfo {year} {1995})}\BibitemShut {NoStop}%
\bibitem [{\citenamefont {McCormack}\ \emph {et~al.}(1998)\citenamefont {McCormack}, \citenamefont {Kroes}, \citenamefont {Baerends},\ and\ \citenamefont {Mowrey}}]{McCormackH21998}%
  \BibitemOpen
  \bibfield  {author} {\bibinfo {author} {\bibfnamefont {D.~A.}\ \bibnamefont {McCormack}}, \bibinfo {author} {\bibfnamefont {G.-J.}\ \bibnamefont {Kroes}}, \bibinfo {author} {\bibfnamefont {E.-J.}\ \bibnamefont {Baerends}},\ and\ \bibinfo {author} {\bibfnamefont {R.~C.}\ \bibnamefont {Mowrey}},\ }\bibfield  {title} {\bibinfo {title} {{Six-dimensional quantum dynamics of dissociation of rotationally excited H$_2$ on Cu(100)}},\ }\href@noop {} {\bibfield  {journal} {\bibinfo  {journal} {Faraday Discuss.}\ }\textbf {\bibinfo {volume} {110}},\ \bibinfo {pages} {267} (\bibinfo {year} {1998})}\BibitemShut {NoStop}%
\bibitem [{\citenamefont {Gao}\ \emph {et~al.}(2001)\citenamefont {Gao}, \citenamefont {Str{\"o}mquist},\ and\ \citenamefont {Lundqvist}}]{BranchDCGao2001}%
  \BibitemOpen
  \bibfield  {author} {\bibinfo {author} {\bibfnamefont {S.}~\bibnamefont {Gao}}, \bibinfo {author} {\bibfnamefont {J.}~\bibnamefont {Str{\"o}mquist}},\ and\ \bibinfo {author} {\bibfnamefont {B.~I.}\ \bibnamefont {Lundqvist}},\ }\bibfield  {title} {\bibinfo {title} {{Dissipative Quantum Dynamics in 2D: Anisotropic Dissipation and Selective Bond Breaking in Surface Photochemistry}},\ }\href {https://doi.org/10.1103/PhysRevLett.86.1805} {\bibfield  {journal} {\bibinfo  {journal} {Phys. Rev. Lett.}\ }\textbf {\bibinfo {volume} {86}},\ \bibinfo {pages} {1805} (\bibinfo {year} {2001})}\BibitemShut {NoStop}%
\bibitem [{\citenamefont {Kroes}(2008)}]{KroesScatter2008}%
  \BibitemOpen
  \bibfield  {author} {\bibinfo {author} {\bibfnamefont {G.-J.}\ \bibnamefont {Kroes}},\ }\bibfield  {title} {\bibinfo {title} {{Frontiers in Surface Scattering Simulations}},\ }\href {https://doi.org/10.1125/science.1157717} {\bibfield  {journal} {\bibinfo  {journal} {Science}\ }\textbf {\bibinfo {volume} {321}},\ \bibinfo {pages} {794} (\bibinfo {year} {2008})}\BibitemShut {NoStop}%
\bibitem [{\citenamefont {Auerbach}\ \emph {et~al.}(2024)\citenamefont {Auerbach}, \citenamefont {Babikov}, \citenamefont {Butler}, \citenamefont {Chandler}, \citenamefont {Fingerhut}, \citenamefont {Guo}, \citenamefont {Harding}, \citenamefont {Heathcote}, \citenamefont {Hertl}, \citenamefont {Jiang}, \citenamefont {Kroes}, \citenamefont {Lane}, \citenamefont {Loreau}, \citenamefont {Mackenzie}, \citenamefont {McKendrick}, \citenamefont {Moon}, \citenamefont {Nathanson}, \citenamefont {Neumark}, \citenamefont {Pandey}, \citenamefont {Schatz}, \citenamefont {Sibener}, \citenamefont {Srivastav}, \citenamefont {Vallance}, \citenamefont {van Bree}, \citenamefont {Wagner}, \citenamefont {Walker}, \citenamefont {Watson}, \citenamefont {Willitsch}, \citenamefont {Wodtke},\ and\ \citenamefont {Zhao}}]{AuerbachScatter24}%
  \BibitemOpen
  \bibfield  {author} {\bibinfo {author} {\bibfnamefont {D.~J.}\ \bibnamefont {Auerbach}}, \bibinfo {author} {\bibfnamefont {D.}~\bibnamefont {Babikov}}, \bibinfo {author} {\bibfnamefont {A.}~\bibnamefont {Butler}}, \bibinfo {author} {\bibfnamefont {D.~W.}\ \bibnamefont {Chandler}}, \bibinfo {author} {\bibfnamefont {J.}~\bibnamefont {Fingerhut}}, \bibinfo {author} {\bibfnamefont {H.}~\bibnamefont {Guo}}, \bibinfo {author} {\bibfnamefont {D.~J.}\ \bibnamefont {Harding}}, \bibinfo {author} {\bibfnamefont {D.}~\bibnamefont {Heathcote}}, \bibinfo {author} {\bibfnamefont {N.}~\bibnamefont {Hertl}}, \bibinfo {author} {\bibfnamefont {B.}~\bibnamefont {Jiang}}, \bibinfo {author} {\bibfnamefont {G.-J.}\ \bibnamefont {Kroes}}, \bibinfo {author} {\bibfnamefont {P.~D.}\ \bibnamefont {Lane}}, \bibinfo {author} {\bibfnamefont {J.}~\bibnamefont {Loreau}}, \bibinfo {author} {\bibfnamefont {S.~R.}\ \bibnamefont {Mackenzie}}, \bibinfo {author} {\bibfnamefont {K.~G.}\ \bibnamefont {McKendrick}}, \bibinfo {author} {\bibfnamefont
  {D.~R.}\ \bibnamefont {Moon}}, \bibinfo {author} {\bibfnamefont {G.~M.}\ \bibnamefont {Nathanson}}, \bibinfo {author} {\bibfnamefont {D.~M.}\ \bibnamefont {Neumark}}, \bibinfo {author} {\bibfnamefont {R.}~\bibnamefont {Pandey}}, \bibinfo {author} {\bibfnamefont {G.~C.}\ \bibnamefont {Schatz}}, \bibinfo {author} {\bibfnamefont {S.~J.}\ \bibnamefont {Sibener}}, \bibinfo {author} {\bibfnamefont {A.}~\bibnamefont {Srivastav}}, \bibinfo {author} {\bibfnamefont {C.}~\bibnamefont {Vallance}}, \bibinfo {author} {\bibfnamefont {R.~A.~B.}\ \bibnamefont {van Bree}}, \bibinfo {author} {\bibfnamefont {J.}~\bibnamefont {Wagner}}, \bibinfo {author} {\bibfnamefont {G.~C.}\ \bibnamefont {Walker}}, \bibinfo {author} {\bibfnamefont {P.~D.}\ \bibnamefont {Watson}}, \bibinfo {author} {\bibfnamefont {S.}~\bibnamefont {Willitsch}}, \bibinfo {author} {\bibfnamefont {A.~M.}\ \bibnamefont {Wodtke}},\ and\ \bibinfo {author} {\bibfnamefont {B.~S.}\ \bibnamefont {Zhao}},\ }\bibfield  {title} {\bibinfo {title} {Scattering at
  condensed-phase surfaces: general discussion},\ }\href {https://doi.org/10.1039/D4FD90020K} {\bibfield  {journal} {\bibinfo  {journal} {Faraday Discuss.}\ }\textbf {\bibinfo {volume} {251}},\ \bibinfo {pages} {471} (\bibinfo {year} {2024})}\BibitemShut {NoStop}%
\bibitem [{\citenamefont {Foster}\ and\ \citenamefont {{\it et al}}(2026)}]{Leiden26}%
  \BibitemOpen
  \bibfield  {author} {\bibinfo {author} {\bibfnamefont {A.}~\bibnamefont {Foster}}\ and\ \bibinfo {author} {\bibnamefont {{\it et al}}},\ }\bibfield  {title} {\bibinfo {title} {{Towards the accurate prediction of\\barriers for reactions on metal surfaces}},\ }\href@noop {} {\bibfield  {journal} {\bibinfo  {journal} {Invited perspective, Phys. Chem. Chem. Phys.}\ } (\bibinfo {year} {2026})}\BibitemShut {NoStop}%
\bibitem [{\citenamefont {Lundqvist}\ \emph {et~al.}(1987)\citenamefont {Lundqvist}, \citenamefont {Fond{\'e}n}, \citenamefont {Idiodi}, \citenamefont {Johnsson}, \citenamefont {M{\"a}llo},\ and\ \citenamefont {Papadia}}]{ChemSorb1987}%
  \BibitemOpen
  \bibfield  {author} {\bibinfo {author} {\bibfnamefont {B.~I.}\ \bibnamefont {Lundqvist}}, \bibinfo {author} {\bibfnamefont {T.}~\bibnamefont {Fond{\'e}n}}, \bibinfo {author} {\bibfnamefont {J.}~\bibnamefont {Idiodi}}, \bibinfo {author} {\bibfnamefont {P.}~\bibnamefont {Johnsson}}, \bibinfo {author} {\bibfnamefont {A.}~\bibnamefont {M{\"a}llo}},\ and\ \bibinfo {author} {\bibfnamefont {S.}~\bibnamefont {Papadia}},\ }\bibfield  {title} {\bibinfo {title} {{Theoretical descriptions of atomic and molecular chemisorption on metals}},\ }\href {https://doi.org/10.1016/S0079-6816(87)80014-X} {\bibfield  {journal} {\bibinfo  {journal} {Prog. Surf. Sci.}\ }\textbf {\bibinfo {volume} {25}},\ \bibinfo {pages} {191} (\bibinfo {year} {1987})}\BibitemShut {NoStop}%
\bibitem [{\citenamefont {Wijzenbroek}\ \emph {et~al.}(2015)\citenamefont {Wijzenbroek}, \citenamefont {Klein}, \citenamefont {Smits}, \citenamefont {Somers},\ and\ \citenamefont {Kroes}}]{catalysisvdW15}%
  \BibitemOpen
  \bibfield  {author} {\bibinfo {author} {\bibfnamefont {M.}~\bibnamefont {Wijzenbroek}}, \bibinfo {author} {\bibfnamefont {D.~M.}\ \bibnamefont {Klein}}, \bibinfo {author} {\bibfnamefont {B.}~\bibnamefont {Smits}}, \bibinfo {author} {\bibfnamefont {M.~F.}\ \bibnamefont {Somers}},\ and\ \bibinfo {author} {\bibfnamefont {G.-J.}\ \bibnamefont {Kroes}},\ }\bibfield  {title} {\bibinfo {title} {{Performance of a Non-Local {van der Waals} Density Functional on the Dissociation of {H}$_2$ on Metal Surfaces}},\ }\href {https://doi.org/10.1021/acs,jpca.5b06008} {\bibfield  {journal} {\bibinfo  {journal} {J. Phys. Chem. A}\ }\textbf {\bibinfo {volume} {119}},\ \bibinfo {pages} {12146} (\bibinfo {year} {2015})}\BibitemShut {NoStop}%
\bibitem [{\citenamefont {Andersson}\ \emph {et~al.}(1988)\citenamefont {Andersson}, \citenamefont {Wilz\'en},\ and\ \citenamefont {Persson}}]{anderssonetal1988}%
  \BibitemOpen
  \bibfield  {author} {\bibinfo {author} {\bibfnamefont {S.}~\bibnamefont {Andersson}}, \bibinfo {author} {\bibfnamefont {L.}~\bibnamefont {Wilz\'en}},\ and\ \bibinfo {author} {\bibfnamefont {M.}~\bibnamefont {Persson}},\ }\bibfield  {title} {\bibinfo {title} {Physisorption interaction of {H}$_2$ with noble-metal surfaces: A new {H}$_2$-{C}u potential},\ }\href@noop {} {\bibfield  {journal} {\bibinfo  {journal} {Phys. Rev. B}\ }\textbf {\bibinfo {volume} {38}},\ \bibinfo {pages} {2967} (\bibinfo {year} {1988})}\BibitemShut {NoStop}%
\bibitem [{\citenamefont {Andersson}\ \emph {et~al.}(1996{\natexlab{a}})\citenamefont {Andersson}, \citenamefont {Persson},\ and\ \citenamefont {Harris}}]{anderssonpeha96}%
  \BibitemOpen
  \bibfield  {author} {\bibinfo {author} {\bibfnamefont {S.}~\bibnamefont {Andersson}}, \bibinfo {author} {\bibfnamefont {M.}~\bibnamefont {Persson}},\ and\ \bibinfo {author} {\bibfnamefont {J.}~\bibnamefont {Harris}},\ }\bibfield  {title} {\bibinfo {title} {Physisorption energies: Influence of surface structure},\ }\href@noop {} {\bibfield  {journal} {\bibinfo  {journal} {Surf. Sci.}\ }\textbf {\bibinfo {volume} {360}},\ \bibinfo {pages} {L499} (\bibinfo {year} {1996}{\natexlab{a}})}\BibitemShut {NoStop}%
\bibitem [{\citenamefont {Lee}\ \emph {et~al.}(2011)\citenamefont {Lee}, \citenamefont {Kelkkanen}, \citenamefont {Berland}, \citenamefont {Andersson}, \citenamefont {Langreth}, \citenamefont {Schr{\"o}der}, \citenamefont {Lundqvist},\ and\ \citenamefont {Hyldgaard}}]{lee11p193408}%
  \BibitemOpen
  \bibfield  {author} {\bibinfo {author} {\bibfnamefont {K.}~\bibnamefont {Lee}}, \bibinfo {author} {\bibfnamefont {A.~K.}\ \bibnamefont {Kelkkanen}}, \bibinfo {author} {\bibfnamefont {K.}~\bibnamefont {Berland}}, \bibinfo {author} {\bibfnamefont {S.}~\bibnamefont {Andersson}}, \bibinfo {author} {\bibfnamefont {D.~C.}\ \bibnamefont {Langreth}}, \bibinfo {author} {\bibfnamefont {E.}~\bibnamefont {Schr{\"o}der}}, \bibinfo {author} {\bibfnamefont {B.~I.}\ \bibnamefont {Lundqvist}},\ and\ \bibinfo {author} {\bibfnamefont {P.}~\bibnamefont {Hyldgaard}},\ }\bibfield  {title} {\bibinfo {title} {Evaluation of a density functional with account of van der {W}aals forces using experimental data of {H}${}_{2}$ physisorption on {C}u(111)},\ }\href@noop {} {\bibfield  {journal} {\bibinfo  {journal} {Phys. Rev. B}\ }\textbf {\bibinfo {volume} {84}},\ \bibinfo {pages} {193408} (\bibinfo {year} {2011})}\BibitemShut {NoStop}%
\bibitem [{\citenamefont {Arrhenius}(1889)}]{Arrhenius1889}%
  \BibitemOpen
  \bibfield  {author} {\bibinfo {author} {\bibfnamefont {S.}~\bibnamefont {Arrhenius}},\ }\bibfield  {title} {\bibinfo {title} {{{\"U}ber die Dissociationsw{\"a}rme und den Einfluss der Temperatur auf den Dissociationsgrad der Elektrolyte (On the heat of dissociation and the influence of temperature on the degree of dissociation of the electrolytes)}},\ }\href {https://doi.org/10.1515/zpch-1889-0408} {\bibfield  {journal} {\bibinfo  {journal} {Z. Phys. Chem.}\ }\textbf {\bibinfo {volume} {4}},\ \bibinfo {pages} {96} (\bibinfo {year} {1889})}\BibitemShut {NoStop}%
\bibitem [{\citenamefont {Hohenberg}\ and\ \citenamefont {Kohn}(1964)}]{hoko64}%
  \BibitemOpen
  \bibfield  {author} {\bibinfo {author} {\bibfnamefont {P.}~\bibnamefont {Hohenberg}}\ and\ \bibinfo {author} {\bibfnamefont {W.}~\bibnamefont {Kohn}},\ }\bibfield  {title} {\bibinfo {title} {Inhomogeneous electron gas},\ }\href@noop {} {\bibfield  {journal} {\bibinfo  {journal} {Phys. Rev.}\ }\textbf {\bibinfo {volume} {136}},\ \bibinfo {pages} {B864} (\bibinfo {year} {1964})}\BibitemShut {NoStop}%
\bibitem [{\citenamefont {Blyholder}(1964)}]{Blyholder}%
  \BibitemOpen
  \bibfield  {author} {\bibinfo {author} {\bibfnamefont {G.}~\bibnamefont {Blyholder}},\ }\bibfield  {title} {\bibinfo {title} {{Molecular Orbital View of Chemisorbed Carbon Monoxide}},\ }\href@noop {} {\bibfield  {journal} {\bibinfo  {journal} {J. Phys. Chem.}\ }\textbf {\bibinfo {volume} {68}},\ \bibinfo {pages} {2772} (\bibinfo {year} {1964})}\BibitemShut {NoStop}%
\bibitem [{\citenamefont {Hedin}(1965)}]{Hedin65}%
  \BibitemOpen
  \bibfield  {author} {\bibinfo {author} {\bibfnamefont {L.}~\bibnamefont {Hedin}},\ }\bibfield  {title} {\bibinfo {title} {New method for calculating the one-particle {G}reen's function with application to the electron-gas problem},\ }\href {https://doi.org/10.1103/PhysRev.139.A796} {\bibfield  {journal} {\bibinfo  {journal} {Phys. Rev.}\ }\textbf {\bibinfo {volume} {139}},\ \bibinfo {pages} {A796} (\bibinfo {year} {1965})}\BibitemShut {NoStop}%
\bibitem [{\citenamefont {Mahan}(1965)}]{jerry65}%
  \BibitemOpen
  \bibfield  {author} {\bibinfo {author} {\bibfnamefont {G.~D.}\ \bibnamefont {Mahan}},\ }\bibfield  {title} {\bibinfo {title} {{Van der Waals Forces in Solids}},\ }\href@noop {} {\bibfield  {journal} {\bibinfo  {journal} {J. Chem. Phys.}\ }\textbf {\bibinfo {volume} {43}},\ \bibinfo {pages} {1569} (\bibinfo {year} {1965})}\BibitemShut {NoStop}%
\bibitem [{\citenamefont {Lundqvist}(1967)}]{lu67}%
  \BibitemOpen
  \bibfield  {author} {\bibinfo {author} {\bibfnamefont {B.~I.}\ \bibnamefont {Lundqvist}},\ }\bibfield  {title} {\bibinfo {title} {{Single-Particle Spectrum of the Degenerate Electron Gas. {I}. The Structure of the Spectral Weight Function}},\ }\href@noop {} {\bibfield  {journal} {\bibinfo  {journal} {Phys. Kondens. Materie}\ }\textbf {\bibinfo {volume} {6}},\ \bibinfo {pages} {193} (\bibinfo {year} {1967})}\BibitemShut {NoStop}%
\bibitem [{\citenamefont {Langreth}(1970)}]{la70}%
  \BibitemOpen
  \bibfield  {author} {\bibinfo {author} {\bibfnamefont {D.~C.}\ \bibnamefont {Langreth}},\ }\bibfield  {title} {\bibinfo {title} {{Singularities in the X-Ray Spectra of Metals}},\ }\href@noop {} {\bibfield  {journal} {\bibinfo  {journal} {Phys. Rev. B}\ }\textbf {\bibinfo {volume} {1}},\ \bibinfo {pages} {471} (\bibinfo {year} {1970})}\BibitemShut {NoStop}%
\bibitem [{\citenamefont {Hedin}\ and\ \citenamefont {Lundqvist}(1971)}]{helujpc1971}%
  \BibitemOpen
  \bibfield  {author} {\bibinfo {author} {\bibfnamefont {L.}~\bibnamefont {Hedin}}\ and\ \bibinfo {author} {\bibfnamefont {B.~I.}\ \bibnamefont {Lundqvist}},\ }\bibfield  {title} {\bibinfo {title} {Explicit local exchange-correlation potentials},\ }\href@noop {} {\bibfield  {journal} {\bibinfo  {journal} {J. Phys. C}\ }\textbf {\bibinfo {volume} {4}},\ \bibinfo {pages} {2064} (\bibinfo {year} {1971})}\BibitemShut {NoStop}%
\bibitem [{\citenamefont {Gunnarsson}\ and\ \citenamefont {Lundqvist}(1976)}]{gulu76}%
  \BibitemOpen
  \bibfield  {author} {\bibinfo {author} {\bibfnamefont {O.}~\bibnamefont {Gunnarsson}}\ and\ \bibinfo {author} {\bibfnamefont {B.~I.}\ \bibnamefont {Lundqvist}},\ }\bibfield  {title} {\bibinfo {title} {Exchange and correlation in atoms, molecules, and solids by the spin-density-functional formalism},\ }\href@noop {} {\bibfield  {journal} {\bibinfo  {journal} {Phys. Rev. B}\ }\textbf {\bibinfo {volume} {13}},\ \bibinfo {pages} {4274} (\bibinfo {year} {1976})}\BibitemShut {NoStop}%
\bibitem [{\citenamefont {Langreth}\ and\ \citenamefont {Perdew}(1977)}]{lape77}%
  \BibitemOpen
  \bibfield  {author} {\bibinfo {author} {\bibfnamefont {D.~C.}\ \bibnamefont {Langreth}}\ and\ \bibinfo {author} {\bibfnamefont {J.~P.}\ \bibnamefont {Perdew}},\ }\bibfield  {title} {\bibinfo {title} {Exchange-correlation energy of a metallic surface: {W}ave-vector analysis},\ }\href@noop {} {\bibfield  {journal} {\bibinfo  {journal} {Phys. Rev. B}\ }\textbf {\bibinfo {volume} {15}},\ \bibinfo {pages} {2884} (\bibinfo {year} {1977})}\BibitemShut {NoStop}%
\bibitem [{\citenamefont {N{\o}rskov}\ and\ \citenamefont {Lundqvist}(1979)}]{JKNBILloss1979}%
  \BibitemOpen
  \bibfield  {author} {\bibinfo {author} {\bibfnamefont {J.~K.}\ \bibnamefont {N{\o}rskov}}\ and\ \bibinfo {author} {\bibfnamefont {B.~I.}\ \bibnamefont {Lundqvist}},\ }\bibfield  {title} {\bibinfo {title} {{Correlation between sticking probability and adsorbate-induced electron structure}},\ }\href@noop {} {\bibfield  {journal} {\bibinfo  {journal} {Surf. Sci.}\ }\textbf {\bibinfo {volume} {89}},\ \bibinfo {pages} {251} (\bibinfo {year} {1979})}\BibitemShut {NoStop}%
\bibitem [{\citenamefont {Langreth}\ and\ \citenamefont {Perdew}(1980)}]{lape80}%
  \BibitemOpen
  \bibfield  {author} {\bibinfo {author} {\bibfnamefont {D.~C.}\ \bibnamefont {Langreth}}\ and\ \bibinfo {author} {\bibfnamefont {J.~P.}\ \bibnamefont {Perdew}},\ }\bibfield  {title} {\bibinfo {title} {Theory of nonuniform electronic systems. {I}. {A}nalysis of the gradient approximation and a generalization that works},\ }\href@noop {} {\bibfield  {journal} {\bibinfo  {journal} {Phys. Rev. B}\ }\textbf {\bibinfo {volume} {21}},\ \bibinfo {pages} {5469} (\bibinfo {year} {1980})}\BibitemShut {NoStop}%
\bibitem [{\citenamefont {Hedin}(1980)}]{Hedin80}%
  \BibitemOpen
  \bibfield  {author} {\bibinfo {author} {\bibfnamefont {L.}~\bibnamefont {Hedin}},\ }\bibfield  {title} {\bibinfo {title} {Effects of recoil on shake-up spectra in metals},\ }\href@noop {} {\bibfield  {journal} {\bibinfo  {journal} {Phys. Scr.}\ }\textbf {\bibinfo {volume} {21}},\ \bibinfo {pages} {477} (\bibinfo {year} {1980})}\BibitemShut {NoStop}%
\bibitem [{\citenamefont {Perdew}\ \emph {et~al.}(1982)\citenamefont {Perdew}, \citenamefont {Parr}, \citenamefont {Levy},\ and\ \citenamefont {J.~L.~Balduz}}]{PePaLe82}%
  \BibitemOpen
  \bibfield  {author} {\bibinfo {author} {\bibfnamefont {J.~P.}\ \bibnamefont {Perdew}}, \bibinfo {author} {\bibfnamefont {R.~G.}\ \bibnamefont {Parr}}, \bibinfo {author} {\bibfnamefont {M.}~\bibnamefont {Levy}},\ and\ \bibinfo {author} {\bibfnamefont {J.}~\bibnamefont {J.~L.~Balduz}},\ }\bibfield  {title} {\bibinfo {title} {{Density-Functional Theory for Fractional Particle Number: Derivative Discontinutities of the Energy}},\ }\href@noop {} {\bibfield  {journal} {\bibinfo  {journal} {Phys. Rev. Lett.}\ }\textbf {\bibinfo {volume} {49}},\ \bibinfo {pages} {1691} (\bibinfo {year} {1982})}\BibitemShut {NoStop}%
\bibitem [{\citenamefont {Levy}\ \emph {et~al.}(1984)\citenamefont {Levy}, \citenamefont {Perdew},\ and\ \citenamefont {Sahni}}]{LePeSa84}%
  \BibitemOpen
  \bibfield  {author} {\bibinfo {author} {\bibfnamefont {M.}~\bibnamefont {Levy}}, \bibinfo {author} {\bibfnamefont {J.~P.}\ \bibnamefont {Perdew}},\ and\ \bibinfo {author} {\bibfnamefont {J.~P.}\ \bibnamefont {Sahni}},\ }\bibfield  {title} {\bibinfo {title} {{Exact differential equation for the density and the ionization energy of a many-particle system}},\ }\href@noop {} {\bibfield  {journal} {\bibinfo  {journal} {Phys. Rev. A}\ }\textbf {\bibinfo {volume} {30}},\ \bibinfo {pages} {2745} (\bibinfo {year} {1984})}\BibitemShut {NoStop}%
\bibitem [{\citenamefont {Hybertsen}\ and\ \citenamefont {Louie}(1986)}]{HybersenLouieGPP}%
  \BibitemOpen
  \bibfield  {author} {\bibinfo {author} {\bibfnamefont {M.~S.}\ \bibnamefont {Hybertsen}}\ and\ \bibinfo {author} {\bibfnamefont {S.~G.}\ \bibnamefont {Louie}},\ }\bibfield  {title} {\bibinfo {title} {Electron correlation in semiconductors and insulators: {B}and gaps and quasiparticle energies},\ }\href@noop {} {\bibfield  {journal} {\bibinfo  {journal} {Phys. Rev. B}\ }\textbf {\bibinfo {volume} {34}},\ \bibinfo {pages} {5390} (\bibinfo {year} {1986})}\BibitemShut {NoStop}%
\bibitem [{\citenamefont {Jones}\ and\ \citenamefont {Gunnarsson}(1989)}]{jogu89}%
  \BibitemOpen
  \bibfield  {author} {\bibinfo {author} {\bibfnamefont {R.~O.}\ \bibnamefont {Jones}}\ and\ \bibinfo {author} {\bibfnamefont {O.}~\bibnamefont {Gunnarsson}},\ }\bibfield  {title} {\bibinfo {title} {The density functional formalism, its applications and prospects},\ }\href@noop {} {\bibfield  {journal} {\bibinfo  {journal} {Rev. Mod. Phys.}\ }\textbf {\bibinfo {volume} {61}},\ \bibinfo {pages} {689} (\bibinfo {year} {1989})}\BibitemShut {NoStop}%
\bibitem [{\citenamefont {Maggs}\ and\ \citenamefont {Ashcroft}(1987)}]{ma}%
  \BibitemOpen
  \bibfield  {author} {\bibinfo {author} {\bibfnamefont {A.~C.}\ \bibnamefont {Maggs}}\ and\ \bibinfo {author} {\bibfnamefont {N.~W.}\ \bibnamefont {Ashcroft}},\ }\bibfield  {title} {\bibinfo {title} {Electronic fluctuation and cohesion in metals},\ }\href@noop {} {\bibfield  {journal} {\bibinfo  {journal} {Phys. Rev. Lett.}\ }\textbf {\bibinfo {volume} {59}},\ \bibinfo {pages} {113} (\bibinfo {year} {1987})}\BibitemShut {NoStop}%
\bibitem [{\citenamefont {Rapcewicz}\ and\ \citenamefont {Ashcroft}(1991)}]{ra}%
  \BibitemOpen
  \bibfield  {author} {\bibinfo {author} {\bibfnamefont {K.}~\bibnamefont {Rapcewicz}}\ and\ \bibinfo {author} {\bibfnamefont {N.~W.}\ \bibnamefont {Ashcroft}},\ }\bibfield  {title} {\bibinfo {title} {Fluctuation attraction in condensed matter: {A} nonlocal functional approach},\ }\href@noop {} {\bibfield  {journal} {\bibinfo  {journal} {Phys. Rev. B}\ }\textbf {\bibinfo {volume} {44}},\ \bibinfo {pages} {4032} (\bibinfo {year} {1991})}\BibitemShut {NoStop}%
\bibitem [{\citenamefont {Hammer}\ and\ \citenamefont {N{\o}rskov}(1995{\natexlab{a}})}]{Noblest}%
  \BibitemOpen
  \bibfield  {author} {\bibinfo {author} {\bibfnamefont {B.}~\bibnamefont {Hammer}}\ and\ \bibinfo {author} {\bibfnamefont {J.~K.}\ \bibnamefont {N{\o}rskov}},\ }\bibfield  {title} {\bibinfo {title} {Why gold is the noblest of all the metals},\ }\href@noop {} {\bibfield  {journal} {\bibinfo  {journal} {Nature}\ }\textbf {\bibinfo {volume} {376}},\ \bibinfo {pages} {238} (\bibinfo {year} {1995}{\natexlab{a}})}\BibitemShut {NoStop}%
\bibitem [{\citenamefont {Hammer}\ and\ \citenamefont {N{\o}rskov}(1995{\natexlab{b}})}]{HaNo95}%
  \BibitemOpen
  \bibfield  {author} {\bibinfo {author} {\bibfnamefont {B.}~\bibnamefont {Hammer}}\ and\ \bibinfo {author} {\bibfnamefont {J.~K.}\ \bibnamefont {N{\o}rskov}},\ }\bibfield  {title} {\bibinfo {title} {Electronic factors determining the reactivity of metal surfaces},\ }\href@noop {} {\bibfield  {journal} {\bibinfo  {journal} {Surf. Sci.}\ }\textbf {\bibinfo {volume} {343}},\ \bibinfo {pages} {211} (\bibinfo {year} {1995}{\natexlab{b}})}\BibitemShut {NoStop}%
\bibitem [{\citenamefont {Baerends}\ and\ \citenamefont {Gritsenko}(1997)}]{Baerends1997}%
  \BibitemOpen
  \bibfield  {author} {\bibinfo {author} {\bibfnamefont {E.~J.}\ \bibnamefont {Baerends}}\ and\ \bibinfo {author} {\bibfnamefont {O.~V.}\ \bibnamefont {Gritsenko}},\ }\bibfield  {title} {\bibinfo {title} {{A Quantum Chemical View of Density Functional Theory}},\ }\href {https://doi.org/10.1021/jp9703768} {\bibfield  {journal} {\bibinfo  {journal} {J. Phys. Chem. A}\ }\textbf {\bibinfo {volume} {101}},\ \bibinfo {pages} {5383} (\bibinfo {year} {1997})}\BibitemShut {NoStop}%
\bibitem [{\citenamefont {Aulbur}\ \emph {et~al.}(2000)\citenamefont {Aulbur}, \citenamefont {J{\"o}nsson},\ and\ \citenamefont {Wilkins}}]{AuJoWi00}%
  \BibitemOpen
  \bibfield  {author} {\bibinfo {author} {\bibfnamefont {W.~G.}\ \bibnamefont {Aulbur}}, \bibinfo {author} {\bibfnamefont {L.}~\bibnamefont {J{\"o}nsson}},\ and\ \bibinfo {author} {\bibfnamefont {J.~W.}\ \bibnamefont {Wilkins}},\ }\bibfield  {title} {\bibinfo {title} {{Quasiparticle Calculations in Solids}},\ }in\ \href@noop {} {\emph {\bibinfo {booktitle} {Solid State Physics}}},\ Vol.~\bibinfo {volume} {54},\ \bibinfo {editor} {edited by\ \bibinfo {editor} {\bibfnamefont {F.}~\bibnamefont {Seitz}}, \bibinfo {editor} {\bibfnamefont {D.}~\bibnamefont {Turnbull}},\ and\ \bibinfo {editor} {\bibfnamefont {H.}~\bibnamefont {Ehrenreich}}}\ (\bibinfo  {publisher} {Academic Press},\ \bibinfo {address} {New York},\ \bibinfo {year} {2000})\ p.~\bibinfo {pages} {1}\BibitemShut {NoStop}%
\bibitem [{\citenamefont {Chong}\ \emph {et~al.}(2002)\citenamefont {Chong}, \citenamefont {Gritsenko},\ and\ \citenamefont {Baerends}}]{Baerends2002}%
  \BibitemOpen
  \bibfield  {author} {\bibinfo {author} {\bibfnamefont {D.~P.}\ \bibnamefont {Chong}}, \bibinfo {author} {\bibfnamefont {O.~V.}\ \bibnamefont {Gritsenko}},\ and\ \bibinfo {author} {\bibfnamefont {E.~J.}\ \bibnamefont {Baerends}},\ }\bibfield  {title} {\bibinfo {title} {{ Interpretation of the Kohn-Sham orbital energies as approximate vertical ionization potentials}},\ }\href {https://doi.org/10.1063/1.1430255} {\bibfield  {journal} {\bibinfo  {journal} {J. Chem. Phys.}\ }\textbf {\bibinfo {volume} {116}},\ \bibinfo {pages} {1760} (\bibinfo {year} {2002})}\BibitemShut {NoStop}%
\bibitem [{\citenamefont {Gritsenko}\ \emph {et~al.}(2003)\citenamefont {Gritsenko}, \citenamefont {Bra{\"i}da},\ and\ \citenamefont {Baerends}}]{Baerends2003}%
  \BibitemOpen
  \bibfield  {author} {\bibinfo {author} {\bibfnamefont {O.~V.}\ \bibnamefont {Gritsenko}}, \bibinfo {author} {\bibfnamefont {B.}~\bibnamefont {Bra{\"i}da}},\ and\ \bibinfo {author} {\bibfnamefont {E.~J.}\ \bibnamefont {Baerends}},\ }\bibfield  {title} {\bibinfo {title} {{Physical interpretation and evaluation of the Kohn-Sham and Dyson components of the $\epsilon-I$ relations between the Kohn-Sham orbital energies and the ionization potentials}},\ }\href {https://doi.org/10.1063/1.1582839} {\bibfield  {journal} {\bibinfo  {journal} {J. Chem. Phys.}\ }\textbf {\bibinfo {volume} {119}},\ \bibinfo {pages} {1937} (\bibinfo {year} {2003})}\BibitemShut {NoStop}%
\bibitem [{\citenamefont {Dabo}\ \emph {et~al.}(2010)\citenamefont {Dabo}, \citenamefont {Ferretti}, \citenamefont {Poilvert}, \citenamefont {Li}, \citenamefont {Marzari},\ and\ \citenamefont {Cococcioni}}]{davo2010}%
  \BibitemOpen
  \bibfield  {author} {\bibinfo {author} {\bibfnamefont {I.}~\bibnamefont {Dabo}}, \bibinfo {author} {\bibfnamefont {A.}~\bibnamefont {Ferretti}}, \bibinfo {author} {\bibfnamefont {N.}~\bibnamefont {Poilvert}}, \bibinfo {author} {\bibfnamefont {Y.}~\bibnamefont {Li}}, \bibinfo {author} {\bibfnamefont {N.}~\bibnamefont {Marzari}},\ and\ \bibinfo {author} {\bibfnamefont {M.}~\bibnamefont {Cococcioni}},\ }\bibfield  {title} {\bibinfo {title} {Koopmans' condition for density-functional theory},\ }\href {https://doi.org/10.1103/PhysRevB.82.115121} {\bibfield  {journal} {\bibinfo  {journal} {Phys. Rev. B}\ }\textbf {\bibinfo {volume} {82}},\ \bibinfo {pages} {115121} (\bibinfo {year} {2010})}\BibitemShut {NoStop}%
\bibitem [{\citenamefont {Refaely-Abramson}\ \emph {et~al.}(2012)\citenamefont {Refaely-Abramson}, \citenamefont {Sharifzadeh}, \citenamefont {Govind}, \citenamefont {Autschbach}, \citenamefont {Neaton}, \citenamefont {Baer},\ and\ \citenamefont {Kronik}}]{OTRSHalga}%
  \BibitemOpen
  \bibfield  {author} {\bibinfo {author} {\bibfnamefont {S.}~\bibnamefont {Refaely-Abramson}}, \bibinfo {author} {\bibfnamefont {S.}~\bibnamefont {Sharifzadeh}}, \bibinfo {author} {\bibfnamefont {N.}~\bibnamefont {Govind}}, \bibinfo {author} {\bibfnamefont {J.}~\bibnamefont {Autschbach}}, \bibinfo {author} {\bibfnamefont {J.~B.}\ \bibnamefont {Neaton}}, \bibinfo {author} {\bibfnamefont {R.}~\bibnamefont {Baer}},\ and\ \bibinfo {author} {\bibfnamefont {L.}~\bibnamefont {Kronik}},\ }\bibfield  {title} {\bibinfo {title} {{Quasiparticle Spectra from a Nonempirical Optimally Tuned Range-Separated Hybrid Density Functional}},\ }\href@noop {} {\bibfield  {journal} {\bibinfo  {journal} {Phys. Rev. Lett.}\ }\textbf {\bibinfo {volume} {109}},\ \bibinfo {pages} {226405} (\bibinfo {year} {2012})}\BibitemShut {NoStop}%
\bibitem [{\citenamefont {Kraisler}\ and\ \citenamefont {Kronik}(2013)}]{KraKro13}%
  \BibitemOpen
  \bibfield  {author} {\bibinfo {author} {\bibfnamefont {E.}~\bibnamefont {Kraisler}}\ and\ \bibinfo {author} {\bibfnamefont {L.}~\bibnamefont {Kronik}},\ }\bibfield  {title} {\bibinfo {title} {{Piecewise Linearity of Approximate Density Functionals Revisited: Implications for Frontier Orbital Energies}},\ }\href@noop {} {\bibfield  {journal} {\bibinfo  {journal} {Phys. Rev. Lett.}\ }\textbf {\bibinfo {volume} {110}},\ \bibinfo {pages} {126403} (\bibinfo {year} {2013})}\BibitemShut {NoStop}%
\bibitem [{\citenamefont {Nguyen}\ \emph {et~al.}(2015)\citenamefont {Nguyen}, \citenamefont {Borghi}, \citenamefont {Ferretti}, \citenamefont {Dabo},\ and\ \citenamefont {Marzari}}]{nguyen2015}%
  \BibitemOpen
  \bibfield  {author} {\bibinfo {author} {\bibfnamefont {N.~L.}\ \bibnamefont {Nguyen}}, \bibinfo {author} {\bibfnamefont {G.}~\bibnamefont {Borghi}}, \bibinfo {author} {\bibfnamefont {A.}~\bibnamefont {Ferretti}}, \bibinfo {author} {\bibfnamefont {I.}~\bibnamefont {Dabo}},\ and\ \bibinfo {author} {\bibfnamefont {N.}~\bibnamefont {Marzari}},\ }\bibfield  {title} {\bibinfo {title} {{First-Principles Photoemission Spectroscopy and Orbital Tomography in Molecules from Koopmans-Compliant Functionals}},\ }\href {https://doi.org/10.1103/PhysRevLett.114.166405} {\bibfield  {journal} {\bibinfo  {journal} {Phys. Rev. Lett.}\ }\textbf {\bibinfo {volume} {114}},\ \bibinfo {pages} {166405} (\bibinfo {year} {2015})}\BibitemShut {NoStop}%
\bibitem [{\citenamefont {Liu}\ \emph {et~al.}(2017)\citenamefont {Liu}, \citenamefont {Egger}, \citenamefont {Refaely-Abramson}, \citenamefont {Kronik},\ and\ \citenamefont {Neaton}}]{OTRSHadsorb17}%
  \BibitemOpen
  \bibfield  {author} {\bibinfo {author} {\bibfnamefont {Z.-F.}\ \bibnamefont {Liu}}, \bibinfo {author} {\bibfnamefont {D.~A.}\ \bibnamefont {Egger}}, \bibinfo {author} {\bibfnamefont {S.}~\bibnamefont {Refaely-Abramson}}, \bibinfo {author} {\bibfnamefont {L.}~\bibnamefont {Kronik}},\ and\ \bibinfo {author} {\bibfnamefont {J.~B.}\ \bibnamefont {Neaton}},\ }\bibfield  {title} {\bibinfo {title} {Energy level alignment at molecule-metal interfaces from an optimally tuned range-separated hybrid functional},\ }\href@noop {} {\bibfield  {journal} {\bibinfo  {journal} {J. Chem. Phys.}\ }\textbf {\bibinfo {volume} {146}},\ \bibinfo {pages} {092326} (\bibinfo {year} {2017})}\BibitemShut {NoStop}%
\bibitem [{\citenamefont {Wing}\ \emph {et~al.}(2021)\citenamefont {Wing}, \citenamefont {Ohad}, \citenamefont {Haber}, \citenamefont {Filip}, \citenamefont {Gant}, \citenamefont {Neaton},\ and\ \citenamefont {Kronik}}]{WiOhHa21}%
  \BibitemOpen
  \bibfield  {author} {\bibinfo {author} {\bibfnamefont {D.}~\bibnamefont {Wing}}, \bibinfo {author} {\bibfnamefont {G.}~\bibnamefont {Ohad}}, \bibinfo {author} {\bibfnamefont {J.~B.}\ \bibnamefont {Haber}}, \bibinfo {author} {\bibfnamefont {M.~R.}\ \bibnamefont {Filip}}, \bibinfo {author} {\bibfnamefont {S.~E.}\ \bibnamefont {Gant}}, \bibinfo {author} {\bibfnamefont {J.~B.}\ \bibnamefont {Neaton}},\ and\ \bibinfo {author} {\bibfnamefont {L.}~\bibnamefont {Kronik}},\ }\bibfield  {title} {\bibinfo {title} {{Band gaps of crystalline solids from Wannier-localization–based optimal tuning of a screened range-separated hybrid functional}},\ }\href@noop {} {\bibfield  {journal} {\bibinfo  {journal} {PNAS}\ }\textbf {\bibinfo {volume} {118}},\ \bibinfo {pages} {e2104556118} (\bibinfo {year} {2021})}\BibitemShut {NoStop}%
\bibitem [{\citenamefont {Hyldgaard}\ \emph {et~al.}(2020)\citenamefont {Hyldgaard}, \citenamefont {Jiao},\ and\ \citenamefont {Shukla}}]{JPCMreview}%
  \BibitemOpen
  \bibfield  {author} {\bibinfo {author} {\bibfnamefont {P.}~\bibnamefont {Hyldgaard}}, \bibinfo {author} {\bibfnamefont {Y.}~\bibnamefont {Jiao}},\ and\ \bibinfo {author} {\bibfnamefont {V.}~\bibnamefont {Shukla}},\ }\bibfield  {title} {\bibinfo {title} {{Screening nature of the van der Waals density functional method: A review and analysis of the many-body physics foundation}},\ }\href@noop {} {\bibfield  {journal} {\bibinfo  {journal} {J. Phys.: Condens. Matter}\ }\textbf {\bibinfo {volume} {32}},\ \bibinfo {pages} {393001} (\bibinfo {year} {2020})}\BibitemShut {NoStop}%
\bibitem [{\citenamefont {Racioppi}\ \emph {et~al.}(2023)\citenamefont {Racioppi}, \citenamefont {Lolur}, \citenamefont {Hyldgaard},\ and\ \citenamefont {Rahm}}]{ChiDFT23}%
  \BibitemOpen
  \bibfield  {author} {\bibinfo {author} {\bibfnamefont {S.}~\bibnamefont {Racioppi}}, \bibinfo {author} {\bibfnamefont {P.}~\bibnamefont {Lolur}}, \bibinfo {author} {\bibfnamefont {P.}~\bibnamefont {Hyldgaard}},\ and\ \bibinfo {author} {\bibfnamefont {M.}~\bibnamefont {Rahm}},\ }\bibfield  {title} {\bibinfo {title} {{A Density Functional Theory for the Average Electron Energy}},\ }\href@noop {} {\bibfield  {journal} {\bibinfo  {journal} {J. Chem. Theory Comput}\ }\textbf {\bibinfo {volume} {19}},\ \bibinfo {pages} {799} (\bibinfo {year} {2023})}\BibitemShut {NoStop}%
\bibitem [{\citenamefont {Schr{\"o}der}\ \emph {et~al.}(2025)\citenamefont {Schr{\"o}der}, \citenamefont {Quintero-Monsebaiz}, \citenamefont {Jiao},\ and\ \citenamefont {Hyldgaard}}]{AHBRmRSH25}%
  \BibitemOpen
  \bibfield  {author} {\bibinfo {author} {\bibfnamefont {E.}~\bibnamefont {Schr{\"o}der}}, \bibinfo {author} {\bibfnamefont {R.}~\bibnamefont {Quintero-Monsebaiz}}, \bibinfo {author} {\bibfnamefont {Y.}~\bibnamefont {Jiao}},\ and\ \bibinfo {author} {\bibfnamefont {P.}~\bibnamefont {Hyldgaard}},\ }\bibfield  {title} {\bibinfo {title} {{Optimally tuned range-separated hybrid van der Waals density functional for molecular binding and quasiparticle characterizations}},\ }\href@noop {} {\bibfield  {journal} {\bibinfo  {journal} {J. Phys.: Condens. Matter.}\ }\textbf {\bibinfo {volume} {37}},\ \bibinfo {pages} {211501} (\bibinfo {year} {2025})}\BibitemShut {NoStop}%
\bibitem [{\citenamefont {Quintero-Monsebaiz}\ and\ \citenamefont {Hyldgaard}(2026)}]{hBN2026}%
  \BibitemOpen
  \bibfield  {author} {\bibinfo {author} {\bibfnamefont {R.}~\bibnamefont {Quintero-Monsebaiz}}\ and\ \bibinfo {author} {\bibfnamefont {P.}~\bibnamefont {Hyldgaard}},\ }\bibfield  {title} {\bibinfo {title} {{Quasiparticle states of hexagonal BN: A van der Waals density functional study}},\ }\href@noop {} {\bibfield  {journal} {\bibinfo  {journal} {Phys. Rev. B}\ }\textbf {\bibinfo {volume} {113}},\ \bibinfo {pages} {195127} (\bibinfo {year} {2026})}\BibitemShut {NoStop}%
\bibitem [{\citenamefont {Nozi\`eres}\ and\ \citenamefont {Pines}(1958)}]{pinesnozieres}%
  \BibitemOpen
  \bibfield  {author} {\bibinfo {author} {\bibfnamefont {P.}~\bibnamefont {Nozi\`eres}}\ and\ \bibinfo {author} {\bibfnamefont {D.}~\bibnamefont {Pines}},\ }\bibfield  {title} {\bibinfo {title} {Correlation energy of a free electron gas},\ }\href@noop {} {\bibfield  {journal} {\bibinfo  {journal} {Phys. Rev.}\ }\textbf {\bibinfo {volume} {111}},\ \bibinfo {pages} {442} (\bibinfo {year} {1958})}\BibitemShut {NoStop}%
\bibitem [{\citenamefont {Fetter}\ and\ \citenamefont {Walecka}(1971)}]{FW7}%
  \BibitemOpen
  \bibfield  {author} {\bibinfo {author} {\bibfnamefont {A.~L.}\ \bibnamefont {Fetter}}\ and\ \bibinfo {author} {\bibfnamefont {J.~D.}\ \bibnamefont {Walecka}},\ }\href@noop {} {\emph {\bibinfo {title} {{Quantum theory of many-particle systems}}}}\ (\bibinfo  {publisher} {McGraw-Hill Book Company},\ \bibinfo {address} {New York},\ \bibinfo {year} {1971})\ pp.\ \bibinfo {pages} {64--82}\BibitemShut {NoStop}%
\bibitem [{\citenamefont {Mahan}(1990)}]{mahansbok}%
  \BibitemOpen
  \bibfield  {author} {\bibinfo {author} {\bibfnamefont {G.~D.}\ \bibnamefont {Mahan}},\ }\href@noop {} {\emph {\bibinfo {title} {Many-Particle Physics}}},\ \bibinfo {edition} {2nd}\ ed.\ (\bibinfo  {publisher} {Plenum Press},\ \bibinfo {address} {New York},\ \bibinfo {year} {1990})\BibitemShut {NoStop}%
\bibitem [{\citenamefont {Kohn}\ and\ \citenamefont {Sham}(1965)}]{kosh65}%
  \BibitemOpen
  \bibfield  {author} {\bibinfo {author} {\bibfnamefont {W.}~\bibnamefont {Kohn}}\ and\ \bibinfo {author} {\bibfnamefont {L.~J.}\ \bibnamefont {Sham}},\ }\bibfield  {title} {\bibinfo {title} {Self-consistent equations including exchange and correlation effects},\ }\href@noop {} {\bibfield  {journal} {\bibinfo  {journal} {Phys. Rev.}\ }\textbf {\bibinfo {volume} {140}},\ \bibinfo {pages} {A1133} (\bibinfo {year} {1965})}\BibitemShut {NoStop}%
\bibitem [{\citenamefont {Racioppi}\ \emph {et~al.}(2024)\citenamefont {Racioppi}, \citenamefont {Hyldgaard},\ and\ \citenamefont {Rahm}}]{ChiDFT24}%
  \BibitemOpen
  \bibfield  {author} {\bibinfo {author} {\bibfnamefont {S.}~\bibnamefont {Racioppi}}, \bibinfo {author} {\bibfnamefont {P.}~\bibnamefont {Hyldgaard}},\ and\ \bibinfo {author} {\bibfnamefont {M.}~\bibnamefont {Rahm}},\ }\bibfield  {title} {\bibinfo {title} {{Quantifying Atomic Volume, Partial Charge, and Electronegativity in Condensed Phases}},\ }\href {https://doi.org/10.1021/acs.jpcc.3c07677} {\bibfield  {journal} {\bibinfo  {journal} {J. Phys. Chem. C}\ }\textbf {\bibinfo {volume} {128}},\ \bibinfo {pages} {4009} (\bibinfo {year} {2024})}\BibitemShut {NoStop}%
\bibitem [{\citenamefont {Lundqvist}\ \emph {et~al.}(1983)\citenamefont {Lundqvist}, \citenamefont {Hellsing}, \citenamefont {Holmstr{\"o}m}, \citenamefont {Nordlander}, \citenamefont {Persson},\ and\ \citenamefont {N{\o}rskov}}]{BILAdsRev1983}%
  \BibitemOpen
  \bibfield  {author} {\bibinfo {author} {\bibfnamefont {B.~I.}\ \bibnamefont {Lundqvist}}, \bibinfo {author} {\bibfnamefont {B.}~\bibnamefont {Hellsing}}, \bibinfo {author} {\bibfnamefont {S.}~\bibnamefont {Holmstr{\"o}m}}, \bibinfo {author} {\bibfnamefont {P.}~\bibnamefont {Nordlander}}, \bibinfo {author} {\bibfnamefont {M.}~\bibnamefont {Persson}},\ and\ \bibinfo {author} {\bibfnamefont {J.~K.}\ \bibnamefont {N{\o}rskov}},\ }\bibfield  {title} {\bibinfo {title} {{Theoretical studies of molecular adsorption on metal surfaces}},\ }\href {https://doi.org/10.1002/qua560230332} {\bibfield  {journal} {\bibinfo  {journal} {Intl. J. Quant. Chem.}\ }\textbf {\bibinfo {volume} {23}},\ \bibinfo {pages} {1083} (\bibinfo {year} {1983})}\BibitemShut {NoStop}%
\bibitem [{\citenamefont {Kelkkanen}\ \emph {et~al.}(2011)\citenamefont {Kelkkanen}, \citenamefont {Lundqvist},\ and\ \citenamefont {N\o{}rskov}}]{kelkkanen11p113401}%
  \BibitemOpen
  \bibfield  {author} {\bibinfo {author} {\bibfnamefont {A.~K.}\ \bibnamefont {Kelkkanen}}, \bibinfo {author} {\bibfnamefont {B.~I.}\ \bibnamefont {Lundqvist}},\ and\ \bibinfo {author} {\bibfnamefont {J.~K.}\ \bibnamefont {N\o{}rskov}},\ }\bibfield  {title} {\bibinfo {title} {{Van der Waals effect in weak adsorption affecting trends in adsorption, reactivity, and the view of substrate nobility}},\ }\href@noop {} {\bibfield  {journal} {\bibinfo  {journal} {Phys. Rev. B}\ }\textbf {\bibinfo {volume} {83}},\ \bibinfo {pages} {113401} (\bibinfo {year} {2011})}\BibitemShut {NoStop}%
\bibitem [{\citenamefont {Hellsing}\ \emph {et~al.}(1983)\citenamefont {Hellsing}, \citenamefont {Persson},\ and\ \citenamefont {Lundqvist}}]{DampingCTH1983}%
  \BibitemOpen
  \bibfield  {author} {\bibinfo {author} {\bibfnamefont {B.}~\bibnamefont {Hellsing}}, \bibinfo {author} {\bibfnamefont {M.}~\bibnamefont {Persson}},\ and\ \bibinfo {author} {\bibfnamefont {B.~I.}\ \bibnamefont {Lundqvist}},\ }\bibfield  {title} {\bibinfo {title} {{Electronic damping mechanism for vibrations, rotations, and translations of adsorbates on metal surfaces}},\ }\href@noop {} {\bibfield  {journal} {\bibinfo  {journal} {Surf. Sci}\ }\textbf {\bibinfo {volume} {126}},\ \bibinfo {pages} {147} (\bibinfo {year} {1983})}\BibitemShut {NoStop}%
\bibitem [{\citenamefont {Hellsing}\ and\ \citenamefont {Persson}(1984)}]{HellsingPersson84}%
  \BibitemOpen
  \bibfield  {author} {\bibinfo {author} {\bibfnamefont {B.}~\bibnamefont {Hellsing}}\ and\ \bibinfo {author} {\bibfnamefont {M.}~\bibnamefont {Persson}},\ }\bibfield  {title} {\bibinfo {title} {{Electronic Damping of Atomic and Molecular Vibrations at Metal Surfaces}},\ }\href@noop {} {\bibfield  {journal} {\bibinfo  {journal} {Phys. Scr.}\ }\textbf {\bibinfo {volume} {29}},\ \bibinfo {pages} {306} (\bibinfo {year} {1984})}\BibitemShut {NoStop}%
\bibitem [{\citenamefont {Head-Gordon}\ and\ \citenamefont {Tully}(1995)}]{HeadGordonTully95}%
  \BibitemOpen
  \bibfield  {author} {\bibinfo {author} {\bibfnamefont {M.}~\bibnamefont {Head-Gordon}}\ and\ \bibinfo {author} {\bibfnamefont {J.~C.}\ \bibnamefont {Tully}},\ }\bibfield  {title} {\bibinfo {title} {{Molecular dynamics with electronic frictions}},\ }\href {https://doi.org/10.1063/1.469915} {\bibfield  {journal} {\bibinfo  {journal} {J. Chem. Phys.}\ }\textbf {\bibinfo {volume} {103}},\ \bibinfo {pages} {10137} (\bibinfo {year} {1995})}\BibitemShut {NoStop}%
\bibitem [{\citenamefont {Huang}\ \emph {et~al.}(2000)\citenamefont {Huang}, \citenamefont {Rettner}, \citenamefont {Auerbach},\ and\ \citenamefont {Wodtke}}]{Huang2000}%
  \BibitemOpen
  \bibfield  {author} {\bibinfo {author} {\bibfnamefont {Y.}~\bibnamefont {Huang}}, \bibinfo {author} {\bibfnamefont {C.~T.}\ \bibnamefont {Rettner}}, \bibinfo {author} {\bibfnamefont {D.~J.}\ \bibnamefont {Auerbach}},\ and\ \bibinfo {author} {\bibfnamefont {A.~M.}\ \bibnamefont {Wodtke}},\ }\bibfield  {title} {\bibinfo {title} {{Vibrational Promotion of Electron Transfer}},\ }\href {https://doi.org/10.1125/science.290.5489.111} {\bibfield  {journal} {\bibinfo  {journal} {Science}\ }\textbf {\bibinfo {volume} {290}},\ \bibinfo {pages} {111} (\bibinfo {year} {2000})}\BibitemShut {NoStop}%
\bibitem [{\citenamefont {Gerrits}\ \emph {et~al.}(2020)\citenamefont {Gerrits}, \citenamefont {Smeets}, \citenamefont {Vuckovic}, \citenamefont {Powell}, \citenamefont {Doblhoff-Dier},\ and\ \citenamefont {Kroes}}]{SurfChallenge2020}%
  \BibitemOpen
  \bibfield  {author} {\bibinfo {author} {\bibfnamefont {N.}~\bibnamefont {Gerrits}}, \bibinfo {author} {\bibfnamefont {E.~W.~F.}\ \bibnamefont {Smeets}}, \bibinfo {author} {\bibfnamefont {S.}~\bibnamefont {Vuckovic}}, \bibinfo {author} {\bibfnamefont {A.~D.}\ \bibnamefont {Powell}}, \bibinfo {author} {\bibfnamefont {K.}~\bibnamefont {Doblhoff-Dier}},\ and\ \bibinfo {author} {\bibfnamefont {G.-J.}\ \bibnamefont {Kroes}},\ }\bibfield  {title} {\bibinfo {title} {{Density Functional Theory for Molecule-Metal Surface Reactions: When Does the Generalized Gradient Approximation Get It Right, and What to Do If It Does Not}},\ }\href@noop {} {\bibfield  {journal} {\bibinfo  {journal} {J. Phys. Chem. Lett.}\ }\textbf {\bibinfo {volume} {11}},\ \bibinfo {pages} {10552} (\bibinfo {year} {2020})}\BibitemShut {NoStop}%
\bibitem [{\citenamefont {Zaremba}\ and\ \citenamefont {Kohn}(1976)}]{zarembakohn1976}%
  \BibitemOpen
  \bibfield  {author} {\bibinfo {author} {\bibfnamefont {E.}~\bibnamefont {Zaremba}}\ and\ \bibinfo {author} {\bibfnamefont {W.}~\bibnamefont {Kohn}},\ }\bibfield  {title} {\bibinfo {title} {{v}an der {W}aals interaction between an atom and a solid surface},\ }\href@noop {} {\bibfield  {journal} {\bibinfo  {journal} {Phys. Rev. B}\ }\textbf {\bibinfo {volume} {13}},\ \bibinfo {pages} {2270} (\bibinfo {year} {1976})}\BibitemShut {NoStop}%
\bibitem [{\citenamefont {Zaremba}\ and\ \citenamefont {Kohn}(1977)}]{zarembakohn1977}%
  \BibitemOpen
  \bibfield  {author} {\bibinfo {author} {\bibfnamefont {E.}~\bibnamefont {Zaremba}}\ and\ \bibinfo {author} {\bibfnamefont {W.}~\bibnamefont {Kohn}},\ }\bibfield  {title} {\bibinfo {title} {Theory of helium adsorption on simple and noble-metal surfaces},\ }\href@noop {} {\bibfield  {journal} {\bibinfo  {journal} {Phys. Rev. B}\ }\textbf {\bibinfo {volume} {15}},\ \bibinfo {pages} {1769} (\bibinfo {year} {1977})}\BibitemShut {NoStop}%
\bibitem [{\citenamefont {Harris}\ and\ \citenamefont {Liebsch}(1982)}]{harrisandliebsch1982b}%
  \BibitemOpen
  \bibfield  {author} {\bibinfo {author} {\bibfnamefont {J.}~\bibnamefont {Harris}}\ and\ \bibinfo {author} {\bibfnamefont {A.}~\bibnamefont {Liebsch}},\ }\bibfield  {title} {\bibinfo {title} {Interaction of helium with a metal surface},\ }\href@noop {} {\bibfield  {journal} {\bibinfo  {journal} {J. Phys. C}\ }\textbf {\bibinfo {volume} {15}},\ \bibinfo {pages} {2275} (\bibinfo {year} {1982})}\BibitemShut {NoStop}%
\bibitem [{\citenamefont {Nordlander}\ and\ \citenamefont {Harris}(1984)}]{harrisnordlander1984}%
  \BibitemOpen
  \bibfield  {author} {\bibinfo {author} {\bibfnamefont {P.}~\bibnamefont {Nordlander}}\ and\ \bibinfo {author} {\bibfnamefont {J.}~\bibnamefont {Harris}},\ }\bibfield  {title} {\bibinfo {title} {The interaction of helium with smooth metal surfaces},\ }\href@noop {} {\bibfield  {journal} {\bibinfo  {journal} {J. Phys. C}\ }\textbf {\bibinfo {volume} {17}},\ \bibinfo {pages} {1141} (\bibinfo {year} {1984})}\BibitemShut {NoStop}%
\bibitem [{\citenamefont {Andersson}\ and\ \citenamefont {Persson}(1993)}]{andersson1993}%
  \BibitemOpen
  \bibfield  {author} {\bibinfo {author} {\bibfnamefont {S.}~\bibnamefont {Andersson}}\ and\ \bibinfo {author} {\bibfnamefont {M.}~\bibnamefont {Persson}},\ }\bibfield  {title} {\bibinfo {title} {Sticking in the physisorption well: Influence of surface structure},\ }\href@noop {} {\bibfield  {journal} {\bibinfo  {journal} {Phys. Rev. Lett.}\ }\textbf {\bibinfo {volume} {70}},\ \bibinfo {pages} {202} (\bibinfo {year} {1993})}\BibitemShut {NoStop}%
\bibitem [{\citenamefont {Lee}\ \emph {et~al.}(2012{\natexlab{a}})\citenamefont {Lee}, \citenamefont {Berland}, \citenamefont {Yoon}, \citenamefont {Andersson}, \citenamefont {Schr{\"o}der}, \citenamefont {Hyldgaard},\ and\ \citenamefont {Lundqvist}}]{lee12p424213}%
  \BibitemOpen
  \bibfield  {author} {\bibinfo {author} {\bibfnamefont {K.}~\bibnamefont {Lee}}, \bibinfo {author} {\bibfnamefont {K.}~\bibnamefont {Berland}}, \bibinfo {author} {\bibfnamefont {M.}~\bibnamefont {Yoon}}, \bibinfo {author} {\bibfnamefont {S.}~\bibnamefont {Andersson}}, \bibinfo {author} {\bibfnamefont {E.}~\bibnamefont {Schr{\"o}der}}, \bibinfo {author} {\bibfnamefont {P.}~\bibnamefont {Hyldgaard}},\ and\ \bibinfo {author} {\bibfnamefont {B.~I.}\ \bibnamefont {Lundqvist}},\ }\bibfield  {title} {\bibinfo {title} {Benchmarking van der {W}aals density functionals with experimental data: potential-energy curves for {H}$_2$ molecules on {C}u(111), (100), and (110) surfaces},\ }\href@noop {} {\bibfield  {journal} {\bibinfo  {journal} {J. Phys.: Condens. Matter}\ }\textbf {\bibinfo {volume} {24}},\ \bibinfo {pages} {424213} (\bibinfo {year} {2012}{\natexlab{a}})}\BibitemShut {NoStop}%
\bibitem [{\citenamefont {Lundqvist}(1991)}]{AdsAspectBIL1991}%
  \BibitemOpen
  \bibfield  {author} {\bibinfo {author} {\bibfnamefont {B.~I.}\ \bibnamefont {Lundqvist}},\ }\bibfield  {title} {\bibinfo {title} {{Aspects of molecule-surface interactions}},\ }\href@noop {} {\bibfield  {journal} {\bibinfo  {journal} {Surf. Sci}\ }\textbf {\bibinfo {volume} {242}},\ \bibinfo {pages} {365} (\bibinfo {year} {1991})}\BibitemShut {NoStop}%
\bibitem [{\citenamefont {{\"O}sterlund}\ \emph {et~al.}(1997)\citenamefont {{\"O}sterlund}, \citenamefont {Zori{\'c}},\ and\ \citenamefont {Kasemo}}]{OsterlundZoricKasemo1997}%
  \BibitemOpen
  \bibfield  {author} {\bibinfo {author} {\bibfnamefont {L.}~\bibnamefont {{\"O}sterlund}}, \bibinfo {author} {\bibfnamefont {I.}~\bibnamefont {Zori{\'c}}},\ and\ \bibinfo {author} {\bibfnamefont {B.}~\bibnamefont {Kasemo}},\ }\bibfield  {title} {\bibinfo {title} {{Dissociative sticking of O$_2$ on Al(111)}},\ }\href {https://doi.org/10.1103/Phys.Rev.B.55.15452} {\bibfield  {journal} {\bibinfo  {journal} {Phys. Rev. B}\ }\textbf {\bibinfo {volume} {55}},\ \bibinfo {pages} {15452} (\bibinfo {year} {1997})}\BibitemShut {NoStop}%
\bibitem [{\citenamefont {Lauhon}\ and\ \citenamefont {Ho}(2000)}]{LauhonDyn2000}%
  \BibitemOpen
  \bibfield  {author} {\bibinfo {author} {\bibfnamefont {L.~J.}\ \bibnamefont {Lauhon}}\ and\ \bibinfo {author} {\bibfnamefont {W.}~\bibnamefont {Ho}},\ }\bibfield  {title} {\bibinfo {title} {{Direct Observation of the Quantum Tunneling of Single Hydrogen Atoms with a Scanning Tunneling Microscope}},\ }\href {https://doi.org/10.1103/Phys.Rev.Lett.85.4556} {\bibfield  {journal} {\bibinfo  {journal} {Phys. Rev. Lett}\ }\textbf {\bibinfo {volume} {85}},\ \bibinfo {pages} {4566} (\bibinfo {year} {2000})}\BibitemShut {NoStop}%
\bibitem [{\citenamefont {Lauhon}\ and\ \citenamefont {Ho}(2002)}]{LauhonDyn2002err}%
  \BibitemOpen
  \bibfield  {author} {\bibinfo {author} {\bibfnamefont {L.~J.}\ \bibnamefont {Lauhon}}\ and\ \bibinfo {author} {\bibfnamefont {W.}~\bibnamefont {Ho}},\ }\bibfield  {title} {\bibinfo {title} {{Erratum: Direct Observation of the Quantum Tunneling of Single Hydrogen Atoms with a Scanning Tunneling Microscope [Phys. Rev. Lett.PRLTAO0031-9007 \textbf{85}, 4566 (2000)]}},\ }\href {https://doi.org/10.1103/Phys.Rev.Lett.89.079901} {\bibfield  {journal} {\bibinfo  {journal} {Phys. Rev. Lett}\ }\textbf {\bibinfo {volume} {89}},\ \bibinfo {pages} {079901} (\bibinfo {year} {2002})}\BibitemShut {NoStop}%
\bibitem [{\citenamefont {Repp}\ \emph {et~al.}(2003)\citenamefont {Repp}, \citenamefont {Meyer}, \citenamefont {Rieder},\ and\ \citenamefont {Hyldgaard}}]{ReppDyn2003}%
  \BibitemOpen
  \bibfield  {author} {\bibinfo {author} {\bibfnamefont {J.}~\bibnamefont {Repp}}, \bibinfo {author} {\bibfnamefont {G.}~\bibnamefont {Meyer}}, \bibinfo {author} {\bibfnamefont {K.-H.}\ \bibnamefont {Rieder}},\ and\ \bibinfo {author} {\bibfnamefont {P.}~\bibnamefont {Hyldgaard}},\ }\bibfield  {title} {\bibinfo {title} {{Site Determination and Thermally Assisted Tunneling in Homogeneous Nucleation}},\ }\href {https://doi.org/10.1103/Phys.Rev.Lett.91.206102} {\bibfield  {journal} {\bibinfo  {journal} {Phys. Rev. Lett}\ }\textbf {\bibinfo {volume} {91}},\ \bibinfo {pages} {206102} (\bibinfo {year} {2003})}\BibitemShut {NoStop}%
\bibitem [{\citenamefont {Seidl}\ \emph {et~al.}(1996)\citenamefont {Seidl}, \citenamefont {G{\"o}rling}, \citenamefont {Vogl}, \citenamefont {Majewski},\ and\ \citenamefont {Levy}}]{GKSstart}%
  \BibitemOpen
  \bibfield  {author} {\bibinfo {author} {\bibfnamefont {A.}~\bibnamefont {Seidl}}, \bibinfo {author} {\bibfnamefont {A.}~\bibnamefont {G{\"o}rling}}, \bibinfo {author} {\bibfnamefont {P.}~\bibnamefont {Vogl}}, \bibinfo {author} {\bibfnamefont {J.~A.}\ \bibnamefont {Majewski}},\ and\ \bibinfo {author} {\bibfnamefont {M.}~\bibnamefont {Levy}},\ }\bibfield  {title} {\bibinfo {title} {{Generalized Kohn-Sham schemes and the band-gap problem}},\ }\href@noop {} {\bibfield  {journal} {\bibinfo  {journal} {Phys. Rev. B}\ }\textbf {\bibinfo {volume} {53}},\ \bibinfo {pages} {3764} (\bibinfo {year} {1996})}\BibitemShut {NoStop}%
\bibitem [{\citenamefont {Burke}(2012)}]{BurkePerspective}%
  \BibitemOpen
  \bibfield  {author} {\bibinfo {author} {\bibfnamefont {K.}~\bibnamefont {Burke}},\ }\bibfield  {title} {\bibinfo {title} {Perspective on density functional theory},\ }\href@noop {} {\bibfield  {journal} {\bibinfo  {journal} {J. Chem. Phys.}\ }\textbf {\bibinfo {volume} {136}},\ \bibinfo {pages} {150901} (\bibinfo {year} {2012})}\BibitemShut {NoStop}%
\bibitem [{\citenamefont {Becke}(2014)}]{beckeperspective}%
  \BibitemOpen
  \bibfield  {author} {\bibinfo {author} {\bibfnamefont {A.~D.}\ \bibnamefont {Becke}},\ }\bibfield  {title} {\bibinfo {title} {Perspective: Fifty years of density-functional theory in chemical physics},\ }\href@noop {} {\bibfield  {journal} {\bibinfo  {journal} {J. Chem. Phys.}\ }\textbf {\bibinfo {volume} {140}},\ \bibinfo {pages} {18A301} (\bibinfo {year} {2014})}\BibitemShut {NoStop}%
\bibitem [{\citenamefont {Burke}\ and\ \citenamefont {Wagner}(2013)}]{burke}%
  \BibitemOpen
  \bibfield  {author} {\bibinfo {author} {\bibfnamefont {K.}~\bibnamefont {Burke}}\ and\ \bibinfo {author} {\bibfnamefont {L.~O.}\ \bibnamefont {Wagner}},\ }\bibfield  {title} {\bibinfo {title} {{DFT} in a nutshell},\ }\href@noop {} {\bibfield  {journal} {\bibinfo  {journal} {Int. J. Quantum Chem.}\ }\textbf {\bibinfo {volume} {113}},\ \bibinfo {pages} {96} (\bibinfo {year} {2013})}\BibitemShut {NoStop}%
\bibitem [{\citenamefont {Chuang}\ \emph {et~al.}(1999)\citenamefont {Chuang}, \citenamefont {Radhakrishnan}, \citenamefont {Fast}, \citenamefont {Cramer},\ and\ \citenamefont {Truhlar}}]{SRPdef1999}%
  \BibitemOpen
  \bibfield  {author} {\bibinfo {author} {\bibfnamefont {Y.~Y.}\ \bibnamefont {Chuang}}, \bibinfo {author} {\bibfnamefont {M.~L.}\ \bibnamefont {Radhakrishnan}}, \bibinfo {author} {\bibfnamefont {P.~L.}\ \bibnamefont {Fast}}, \bibinfo {author} {\bibfnamefont {C.~J.}\ \bibnamefont {Cramer}},\ and\ \bibinfo {author} {\bibfnamefont {D.~G.}\ \bibnamefont {Truhlar}},\ }\bibfield  {title} {\bibinfo {title} {{Direct Dynamics for Free Radical Kinetics in Solution: Solvent Effect on the Rate Constant for the Reaction of Methanol with Atomic Hydrogen}},\ }\href@noop {} {\bibfield  {journal} {\bibinfo  {journal} {J. Phys. Chem. A}\ }\textbf {\bibinfo {volume} {103}},\ \bibinfo {pages} {4893–4909} (\bibinfo {year} {1999})}\BibitemShut {NoStop}%
\bibitem [{\citenamefont {Berland}\ \emph {et~al.}(2014)\citenamefont {Berland}, \citenamefont {Arter}, \citenamefont {Cooper}, \citenamefont {Lee}, \citenamefont {Lundqvist}, \citenamefont {Schr{\"o}der}, \citenamefont {Thonhauser},\ and\ \citenamefont {Hyldgaard}}]{bearcoleluscthhy14}%
  \BibitemOpen
  \bibfield  {author} {\bibinfo {author} {\bibfnamefont {K.}~\bibnamefont {Berland}}, \bibinfo {author} {\bibfnamefont {C.~A.}\ \bibnamefont {Arter}}, \bibinfo {author} {\bibfnamefont {V.~R.}\ \bibnamefont {Cooper}}, \bibinfo {author} {\bibfnamefont {K.}~\bibnamefont {Lee}}, \bibinfo {author} {\bibfnamefont {B.~I.}\ \bibnamefont {Lundqvist}}, \bibinfo {author} {\bibfnamefont {E.}~\bibnamefont {Schr{\"o}der}}, \bibinfo {author} {\bibfnamefont {T.}~\bibnamefont {Thonhauser}},\ and\ \bibinfo {author} {\bibfnamefont {P.}~\bibnamefont {Hyldgaard}},\ }\bibfield  {title} {\bibinfo {title} {{v}an der {W}aals density functionals built upon the electron-gas tradition: Facing the challenge of competing interactions},\ }\href@noop {} {\bibfield  {journal} {\bibinfo  {journal} {J. Chem. Phys.}\ }\textbf {\bibinfo {volume} {140}},\ \bibinfo {pages} {18A539} (\bibinfo {year} {2014})}\BibitemShut {NoStop}%
\bibitem [{\citenamefont {Shukla}\ \emph {et~al.}(2022{\natexlab{a}})\citenamefont {Shukla}, \citenamefont {Jiao}, \citenamefont {Lee}, \citenamefont {Schr{\"o}der}, \citenamefont {Neaton},\ and\ \citenamefont {Hyldgaard}}]{AHBRlaunch}%
  \BibitemOpen
  \bibfield  {author} {\bibinfo {author} {\bibfnamefont {V.}~\bibnamefont {Shukla}}, \bibinfo {author} {\bibfnamefont {Y.}~\bibnamefont {Jiao}}, \bibinfo {author} {\bibfnamefont {J.-H.}\ \bibnamefont {Lee}}, \bibinfo {author} {\bibfnamefont {E.}~\bibnamefont {Schr{\"o}der}}, \bibinfo {author} {\bibfnamefont {J.~B.}\ \bibnamefont {Neaton}},\ and\ \bibinfo {author} {\bibfnamefont {P.}~\bibnamefont {Hyldgaard}},\ }\bibfield  {title} {\bibinfo {title} {{Accurate Nonempirical Range-Separated Hybrid van der Waals Density Functional for Complex Molecular Problems, Solids, and Surfaces}},\ }\href@noop {} {\bibfield  {journal} {\bibinfo  {journal} {Phys.\ Rev.\ X}\ }\textbf {\bibinfo {volume} {12}},\ \bibinfo {pages} {041003} (\bibinfo {year} {2022}{\natexlab{a}})}\BibitemShut {NoStop}%
\bibitem [{\citenamefont {Cococcioni}\ and\ \citenamefont {de~Gironcoli}(2005)}]{cococcioni2005}%
  \BibitemOpen
  \bibfield  {author} {\bibinfo {author} {\bibfnamefont {M.}~\bibnamefont {Cococcioni}}\ and\ \bibinfo {author} {\bibfnamefont {S.}~\bibnamefont {de~Gironcoli}},\ }\bibfield  {title} {\bibinfo {title} {Linear response approach to the calculation of the effective interaction parameters in the $\mathrm{LDA}+\mathrm{U}$ method},\ }\href {https://doi.org/10.1103/PhysRevB.71.035105} {\bibfield  {journal} {\bibinfo  {journal} {Phys. Rev. B}\ }\textbf {\bibinfo {volume} {71}},\ \bibinfo {pages} {035105} (\bibinfo {year} {2005})}\BibitemShut {NoStop}%
\bibitem [{\citenamefont {Kuisma}\ \emph {et~al.}(2010)\citenamefont {Kuisma}, \citenamefont {Ojanen}, \citenamefont {Enkovaara},\ and\ \citenamefont {Rantala}}]{KuismaGB}%
  \BibitemOpen
  \bibfield  {author} {\bibinfo {author} {\bibfnamefont {M.}~\bibnamefont {Kuisma}}, \bibinfo {author} {\bibfnamefont {J.}~\bibnamefont {Ojanen}}, \bibinfo {author} {\bibfnamefont {J.}~\bibnamefont {Enkovaara}},\ and\ \bibinfo {author} {\bibfnamefont {T.~T.}\ \bibnamefont {Rantala}},\ }\bibfield  {title} {\bibinfo {title} {{Kohn-Sham potential with discontinuity for band gap materials}},\ }\href {https://doi.org/10.1103/PhysRevB.82.115106} {\bibfield  {journal} {\bibinfo  {journal} {Phys. Rev. B}\ }\textbf {\bibinfo {volume} {82}},\ \bibinfo {pages} {115106} (\bibinfo {year} {2010})}\BibitemShut {NoStop}%
\bibitem [{\citenamefont {Ma}\ and\ \citenamefont {Wang}(2016)}]{Ma2016}%
  \BibitemOpen
  \bibfield  {author} {\bibinfo {author} {\bibfnamefont {J.}~\bibnamefont {Ma}}\ and\ \bibinfo {author} {\bibfnamefont {L.-W.}\ \bibnamefont {Wang}},\ }\bibfield  {title} {\bibinfo {title} {{Using Wannier functions to improve solid band gap predictions in density functional theory}},\ }\href {https://doi.org/10.1038/srep24924} {\bibfield  {journal} {\bibinfo  {journal} {Sci. Rep.}\ }\textbf {\bibinfo {volume} {6}},\ \bibinfo {pages} {24924} (\bibinfo {year} {2016})}\BibitemShut {NoStop}%
\bibitem [{\citenamefont {Nguyen}\ \emph {et~al.}(2016)\citenamefont {Nguyen}, \citenamefont {Borghi}, \citenamefont {Ferretti},\ and\ \citenamefont {Marzari}}]{nguyen2016}%
  \BibitemOpen
  \bibfield  {author} {\bibinfo {author} {\bibfnamefont {N.~L.}\ \bibnamefont {Nguyen}}, \bibinfo {author} {\bibfnamefont {G.}~\bibnamefont {Borghi}}, \bibinfo {author} {\bibfnamefont {A.}~\bibnamefont {Ferretti}},\ and\ \bibinfo {author} {\bibfnamefont {N.}~\bibnamefont {Marzari}},\ }\bibfield  {title} {\bibinfo {title} {{First-Principles Photoemission Spectroscopy of DNA and RNA Nucleobases from Koopmans-Compliant Functionals}},\ }\href {https://doi.org/10.1021/acs.jctc.6b00145} {\bibfield  {journal} {\bibinfo  {journal} {J. Chem. Theory Comput.}\ }\textbf {\bibinfo {volume} {12}},\ \bibinfo {pages} {3948} (\bibinfo {year} {2016})}\BibitemShut {NoStop}%
\bibitem [{\citenamefont {Colonna}\ \emph {et~al.}(2022)\citenamefont {Colonna}, \citenamefont {De~Gennaro}, \citenamefont {Linscott},\ and\ \citenamefont {Marzari}}]{colonna2022}%
  \BibitemOpen
  \bibfield  {author} {\bibinfo {author} {\bibfnamefont {N.}~\bibnamefont {Colonna}}, \bibinfo {author} {\bibfnamefont {R.}~\bibnamefont {De~Gennaro}}, \bibinfo {author} {\bibfnamefont {E.}~\bibnamefont {Linscott}},\ and\ \bibinfo {author} {\bibfnamefont {N.}~\bibnamefont {Marzari}},\ }\bibfield  {title} {\bibinfo {title} {Koopmans spectral functionals in periodic boundary conditions},\ }\href {https://doi.org/10.1021/acs.jctc.2c00161} {\bibfield  {journal} {\bibinfo  {journal} {J. Chem. Theory Comput.}\ }\textbf {\bibinfo {volume} {18}},\ \bibinfo {pages} {5435} (\bibinfo {year} {2022})}\BibitemShut {NoStop}%
\bibitem [{\citenamefont {Camarasa-G{\'o}mez}\ \emph {et~al.}(2024)\citenamefont {Camarasa-G{\'o}mez}, \citenamefont {Gant}, \citenamefont {Ohad}, \citenamefont {Neaton}, \citenamefont {Ramasubramaniam},\ and\ \citenamefont {Kronik}}]{GoGaOh2024}%
  \BibitemOpen
  \bibfield  {author} {\bibinfo {author} {\bibfnamefont {M.}~\bibnamefont {Camarasa-G{\'o}mez}}, \bibinfo {author} {\bibfnamefont {S.~E.}\ \bibnamefont {Gant}}, \bibinfo {author} {\bibfnamefont {G.}~\bibnamefont {Ohad}}, \bibinfo {author} {\bibfnamefont {J.~B.}\ \bibnamefont {Neaton}}, \bibinfo {author} {\bibfnamefont {A.}~\bibnamefont {Ramasubramaniam}},\ and\ \bibinfo {author} {\bibfnamefont {L.}~\bibnamefont {Kronik}},\ }\bibfield  {title} {\bibinfo {title} {{Excitations in layered materials from a non-empirical Wannier-localized optimally-tuned screened ranged-separated hybrid functional}},\ }\href {https://doi.org/10.1038/s41524-024-01478-1} {\bibfield  {journal} {\bibinfo  {journal} {npj Comput. Mater.}\ }\textbf {\bibinfo {volume} {10}},\ \bibinfo {pages} {1} (\bibinfo {year} {2024})}\BibitemShut {NoStop}%
\bibitem [{\citenamefont {Quintero-Monsebaiz}\ \emph {et~al.}(2026)\citenamefont {Quintero-Monsebaiz}, \citenamefont {Hyldgaard},\ and\ \citenamefont {Schr{\"o}der}}]{NitrogenBasesAHBR-mRSH26}%
  \BibitemOpen
  \bibfield  {author} {\bibinfo {author} {\bibfnamefont {R.}~\bibnamefont {Quintero-Monsebaiz}}, \bibinfo {author} {\bibfnamefont {P.}~\bibnamefont {Hyldgaard}},\ and\ \bibinfo {author} {\bibfnamefont {E.}~\bibnamefont {Schr{\"o}der}},\ }\bibfield  {title} {\bibinfo {title} {{Nature of frontier quasi-particle states in nitrogen-base systems}},\ }\href@noop {} {\bibfield  {journal} {\bibinfo  {journal} {Phys. Chem. Chem. Phys.}\ }\textbf {\bibinfo {volume} {28}},\ \bibinfo {pages} {3336} (\bibinfo {year} {2026})}\BibitemShut {NoStop}%
\bibitem [{\citenamefont {Olsen}\ \emph {et~al.}(2003)\citenamefont {Olsen}, \citenamefont {Philipsen},\ and\ \citenamefont {Baerends}}]{olsen2003co}%
  \BibitemOpen
  \bibfield  {author} {\bibinfo {author} {\bibfnamefont {R.~A.}\ \bibnamefont {Olsen}}, \bibinfo {author} {\bibfnamefont {P.~H.~T.}\ \bibnamefont {Philipsen}},\ and\ \bibinfo {author} {\bibfnamefont {E.~J.}\ \bibnamefont {Baerends}},\ }\bibfield  {title} {\bibinfo {title} {{CO} on {P}t (111): A puzzle revisited},\ }\href@noop {} {\bibfield  {journal} {\bibinfo  {journal} {J. Chem. Phys.}\ }\textbf {\bibinfo {volume} {119}},\ \bibinfo {pages} {4522} (\bibinfo {year} {2003})}\BibitemShut {NoStop}%
\bibitem [{\citenamefont {Shukla}\ \emph {et~al.}(2022{\natexlab{b}})\citenamefont {Shukla}, \citenamefont {Jiao}, \citenamefont {Frostenson},\ and\ \citenamefont {Hyldgaard}}]{DefineAHCX}%
  \BibitemOpen
  \bibfield  {author} {\bibinfo {author} {\bibfnamefont {V.}~\bibnamefont {Shukla}}, \bibinfo {author} {\bibfnamefont {Y.}~\bibnamefont {Jiao}}, \bibinfo {author} {\bibfnamefont {C.~M.}\ \bibnamefont {Frostenson}},\ and\ \bibinfo {author} {\bibfnamefont {P.}~\bibnamefont {Hyldgaard}},\ }\bibfield  {title} {\bibinfo {title} {{vdW-DF-ahcx: a range-separated van der Waals density functional hybrid}},\ }\href@noop {} {\bibfield  {journal} {\bibinfo  {journal} {J. Phys.: Condens. Matter}\ }\textbf {\bibinfo {volume} {34}},\ \bibinfo {pages} {025902} (\bibinfo {year} {2022}{\natexlab{b}})}\BibitemShut {NoStop}%
\bibitem [{\citenamefont {Yourdshahyan}\ \emph {et~al.}(2002)\citenamefont {Yourdshahyan}, \citenamefont {Razaznejad},\ and\ \citenamefont {Lundqvist}}]{OadsYourdBIL2002}%
  \BibitemOpen
  \bibfield  {author} {\bibinfo {author} {\bibfnamefont {Y.}~\bibnamefont {Yourdshahyan}}, \bibinfo {author} {\bibfnamefont {B.}~\bibnamefont {Razaznejad}},\ and\ \bibinfo {author} {\bibfnamefont {B.~I.}\ \bibnamefont {Lundqvist}},\ }\bibfield  {title} {\bibinfo {title} {{Adiabatic potential-energy surfaces for oxygen on Al(111)}},\ }\href {https://doi.org/10.1103/PhysRevB.65.075416} {\bibfield  {journal} {\bibinfo  {journal} {Phys. Rev. B}\ }\textbf {\bibinfo {volume} {65}},\ \bibinfo {pages} {075416} (\bibinfo {year} {2002})}\BibitemShut {NoStop}%
\bibitem [{\citenamefont {Hellman}\ \emph {et~al.}(2005)\citenamefont {Hellman}, \citenamefont {Razaznejad},\ and\ \citenamefont {Lundqvist}}]{adsHellmanBIL2005}%
  \BibitemOpen
  \bibfield  {author} {\bibinfo {author} {\bibfnamefont {A.}~\bibnamefont {Hellman}}, \bibinfo {author} {\bibfnamefont {B.}~\bibnamefont {Razaznejad}},\ and\ \bibinfo {author} {\bibfnamefont {B.~I.}\ \bibnamefont {Lundqvist}},\ }\bibfield  {title} {\bibinfo {title} {{Trends in sticking and adsorption of diatomic molecules on the Al(111) surface}},\ }\href {https://doi.org/10.1103/PhysRevB.71.205424} {\bibfield  {journal} {\bibinfo  {journal} {Phys. Rev. B}\ }\textbf {\bibinfo {volume} {71}},\ \bibinfo {pages} {205424} (\bibinfo {year} {2005})}\BibitemShut {NoStop}%
\bibitem [{\citenamefont {Yin}\ \emph {et~al.}(2018)\citenamefont {Yin}, \citenamefont {Zhang}, \citenamefont {Libisch}, \citenamefont {Carter}, \citenamefont {Guo},\ and\ \citenamefont {Jiang}}]{YinZhang2018}%
  \BibitemOpen
  \bibfield  {author} {\bibinfo {author} {\bibfnamefont {R.}~\bibnamefont {Yin}}, \bibinfo {author} {\bibfnamefont {Y.}~\bibnamefont {Zhang}}, \bibinfo {author} {\bibfnamefont {F.}~\bibnamefont {Libisch}}, \bibinfo {author} {\bibfnamefont {E.~A.}\ \bibnamefont {Carter}}, \bibinfo {author} {\bibfnamefont {H.}~\bibnamefont {Guo}},\ and\ \bibinfo {author} {\bibfnamefont {B.}~\bibnamefont {Jiang}},\ }\bibfield  {title} {\bibinfo {title} {{Dissociative Chemisorption of O$_2$ on Al(111): Dynamics on a Correlated Wave-Function-Based Potential Energy Surface}},\ }\href {https://doi.org/10.1021/acs.jpclett.8b01470} {\bibfield  {journal} {\bibinfo  {journal} {J. Phys. Chem. Lett.}\ }\textbf {\bibinfo {volume} {9}},\ \bibinfo {pages} {3271} (\bibinfo {year} {2018})}\BibitemShut {NoStop}%
\bibitem [{\citenamefont {Schimka}\ \emph {et~al.}(2010)\citenamefont {Schimka}, \citenamefont {Harl}, \citenamefont {Stroppa}, \citenamefont {Gr\"{u}neis}, \citenamefont {Marsman}, \citenamefont {Mittendorfer},\ and\ \citenamefont {Kresse}}]{rpa:ads}%
  \BibitemOpen
  \bibfield  {author} {\bibinfo {author} {\bibfnamefont {L.}~\bibnamefont {Schimka}}, \bibinfo {author} {\bibfnamefont {J.}~\bibnamefont {Harl}}, \bibinfo {author} {\bibfnamefont {A.}~\bibnamefont {Stroppa}}, \bibinfo {author} {\bibfnamefont {A.}~\bibnamefont {Gr\"{u}neis}}, \bibinfo {author} {\bibfnamefont {M.}~\bibnamefont {Marsman}}, \bibinfo {author} {\bibfnamefont {F.}~\bibnamefont {Mittendorfer}},\ and\ \bibinfo {author} {\bibfnamefont {G.}~\bibnamefont {Kresse}},\ }\bibfield  {title} {\bibinfo {title} {Accurate surface and adsorption energies from many-body perturbation theory},\ }\href@noop {} {\bibfield  {journal} {\bibinfo  {journal} {Nat. Mater.}\ }\textbf {\bibinfo {volume} {9}},\ \bibinfo {pages} {741} (\bibinfo {year} {2010})}\BibitemShut {NoStop}%
\bibitem [{\citenamefont {Ren}\ \emph {et~al.}(2011)\citenamefont {Ren}, \citenamefont {Tkatchenko}, \citenamefont {Rinke},\ and\ \citenamefont {Scheffler}}]{rpa_single}%
  \BibitemOpen
  \bibfield  {author} {\bibinfo {author} {\bibfnamefont {X.}~\bibnamefont {Ren}}, \bibinfo {author} {\bibfnamefont {A.}~\bibnamefont {Tkatchenko}}, \bibinfo {author} {\bibfnamefont {P.}~\bibnamefont {Rinke}},\ and\ \bibinfo {author} {\bibfnamefont {M.}~\bibnamefont {Scheffler}},\ }\bibfield  {title} {\bibinfo {title} {Beyond the random-phase approximation for the electron correlation energy: {T}he importance of single excitations},\ }\href@noop {} {\bibfield  {journal} {\bibinfo  {journal} {Phys. Rev. Lett.}\ }\textbf {\bibinfo {volume} {106}},\ \bibinfo {pages} {153003} (\bibinfo {year} {2011})}\BibitemShut {NoStop}%
\bibitem [{\citenamefont {Hyldgaard}\ \emph {et~al.}(2014)\citenamefont {Hyldgaard}, \citenamefont {Berland},\ and\ \citenamefont {Schr{\"o}der}}]{hybesc14}%
  \BibitemOpen
  \bibfield  {author} {\bibinfo {author} {\bibfnamefont {P.}~\bibnamefont {Hyldgaard}}, \bibinfo {author} {\bibfnamefont {K.}~\bibnamefont {Berland}},\ and\ \bibinfo {author} {\bibfnamefont {E.}~\bibnamefont {Schr{\"o}der}},\ }\bibfield  {title} {\bibinfo {title} {Interpretation of van der {W}aals density functionals},\ }\href@noop {} {\bibfield  {journal} {\bibinfo  {journal} {Phys. Rev. B}\ }\textbf {\bibinfo {volume} {90}},\ \bibinfo {pages} {075148} (\bibinfo {year} {2014})}\BibitemShut {NoStop}%
\bibitem [{\citenamefont {Ma}\ and\ \citenamefont {Brueckner}(1968)}]{mabr}%
  \BibitemOpen
  \bibfield  {author} {\bibinfo {author} {\bibfnamefont {S.-K.}\ \bibnamefont {Ma}}\ and\ \bibinfo {author} {\bibfnamefont {K.~A.}\ \bibnamefont {Brueckner}},\ }\bibfield  {title} {\bibinfo {title} {Correlation energy of an electron gas with a slowly varying high density},\ }\href@noop {} {\bibfield  {journal} {\bibinfo  {journal} {Phys. Rev.}\ }\textbf {\bibinfo {volume} {165}},\ \bibinfo {pages} {18} (\bibinfo {year} {1968})}\BibitemShut {NoStop}%
\bibitem [{\citenamefont {Singwi}\ \emph {et~al.}(1968)\citenamefont {Singwi}, \citenamefont {Tosi}, \citenamefont {Land},\ and\ \citenamefont {Sj{\"o}lander}}]{Singwi68}%
  \BibitemOpen
  \bibfield  {author} {\bibinfo {author} {\bibfnamefont {K.~S.}\ \bibnamefont {Singwi}}, \bibinfo {author} {\bibfnamefont {M.~P.}\ \bibnamefont {Tosi}}, \bibinfo {author} {\bibfnamefont {R.~H.}\ \bibnamefont {Land}},\ and\ \bibinfo {author} {\bibfnamefont {A.}~\bibnamefont {Sj{\"o}lander}},\ }\bibfield  {title} {\bibinfo {title} {Electron correlations at metallic densities},\ }\href@noop {} {\bibfield  {journal} {\bibinfo  {journal} {Phys. Rev.}\ }\textbf {\bibinfo {volume} {176}},\ \bibinfo {pages} {589} (\bibinfo {year} {1968})}\BibitemShut {NoStop}%
\bibitem [{\citenamefont {Singwi}\ \emph {et~al.}(1969)\citenamefont {Singwi}, \citenamefont {Sj{\"o}lander}, \citenamefont {Tosi},\ and\ \citenamefont {Land}}]{Singwi69}%
  \BibitemOpen
  \bibfield  {author} {\bibinfo {author} {\bibfnamefont {K.~S.}\ \bibnamefont {Singwi}}, \bibinfo {author} {\bibfnamefont {A.}~\bibnamefont {Sj{\"o}lander}}, \bibinfo {author} {\bibfnamefont {M.~P.}\ \bibnamefont {Tosi}},\ and\ \bibinfo {author} {\bibfnamefont {R.~H.}\ \bibnamefont {Land}},\ }\bibfield  {title} {\bibinfo {title} {{Electron Correlations at Metallic Densities -- III}},\ }\href@noop {} {\bibfield  {journal} {\bibinfo  {journal} {Sol. State Commun.}\ }\textbf {\bibinfo {volume} {7}},\ \bibinfo {pages} {1503} (\bibinfo {year} {1969})}\BibitemShut {NoStop}%
\bibitem [{\citenamefont {Singwi}\ \emph {et~al.}(1970)\citenamefont {Singwi}, \citenamefont {Sj{\"o}lander}, \citenamefont {Tosi},\ and\ \citenamefont {Land}}]{Singwi70}%
  \BibitemOpen
  \bibfield  {author} {\bibinfo {author} {\bibfnamefont {K.~S.}\ \bibnamefont {Singwi}}, \bibinfo {author} {\bibfnamefont {A.}~\bibnamefont {Sj{\"o}lander}}, \bibinfo {author} {\bibfnamefont {M.~P.}\ \bibnamefont {Tosi}},\ and\ \bibinfo {author} {\bibfnamefont {R.~H.}\ \bibnamefont {Land}},\ }\bibfield  {title} {\bibinfo {title} {{Electron Correlations at Metallic Densities. IV}},\ }\href@noop {} {\bibfield  {journal} {\bibinfo  {journal} {Phys. Rev. B}\ }\textbf {\bibinfo {volume} {1}},\ \bibinfo {pages} {1044} (\bibinfo {year} {1970})}\BibitemShut {NoStop}%
\bibitem [{\citenamefont {Rasolt}\ and\ \citenamefont {Geldart}(1975)}]{rasolt}%
  \BibitemOpen
  \bibfield  {author} {\bibinfo {author} {\bibfnamefont {M.}~\bibnamefont {Rasolt}}\ and\ \bibinfo {author} {\bibfnamefont {D.~J.~W.}\ \bibnamefont {Geldart}},\ }\bibfield  {title} {\bibinfo {title} {Gradient corrections in the exchange and correlation energy of an inhomogeneous electron gas},\ }\href@noop {} {\bibfield  {journal} {\bibinfo  {journal} {Phys. Rev. Lett.}\ }\textbf {\bibinfo {volume} {35}},\ \bibinfo {pages} {1234} (\bibinfo {year} {1975})}\BibitemShut {NoStop}%
\bibitem [{\citenamefont {Langreth}\ and\ \citenamefont {Vosko}(1987)}]{lavo87}%
  \BibitemOpen
  \bibfield  {author} {\bibinfo {author} {\bibfnamefont {D.~C.}\ \bibnamefont {Langreth}}\ and\ \bibinfo {author} {\bibfnamefont {S.~H.}\ \bibnamefont {Vosko}},\ }\bibfield  {title} {\bibinfo {title} {Exact electron-gas response functions at high density},\ }\href@noop {} {\bibfield  {journal} {\bibinfo  {journal} {Phys. Rev. Lett.}\ }\textbf {\bibinfo {volume} {59}},\ \bibinfo {pages} {497} (\bibinfo {year} {1987})}\BibitemShut {NoStop}%
\bibitem [{\citenamefont {Thonhauser}\ \emph {et~al.}(2007)\citenamefont {Thonhauser}, \citenamefont {Cooper}, \citenamefont {Li}, \citenamefont {Puzder}, \citenamefont {Hyldgaard},\ and\ \citenamefont {Langreth}}]{thonhauser}%
  \BibitemOpen
  \bibfield  {author} {\bibinfo {author} {\bibfnamefont {T.}~\bibnamefont {Thonhauser}}, \bibinfo {author} {\bibfnamefont {V.~R.}\ \bibnamefont {Cooper}}, \bibinfo {author} {\bibfnamefont {S.}~\bibnamefont {Li}}, \bibinfo {author} {\bibfnamefont {A.}~\bibnamefont {Puzder}}, \bibinfo {author} {\bibfnamefont {P.}~\bibnamefont {Hyldgaard}},\ and\ \bibinfo {author} {\bibfnamefont {D.~C.}\ \bibnamefont {Langreth}},\ }\bibfield  {title} {\bibinfo {title} {{v}an der {W}aals density functional: {S}elf-consistent potential and the nature of the van der {W}aals bond},\ }\href@noop {} {\bibfield  {journal} {\bibinfo  {journal} {Phys. Rev. B.}\ }\textbf {\bibinfo {volume} {76}},\ \bibinfo {pages} {125112} (\bibinfo {year} {2007})}\BibitemShut {NoStop}%
\bibitem [{\citenamefont {Perdew}\ and\ \citenamefont {Wang}(1986)}]{pewa86}%
  \BibitemOpen
  \bibfield  {author} {\bibinfo {author} {\bibfnamefont {J.~P.}\ \bibnamefont {Perdew}}\ and\ \bibinfo {author} {\bibfnamefont {Y.}~\bibnamefont {Wang}},\ }\bibfield  {title} {\bibinfo {title} {Accurate and simple density functional for the electronic exchange energy: {G}eneralized gradient approximation},\ }\href@noop {} {\bibfield  {journal} {\bibinfo  {journal} {Phys. Rev. B}\ }\textbf {\bibinfo {volume} {33}},\ \bibinfo {pages} {8800} (\bibinfo {year} {1986})}\BibitemShut {NoStop}%
\bibitem [{\citenamefont {Perdew}\ \emph {et~al.}(1996{\natexlab{a}})\citenamefont {Perdew}, \citenamefont {Burke},\ and\ \citenamefont {Wang}}]{pebuwa96}%
  \BibitemOpen
  \bibfield  {author} {\bibinfo {author} {\bibfnamefont {J.~P.}\ \bibnamefont {Perdew}}, \bibinfo {author} {\bibfnamefont {K.}~\bibnamefont {Burke}},\ and\ \bibinfo {author} {\bibfnamefont {Y.}~\bibnamefont {Wang}},\ }\bibfield  {title} {\bibinfo {title} {Generalized gradient approximation for the exchange-correlation hole of a many-electron system},\ }\href@noop {} {\bibfield  {journal} {\bibinfo  {journal} {Phys. Rev. B}\ }\textbf {\bibinfo {volume} {54}},\ \bibinfo {pages} {16533} (\bibinfo {year} {1996}{\natexlab{a}})}\BibitemShut {NoStop}%
\bibitem [{\citenamefont {Perdew}\ \emph {et~al.}(1996{\natexlab{b}})\citenamefont {Perdew}, \citenamefont {Burke},\ and\ \citenamefont {Ernzerhof}}]{pebuer96}%
  \BibitemOpen
  \bibfield  {author} {\bibinfo {author} {\bibfnamefont {J.~P.}\ \bibnamefont {Perdew}}, \bibinfo {author} {\bibfnamefont {K.}~\bibnamefont {Burke}},\ and\ \bibinfo {author} {\bibfnamefont {M.}~\bibnamefont {Ernzerhof}},\ }\bibfield  {title} {\bibinfo {title} {Generalized gradient approximation made simple},\ }\href@noop {} {\bibfield  {journal} {\bibinfo  {journal} {Phys. Rev. Lett.}\ }\textbf {\bibinfo {volume} {77}},\ \bibinfo {pages} {3865} (\bibinfo {year} {1996}{\natexlab{b}})}\BibitemShut {NoStop}%
\bibitem [{\citenamefont {Perdew}\ \emph {et~al.}(2008)\citenamefont {Perdew}, \citenamefont {Ruzsinszky}, \citenamefont {Csonka}, \citenamefont {Vydrov}, \citenamefont {Scuseria}, \citenamefont {Constantin}, \citenamefont {Zhou},\ and\ \citenamefont {Burke}}]{PBEsol}%
  \BibitemOpen
  \bibfield  {author} {\bibinfo {author} {\bibfnamefont {J.~P.}\ \bibnamefont {Perdew}}, \bibinfo {author} {\bibfnamefont {A.}~\bibnamefont {Ruzsinszky}}, \bibinfo {author} {\bibfnamefont {G.~I.}\ \bibnamefont {Csonka}}, \bibinfo {author} {\bibfnamefont {O.~A.}\ \bibnamefont {Vydrov}}, \bibinfo {author} {\bibfnamefont {G.~E.}\ \bibnamefont {Scuseria}}, \bibinfo {author} {\bibfnamefont {L.~A.}\ \bibnamefont {Constantin}}, \bibinfo {author} {\bibfnamefont {X.}~\bibnamefont {Zhou}},\ and\ \bibinfo {author} {\bibfnamefont {K.}~\bibnamefont {Burke}},\ }\bibfield  {title} {\bibinfo {title} {Restoring the density-gradient expansion for exchange in solids and surfaces},\ }\href@noop {} {\bibfield  {journal} {\bibinfo  {journal} {Phys. Rev. Lett.}\ }\textbf {\bibinfo {volume} {100}},\ \bibinfo {pages} {136406} (\bibinfo {year} {2008})}\BibitemShut {NoStop}%
\bibitem [{\citenamefont {Sun}\ \emph {et~al.}(2015)\citenamefont {Sun}, \citenamefont {Ruzsinszky},\ and\ \citenamefont {Perdew}}]{SCAN}%
  \BibitemOpen
  \bibfield  {author} {\bibinfo {author} {\bibfnamefont {J.}~\bibnamefont {Sun}}, \bibinfo {author} {\bibfnamefont {A.}~\bibnamefont {Ruzsinszky}},\ and\ \bibinfo {author} {\bibfnamefont {J.~P.}\ \bibnamefont {Perdew}},\ }\bibfield  {title} {\bibinfo {title} {Strongly constrained and appropriately normed semilocal density functional},\ }\href@noop {} {\bibfield  {journal} {\bibinfo  {journal} {Phys. Rev. Lett.}\ }\textbf {\bibinfo {volume} {115}},\ \bibinfo {pages} {036402} (\bibinfo {year} {2015})}\BibitemShut {NoStop}%
\bibitem [{\citenamefont {Peng}\ \emph {et~al.}(2016)\citenamefont {Peng}, \citenamefont {Yang}, \citenamefont {Perdew},\ and\ \citenamefont {Sun}}]{SCANvdW}%
  \BibitemOpen
  \bibfield  {author} {\bibinfo {author} {\bibfnamefont {H.}~\bibnamefont {Peng}}, \bibinfo {author} {\bibfnamefont {Z.-H.}\ \bibnamefont {Yang}}, \bibinfo {author} {\bibfnamefont {J.~P.}\ \bibnamefont {Perdew}},\ and\ \bibinfo {author} {\bibfnamefont {J.}~\bibnamefont {Sun}},\ }\bibfield  {title} {\bibinfo {title} {{Versatile van der Waals Density Functional Based on a Meta-Generalized Gradient Approximation}},\ }\href {https://doi.org/10.1103/PhysRevX.6.041005} {\bibfield  {journal} {\bibinfo  {journal} {Phys. Rev. X}\ }\textbf {\bibinfo {volume} {6}},\ \bibinfo {pages} {041005} (\bibinfo {year} {2016})}\BibitemShut {NoStop}%
\bibitem [{\citenamefont {Andersson}\ \emph {et~al.}(1996{\natexlab{b}})\citenamefont {Andersson}, \citenamefont {Langreth},\ and\ \citenamefont {Lundqvist}}]{anlalu96}%
  \BibitemOpen
  \bibfield  {author} {\bibinfo {author} {\bibfnamefont {Y.}~\bibnamefont {Andersson}}, \bibinfo {author} {\bibfnamefont {D.~C.}\ \bibnamefont {Langreth}},\ and\ \bibinfo {author} {\bibfnamefont {B.~I.}\ \bibnamefont {Lundqvist}},\ }\bibfield  {title} {\bibinfo {title} {{van der Waals Interactions in Density-Functional Theory}},\ }\href@noop {} {\bibfield  {journal} {\bibinfo  {journal} {Phys. Rev. Lett.}\ }\textbf {\bibinfo {volume} {76}},\ \bibinfo {pages} {102} (\bibinfo {year} {1996}{\natexlab{b}})}\BibitemShut {NoStop}%
\bibitem [{\citenamefont {Dion}\ \emph {et~al.}(2004)\citenamefont {Dion}, \citenamefont {Rydberg}, \citenamefont {Schr{\"o}der}, \citenamefont {Langreth},\ and\ \citenamefont {Lundqvist}}]{Dion}%
  \BibitemOpen
  \bibfield  {author} {\bibinfo {author} {\bibfnamefont {M.}~\bibnamefont {Dion}}, \bibinfo {author} {\bibfnamefont {H.}~\bibnamefont {Rydberg}}, \bibinfo {author} {\bibfnamefont {E.}~\bibnamefont {Schr{\"o}der}}, \bibinfo {author} {\bibfnamefont {D.~C.}\ \bibnamefont {Langreth}},\ and\ \bibinfo {author} {\bibfnamefont {B.~I.}\ \bibnamefont {Lundqvist}},\ }\bibfield  {title} {\bibinfo {title} {{Van der Waals Density Functional for General Geometries}},\ }\href@noop {} {\bibfield  {journal} {\bibinfo  {journal} {Phys. Rev. Lett.}\ }\textbf {\bibinfo {volume} {92}},\ \bibinfo {pages} {246401} (\bibinfo {year} {2004})}\BibitemShut {NoStop}%
\bibitem [{\citenamefont {Berland}\ \emph {et~al.}(2015)\citenamefont {Berland}, \citenamefont {Cooper}, \citenamefont {Lee}, \citenamefont {Schr\"{o}der}, \citenamefont {Thonhauser}, \citenamefont {Hyldgaard},\ and\ \citenamefont {Lundqvist}}]{Berland_2015:van_waals}%
  \BibitemOpen
  \bibfield  {author} {\bibinfo {author} {\bibfnamefont {K.}~\bibnamefont {Berland}}, \bibinfo {author} {\bibfnamefont {V.~R.}\ \bibnamefont {Cooper}}, \bibinfo {author} {\bibfnamefont {K.}~\bibnamefont {Lee}}, \bibinfo {author} {\bibfnamefont {E.}~\bibnamefont {Schr\"{o}der}}, \bibinfo {author} {\bibfnamefont {T.}~\bibnamefont {Thonhauser}}, \bibinfo {author} {\bibfnamefont {P.}~\bibnamefont {Hyldgaard}},\ and\ \bibinfo {author} {\bibfnamefont {B.~I.}\ \bibnamefont {Lundqvist}},\ }\bibfield  {title} {\bibinfo {title} {{v}an der {W}aals forces in density functional theory: {A} review of the {vdW-DF} method},\ }\href {https://doi.org/10.1088/0034-4885/78/6/066501} {\bibfield  {journal} {\bibinfo  {journal} {Rep. Prog. Phys.}\ }\textbf {\bibinfo {volume} {78}},\ \bibinfo {pages} {066501} (\bibinfo {year} {2015})}\BibitemShut {NoStop}%
\bibitem [{\citenamefont {Thonhauser}\ \emph {et~al.}(2015)\citenamefont {Thonhauser}, \citenamefont {Zuluaga}, \citenamefont {Arter}, \citenamefont {Berland}, \citenamefont {Schr{\"{o}}der},\ and\ \citenamefont {Hyldgaard}}]{Thonhauser_2015:spin_signature}%
  \BibitemOpen
  \bibfield  {author} {\bibinfo {author} {\bibfnamefont {T.}~\bibnamefont {Thonhauser}}, \bibinfo {author} {\bibfnamefont {S.}~\bibnamefont {Zuluaga}}, \bibinfo {author} {\bibfnamefont {C.~A.}\ \bibnamefont {Arter}}, \bibinfo {author} {\bibfnamefont {K.}~\bibnamefont {Berland}}, \bibinfo {author} {\bibfnamefont {E.}~\bibnamefont {Schr{\"{o}}der}},\ and\ \bibinfo {author} {\bibfnamefont {P.}~\bibnamefont {Hyldgaard}},\ }\bibfield  {title} {\bibinfo {title} {{Spin Signature of Nonlocal Correlation Binding in Metal-Organic Frameworks}},\ }\href {https://doi.org/10.1103/PhysRevLett.115.136402} {\bibfield  {journal} {\bibinfo  {journal} {Phys. Rev. Lett.}\ }\textbf {\bibinfo {volume} {115}},\ \bibinfo {pages} {136402} (\bibinfo {year} {2015})}\BibitemShut {NoStop}%
\bibitem [{\citenamefont {Berland}\ \emph {et~al.}(2017)\citenamefont {Berland}, \citenamefont {Jiao}, \citenamefont {Lee}, \citenamefont {Rangel}, \citenamefont {Neaton},\ and\ \citenamefont {Hyldgaard}}]{DFcx02017}%
  \BibitemOpen
  \bibfield  {author} {\bibinfo {author} {\bibfnamefont {K.}~\bibnamefont {Berland}}, \bibinfo {author} {\bibfnamefont {Y.}~\bibnamefont {Jiao}}, \bibinfo {author} {\bibfnamefont {J.-H.}\ \bibnamefont {Lee}}, \bibinfo {author} {\bibfnamefont {T.}~\bibnamefont {Rangel}}, \bibinfo {author} {\bibfnamefont {J.~B.}\ \bibnamefont {Neaton}},\ and\ \bibinfo {author} {\bibfnamefont {P.}~\bibnamefont {Hyldgaard}},\ }\bibfield  {title} {\bibinfo {title} {Assessment of two hybrid van der {Waals} density functionals for covalent and non-covalent binding of molecules},\ }\href@noop {} {\bibfield  {journal} {\bibinfo  {journal} {J. Chem. Phys.}\ }\textbf {\bibinfo {volume} {146}},\ \bibinfo {pages} {234106} (\bibinfo {year} {2017})}\BibitemShut {NoStop}%
\bibitem [{\citenamefont {Burke}\ \emph {et~al.}(1997)\citenamefont {Burke}, \citenamefont {Ernzerhof},\ and\ \citenamefont {Perdew}}]{Burke97}%
  \BibitemOpen
  \bibfield  {author} {\bibinfo {author} {\bibfnamefont {K.}~\bibnamefont {Burke}}, \bibinfo {author} {\bibfnamefont {M.}~\bibnamefont {Ernzerhof}},\ and\ \bibinfo {author} {\bibfnamefont {J.~P.}\ \bibnamefont {Perdew}},\ }\bibfield  {title} {\bibinfo {title} {The adiabatic connection method: a non-empirical hybrid},\ }\href {https://doi.org/http://dx.doi.org/10.1016/S0009-2614(96)01373-5} {\bibfield  {journal} {\bibinfo  {journal} {Chem. Phys. Lett.}\ }\textbf {\bibinfo {volume} {265}},\ \bibinfo {pages} {115 } (\bibinfo {year} {1997})}\BibitemShut {NoStop}%
\bibitem [{\citenamefont {Adamo}\ and\ \citenamefont {Barone}(1999)}]{PBE0}%
  \BibitemOpen
  \bibfield  {author} {\bibinfo {author} {\bibfnamefont {C.}~\bibnamefont {Adamo}}\ and\ \bibinfo {author} {\bibfnamefont {V.}~\bibnamefont {Barone}},\ }\bibfield  {title} {\bibinfo {title} {Towards reliable density functional methods without adjustable parameters: {T}he {PBE}0 model},\ }\href@noop {} {\bibfield  {journal} {\bibinfo  {journal} {J. Chem. Phys.}\ }\textbf {\bibinfo {volume} {110}},\ \bibinfo {pages} {6158} (\bibinfo {year} {1999})}\BibitemShut {NoStop}%
\bibitem [{\citenamefont {Wei}\ \emph {et~al.}(2023)\citenamefont {Wei}, \citenamefont {Martirez},\ and\ \citenamefont {Carter}}]{Wei2023}%
  \BibitemOpen
  \bibfield  {author} {\bibinfo {author} {\bibfnamefont {Z.}~\bibnamefont {Wei}}, \bibinfo {author} {\bibfnamefont {J.~M.~P.}\ \bibnamefont {Martirez}},\ and\ \bibinfo {author} {\bibfnamefont {E.~A.}\ \bibnamefont {Carter}},\ }\bibfield  {title} {\bibinfo {title} {{Introducing the embedded random phase approximation: H$2$ dissociative adsorption on Cu(111) as an examplar}},\ }\href {https://doi.org/10.1063/5.0181229} {\bibfield  {journal} {\bibinfo  {journal} {J. Chem. Phys.}\ }\textbf {\bibinfo {volume} {159}},\ \bibinfo {pages} {194108} (\bibinfo {year} {2023})}\BibitemShut {NoStop}%
\bibitem [{\citenamefont {Oudot}\ and\ \citenamefont {Doblhoff-Dier}(2024)}]{Oudot2024}%
  \BibitemOpen
  \bibfield  {author} {\bibinfo {author} {\bibfnamefont {B.}~\bibnamefont {Oudot}}\ and\ \bibinfo {author} {\bibfnamefont {K.}~\bibnamefont {Doblhoff-Dier}},\ }\bibfield  {title} {\bibinfo {title} {{Reaction barriers at metal surfaces computed using the random phase approximation: Can we beat DFT in the generalized gradient approximations?}},\ }\href {https://doi.org/10.1063/5.0220465} {\bibfield  {journal} {\bibinfo  {journal} {J. Chem. Phys.}\ }\textbf {\bibinfo {volume} {161}},\ \bibinfo {pages} {054708} (\bibinfo {year} {2024})}\BibitemShut {NoStop}%
\bibitem [{\citenamefont {Tchakoua}\ \emph {et~al.}(2023)\citenamefont {Tchakoua}, \citenamefont {Gerrits}, \citenamefont {Smeets},\ and\ \citenamefont {Kroes}}]{SBH17}%
  \BibitemOpen
  \bibfield  {author} {\bibinfo {author} {\bibfnamefont {T.}~\bibnamefont {Tchakoua}}, \bibinfo {author} {\bibfnamefont {N.}~\bibnamefont {Gerrits}}, \bibinfo {author} {\bibfnamefont {E.~W.~F.}\ \bibnamefont {Smeets}},\ and\ \bibinfo {author} {\bibfnamefont {G.-J.}\ \bibnamefont {Kroes}},\ }\bibfield  {title} {\bibinfo {title} {{SBH17: Benchmark Database of Barrier Heights for Dissociative Chemisorption on Transition Metal Surfaces}},\ }\href {https://doi.org/10.1021/acs.jctc.2c00824} {\bibfield  {journal} {\bibinfo  {journal} {J. Chem. Theory Comput.}\ }\textbf {\bibinfo {volume} {19}},\ \bibinfo {pages} {245} (\bibinfo {year} {2023})}\BibitemShut {NoStop}%
\bibitem [{\citenamefont {Lee}\ \emph {et~al.}(2010{\natexlab{a}})\citenamefont {Lee}, \citenamefont {Murray}, \citenamefont {Kong}, \citenamefont {Lundqvist},\ and\ \citenamefont {Langreth}}]{lee10p081101}%
  \BibitemOpen
  \bibfield  {author} {\bibinfo {author} {\bibfnamefont {K.}~\bibnamefont {Lee}}, \bibinfo {author} {\bibfnamefont {{\`E}.~D.}\ \bibnamefont {Murray}}, \bibinfo {author} {\bibfnamefont {L.}~\bibnamefont {Kong}}, \bibinfo {author} {\bibfnamefont {B.~I.}\ \bibnamefont {Lundqvist}},\ and\ \bibinfo {author} {\bibfnamefont {D.~C.}\ \bibnamefont {Langreth}},\ }\bibfield  {title} {\bibinfo {title} {Higher-accuracy van der {W}aals density functional},\ }\href@noop {} {\bibfield  {journal} {\bibinfo  {journal} {Phys. Rev. B}\ }\textbf {\bibinfo {volume} {82}},\ \bibinfo {pages} {081101(R)} (\bibinfo {year} {2010}{\natexlab{a}})}\BibitemShut {NoStop}%
\bibitem [{\citenamefont {Hamada}(2014)}]{hamada14}%
  \BibitemOpen
  \bibfield  {author} {\bibinfo {author} {\bibfnamefont {I.}~\bibnamefont {Hamada}},\ }\bibfield  {title} {\bibinfo {title} {van der {W}aals density functional made accurate},\ }\href@noop {} {\bibfield  {journal} {\bibinfo  {journal} {Phys. Rev. B}\ }\textbf {\bibinfo {volume} {89}},\ \bibinfo {pages} {121103(R)} (\bibinfo {year} {2014})}\BibitemShut {NoStop}%
\bibitem [{\citenamefont {Becke}(1986)}]{becke1986p7184}%
  \BibitemOpen
  \bibfield  {author} {\bibinfo {author} {\bibfnamefont {A.~D.}\ \bibnamefont {Becke}},\ }\bibfield  {title} {\bibinfo {title} {On the large-gradient behavior of the density functional exchange energy},\ }\href@noop {} {\bibfield  {journal} {\bibinfo  {journal} {J. Chem. Phys.}\ }\textbf {\bibinfo {volume} {85}},\ \bibinfo {pages} {7184} (\bibinfo {year} {1986})}\BibitemShut {NoStop}%
\bibitem [{\citenamefont {Feibelman}\ \emph {et~al.}(2001)\citenamefont {Feibelman}, \citenamefont {Hammer}, \citenamefont {N{\o}rskov}, \citenamefont {Wagner}, \citenamefont {Scheffler}, \citenamefont {Stumpf}, \citenamefont {Watwe},\ and\ \citenamefont {Dumesic}}]{Feibelman01p4018}%
  \BibitemOpen
  \bibfield  {author} {\bibinfo {author} {\bibfnamefont {P.~J.}\ \bibnamefont {Feibelman}}, \bibinfo {author} {\bibfnamefont {B.}~\bibnamefont {Hammer}}, \bibinfo {author} {\bibfnamefont {J.~K.}\ \bibnamefont {N{\o}rskov}}, \bibinfo {author} {\bibfnamefont {F.}~\bibnamefont {Wagner}}, \bibinfo {author} {\bibfnamefont {M.}~\bibnamefont {Scheffler}}, \bibinfo {author} {\bibfnamefont {R.}~\bibnamefont {Stumpf}}, \bibinfo {author} {\bibfnamefont {R.}~\bibnamefont {Watwe}},\ and\ \bibinfo {author} {\bibfnamefont {J.}~\bibnamefont {Dumesic}},\ }\bibfield  {title} {\bibinfo {title} {The {CO/P}t(111) puzzle},\ }\href@noop {} {\bibfield  {journal} {\bibinfo  {journal} {J. Phys. Chem. B}\ }\textbf {\bibinfo {volume} {105}},\ \bibinfo {pages} {4018} (\bibinfo {year} {2001})}\BibitemShut {NoStop}%
\bibitem [{\citenamefont {Heyd}\ \emph {et~al.}(2003)\citenamefont {Heyd}, \citenamefont {Scuseria},\ and\ \citenamefont {Ernzerhof}}]{HSE03}%
  \BibitemOpen
  \bibfield  {author} {\bibinfo {author} {\bibfnamefont {J.}~\bibnamefont {Heyd}}, \bibinfo {author} {\bibfnamefont {G.~E.}\ \bibnamefont {Scuseria}},\ and\ \bibinfo {author} {\bibfnamefont {M.}~\bibnamefont {Ernzerhof}},\ }\bibfield  {title} {\bibinfo {title} {{Hybrid} functionals based on a screened {C}oulomb potential},\ }\href {https://doi.org/https://doi.org/10.1063/1.1564060} {\bibfield  {journal} {\bibinfo  {journal} {J. Chem. Phys.}\ }\textbf {\bibinfo {volume} {118}},\ \bibinfo {pages} {8207} (\bibinfo {year} {2003})}\BibitemShut {NoStop}%
\bibitem [{\citenamefont {Heyd}\ \emph {et~al.}(2006)\citenamefont {Heyd}, \citenamefont {Scuseria},\ and\ \citenamefont {Ernzerhof}}]{HSE06}%
  \BibitemOpen
  \bibfield  {author} {\bibinfo {author} {\bibfnamefont {J.}~\bibnamefont {Heyd}}, \bibinfo {author} {\bibfnamefont {G.~E.}\ \bibnamefont {Scuseria}},\ and\ \bibinfo {author} {\bibfnamefont {M.}~\bibnamefont {Ernzerhof}},\ }\bibfield  {title} {\bibinfo {title} {{Erratum:} "{Hybrid} functionals based on a screened {C}oulomb potential" [{J. Chem. Phys. 118, 8207 (2003)}]},\ }\href {https://doi.org/https://doi.org/10.1063/1.2204597} {\bibfield  {journal} {\bibinfo  {journal} {J. Chem. Phys.}\ }\textbf {\bibinfo {volume} {124}},\ \bibinfo {pages} {219906} (\bibinfo {year} {2006})}\BibitemShut {NoStop}%
\bibitem [{\citenamefont {Henderson}\ \emph {et~al.}(2008)\citenamefont {Henderson}, \citenamefont {Janesko},\ and\ \citenamefont {Scusseria}}]{HJS08}%
  \BibitemOpen
  \bibfield  {author} {\bibinfo {author} {\bibfnamefont {T.~M.}\ \bibnamefont {Henderson}}, \bibinfo {author} {\bibfnamefont {B.~G.}\ \bibnamefont {Janesko}},\ and\ \bibinfo {author} {\bibfnamefont {G.~E.}\ \bibnamefont {Scusseria}},\ }\bibfield  {title} {\bibinfo {title} {Generalized gradient approximation model exchange holes for range-separated hybrids},\ }\href@noop {} {\bibfield  {journal} {\bibinfo  {journal} {J. Chem. Phys}\ }\textbf {\bibinfo {volume} {128}},\ \bibinfo {pages} {194105} (\bibinfo {year} {2008})}\BibitemShut {NoStop}%
\bibitem [{\citenamefont {Lee}\ \emph {et~al.}(2010{\natexlab{b}})\citenamefont {Lee}, \citenamefont {Furche},\ and\ \citenamefont {Burke}}]{BurkeSIE}%
  \BibitemOpen
  \bibfield  {author} {\bibinfo {author} {\bibfnamefont {D.}~\bibnamefont {Lee}}, \bibinfo {author} {\bibfnamefont {F.}~\bibnamefont {Furche}},\ and\ \bibinfo {author} {\bibfnamefont {K.}~\bibnamefont {Burke}},\ }\bibfield  {title} {\bibinfo {title} {{Accuracy of Electron Affinities of Atoms in Approximate Density Functional Theory}},\ }\href@noop {} {\bibfield  {journal} {\bibinfo  {journal} {J. Phys. Chem. Lett.}\ }\textbf {\bibinfo {volume} {1}},\ \bibinfo {pages} {2124} (\bibinfo {year} {2010}{\natexlab{b}})}\BibitemShut {NoStop}%
\bibitem [{\citenamefont {Kronik}\ \emph {et~al.}(2012)\citenamefont {Kronik}, \citenamefont {Stein}, \citenamefont {Refaely-Abramson},\ and\ \citenamefont {Baer}}]{kronik2012}%
  \BibitemOpen
  \bibfield  {author} {\bibinfo {author} {\bibfnamefont {L.}~\bibnamefont {Kronik}}, \bibinfo {author} {\bibfnamefont {T.}~\bibnamefont {Stein}}, \bibinfo {author} {\bibfnamefont {S.}~\bibnamefont {Refaely-Abramson}},\ and\ \bibinfo {author} {\bibfnamefont {R.}~\bibnamefont {Baer}},\ }\bibfield  {title} {\bibinfo {title} {{Excitation Gaps of Finite-Sized Systems from Optimally Tuned Range-Separated Hybrid Functionals}},\ }\href {https://doi.org/10.1021/ct2009363} {\bibfield  {journal} {\bibinfo  {journal} {J. Chem. Theory Comput.}\ }\textbf {\bibinfo {volume} {8}},\ \bibinfo {pages} {1515} (\bibinfo {year} {2012})}\BibitemShut {NoStop}%
\bibitem [{\citenamefont {Roy}(1976)}]{roy}%
  \BibitemOpen
  \bibfield  {author} {\bibinfo {author} {\bibfnamefont {R.~J.~L.}\ \bibnamefont {Roy}},\ }\bibfield  {title} {\bibinfo {title} {Determining potential energy constants for atom- and molecule-surface interactions},\ }\href@noop {} {\bibfield  {journal} {\bibinfo  {journal} {Surf. Sci.}\ }\textbf {\bibinfo {volume} {59}},\ \bibinfo {pages} {541} (\bibinfo {year} {1976})}\BibitemShut {NoStop}%
\bibitem [{\citenamefont {Lee}\ \emph {et~al.}(2012{\natexlab{b}})\citenamefont {Lee}, \citenamefont {Kolb}, \citenamefont {Thonhauser}, \citenamefont {Vanderbilt},\ and\ \citenamefont {Langreth}}]{lee12p104102}%
  \BibitemOpen
  \bibfield  {author} {\bibinfo {author} {\bibfnamefont {K.}~\bibnamefont {Lee}}, \bibinfo {author} {\bibfnamefont {B.}~\bibnamefont {Kolb}}, \bibinfo {author} {\bibfnamefont {T.}~\bibnamefont {Thonhauser}}, \bibinfo {author} {\bibfnamefont {D.}~\bibnamefont {Vanderbilt}},\ and\ \bibinfo {author} {\bibfnamefont {D.~C.}\ \bibnamefont {Langreth}},\ }\bibfield  {title} {\bibinfo {title} {Structure and energetics of a ferroelectric organic crystal of phenazine and chloranilic acid},\ }\href@noop {} {\bibfield  {journal} {\bibinfo  {journal} {Phys. Rev. B}\ }\textbf {\bibinfo {volume} {86}},\ \bibinfo {pages} {104102} (\bibinfo {year} {2012}{\natexlab{b}})}\BibitemShut {NoStop}%
\bibitem [{\citenamefont {Berland}\ and\ \citenamefont {Hyldgaard}(2014)}]{behy14}%
  \BibitemOpen
  \bibfield  {author} {\bibinfo {author} {\bibfnamefont {K.}~\bibnamefont {Berland}}\ and\ \bibinfo {author} {\bibfnamefont {P.}~\bibnamefont {Hyldgaard}},\ }\bibfield  {title} {\bibinfo {title} {Exchange functional that tests the robustness of the plasmon description of the van der {W}aals density functional},\ }\href@noop {} {\bibfield  {journal} {\bibinfo  {journal} {Phys. Rev. B}\ }\textbf {\bibinfo {volume} {89}},\ \bibinfo {pages} {035412} (\bibinfo {year} {2014})}\BibitemShut {NoStop}%
\bibitem [{\citenamefont {Perrichon}\ \emph {et~al.}(2020)\citenamefont {Perrichon}, \citenamefont {Granhed}, \citenamefont {Romanelli}, \citenamefont {Piovano}, \citenamefont {Lindman}, \citenamefont {Hyldgaard}, \citenamefont {Wahnstr{\"o}m},\ and\ \citenamefont {Karlsson}}]{PeGrRo20}%
  \BibitemOpen
  \bibfield  {author} {\bibinfo {author} {\bibfnamefont {A.}~\bibnamefont {Perrichon}}, \bibinfo {author} {\bibfnamefont {E.~J.}\ \bibnamefont {Granhed}}, \bibinfo {author} {\bibfnamefont {G.}~\bibnamefont {Romanelli}}, \bibinfo {author} {\bibfnamefont {A.}~\bibnamefont {Piovano}}, \bibinfo {author} {\bibfnamefont {A.}~\bibnamefont {Lindman}}, \bibinfo {author} {\bibfnamefont {P.}~\bibnamefont {Hyldgaard}}, \bibinfo {author} {\bibfnamefont {G.}~\bibnamefont {Wahnstr{\"o}m}},\ and\ \bibinfo {author} {\bibfnamefont {M.}~\bibnamefont {Karlsson}},\ }\bibfield  {title} {\bibinfo {title} {{Unraveling the Ground-State Structure of BaZrO$_3$ by Neutron Scattering Experiments and First-Principles Calculations}},\ }\href@noop {} {\bibfield  {journal} {\bibinfo  {journal} {Chem. Mater.}\ }\textbf {\bibinfo {volume} {32}},\ \bibinfo {pages} {2824} (\bibinfo {year} {2020})}\BibitemShut {NoStop}%
\bibitem [{\citenamefont {Granhed}\ \emph {et~al.}(2020)\citenamefont {Granhed}, \citenamefont {Wahnstr{\"o}m},\ and\ \citenamefont {Hyldgaard}}]{jewahy20}%
  \BibitemOpen
  \bibfield  {author} {\bibinfo {author} {\bibfnamefont {E.~J.}\ \bibnamefont {Granhed}}, \bibinfo {author} {\bibfnamefont {G.}~\bibnamefont {Wahnstr{\"o}m}},\ and\ \bibinfo {author} {\bibfnamefont {P.}~\bibnamefont {Hyldgaard}},\ }\bibfield  {title} {\bibinfo {title} {{BaZrO$_3$ stability under pressure: The role of nonlocal exchange and correlation}},\ }\href@noop {} {\bibfield  {journal} {\bibinfo  {journal} {Phys. Rev. B}\ }\textbf {\bibinfo {volume} {101}},\ \bibinfo {pages} {224105} (\bibinfo {year} {2020})}\BibitemShut {NoStop}%
\bibitem [{\citenamefont {Frostenson}\ \emph {et~al.}(2022)\citenamefont {Frostenson}, \citenamefont {Granhed}, \citenamefont {Shukla}, \citenamefont {Olsson}, \citenamefont {Schr{\"o}der},\ and\ \citenamefont {Hyldgaard}}]{Hard2Soft}%
  \BibitemOpen
  \bibfield  {author} {\bibinfo {author} {\bibfnamefont {C.~M.}\ \bibnamefont {Frostenson}}, \bibinfo {author} {\bibfnamefont {E.~J.}\ \bibnamefont {Granhed}}, \bibinfo {author} {\bibfnamefont {V.}~\bibnamefont {Shukla}}, \bibinfo {author} {\bibfnamefont {P.~A.~T.}\ \bibnamefont {Olsson}}, \bibinfo {author} {\bibfnamefont {E.}~\bibnamefont {Schr{\"o}der}},\ and\ \bibinfo {author} {\bibfnamefont {P.}~\bibnamefont {Hyldgaard}},\ }\bibfield  {title} {\bibinfo {title} {{Hard and soft materials: Putting consistent van der Waals density functionals to work}},\ }\href@noop {} {\bibfield  {journal} {\bibinfo  {journal} {Electron. Struct.}\ }\textbf {\bibinfo {volume} {4}},\ \bibinfo {pages} {014001} (\bibinfo {year} {2022})}\BibitemShut {NoStop}%
\bibitem [{\citenamefont {Perdew}\ and\ \citenamefont {Zunger}(1981)}]{PerZun81}%
  \BibitemOpen
  \bibfield  {author} {\bibinfo {author} {\bibfnamefont {J.~P.}\ \bibnamefont {Perdew}}\ and\ \bibinfo {author} {\bibfnamefont {A.}~\bibnamefont {Zunger}},\ }\bibfield  {title} {\bibinfo {title} {{Self-interaction correction to density-functional approximations for many-electron systems}},\ }\href@noop {} {\bibfield  {journal} {\bibinfo  {journal} {Phys. Rev. B}\ }\textbf {\bibinfo {volume} {23}},\ \bibinfo {pages} {5048} (\bibinfo {year} {1981})}\BibitemShut {NoStop}%
\bibitem [{\citenamefont {Jiao}\ \emph {et~al.}(2018{\natexlab{a}})\citenamefont {Jiao}, \citenamefont {Schr{\"o}der},\ and\ \citenamefont {Hyldgaard}}]{JiScHy18a}%
  \BibitemOpen
  \bibfield  {author} {\bibinfo {author} {\bibfnamefont {Y.}~\bibnamefont {Jiao}}, \bibinfo {author} {\bibfnamefont {E.}~\bibnamefont {Schr{\"o}der}},\ and\ \bibinfo {author} {\bibfnamefont {P.}~\bibnamefont {Hyldgaard}},\ }\bibfield  {title} {\bibinfo {title} {{Signatures of van der Waals binding: a coupling-constant scaling analysis}},\ }\href@noop {} {\bibfield  {journal} {\bibinfo  {journal} {Phys. Rev. B}\ }\textbf {\bibinfo {volume} {97}},\ \bibinfo {pages} {085115} (\bibinfo {year} {2018}{\natexlab{a}})}\BibitemShut {NoStop}%
\bibitem [{\citenamefont {Ferretti}\ \emph {et~al.}(2014)\citenamefont {Ferretti}, \citenamefont {Dabo}, \citenamefont {Cococcioni},\ and\ \citenamefont {Marzari}}]{ferretti2014}%
  \BibitemOpen
  \bibfield  {author} {\bibinfo {author} {\bibfnamefont {A.}~\bibnamefont {Ferretti}}, \bibinfo {author} {\bibfnamefont {I.}~\bibnamefont {Dabo}}, \bibinfo {author} {\bibfnamefont {M.}~\bibnamefont {Cococcioni}},\ and\ \bibinfo {author} {\bibfnamefont {N.}~\bibnamefont {Marzari}},\ }\bibfield  {title} {\bibinfo {title} {Bridging density-functional and many-body perturbation theory: Orbital-density dependence in electronic-structure functionals},\ }\href {https://doi.org/10.1103/PhysRevB.89.195134} {\bibfield  {journal} {\bibinfo  {journal} {Phys. Rev. B}\ }\textbf {\bibinfo {volume} {89}},\ \bibinfo {pages} {195134} (\bibinfo {year} {2014})}\BibitemShut {NoStop}%
\bibitem [{\citenamefont {Colonna}\ \emph {et~al.}(2018)\citenamefont {Colonna}, \citenamefont {Nguyen}, \citenamefont {Ferretti},\ and\ \citenamefont {Marzari}}]{colonna2018}%
  \BibitemOpen
  \bibfield  {author} {\bibinfo {author} {\bibfnamefont {N.}~\bibnamefont {Colonna}}, \bibinfo {author} {\bibfnamefont {N.~L.}\ \bibnamefont {Nguyen}}, \bibinfo {author} {\bibfnamefont {A.}~\bibnamefont {Ferretti}},\ and\ \bibinfo {author} {\bibfnamefont {N.}~\bibnamefont {Marzari}},\ }\bibfield  {title} {\bibinfo {title} {{Screening in Orbital-Density-Dependent Functionals}},\ }\href {https://doi.org/10.1021/acs.jctc.7b01116} {\bibfield  {journal} {\bibinfo  {journal} {J. Chem. Theory Comput.}\ }\textbf {\bibinfo {volume} {14}},\ \bibinfo {pages} {2549} (\bibinfo {year} {2018})}\BibitemShut {NoStop}%
\bibitem [{\citenamefont {Nguyen}\ \emph {et~al.}(2018)\citenamefont {Nguyen}, \citenamefont {Colonna}, \citenamefont {Ferretti},\ and\ \citenamefont {Marzari}}]{nguyen2018}%
  \BibitemOpen
  \bibfield  {author} {\bibinfo {author} {\bibfnamefont {N.~L.}\ \bibnamefont {Nguyen}}, \bibinfo {author} {\bibfnamefont {N.}~\bibnamefont {Colonna}}, \bibinfo {author} {\bibfnamefont {A.}~\bibnamefont {Ferretti}},\ and\ \bibinfo {author} {\bibfnamefont {N.}~\bibnamefont {Marzari}},\ }\bibfield  {title} {\bibinfo {title} {{Koopmans-Compliant Spectral Functionals for Extended Systems}},\ }\href {https://doi.org/10.1103/PhysRevX.8.021051} {\bibfield  {journal} {\bibinfo  {journal} {Phys. Rev. X}\ }\textbf {\bibinfo {volume} {8}},\ \bibinfo {pages} {021051} (\bibinfo {year} {2018})}\BibitemShut {NoStop}%
\bibitem [{\citenamefont {De~Gennaro}\ \emph {et~al.}(2022)\citenamefont {De~Gennaro}, \citenamefont {Colonna}, \citenamefont {Linscott},\ and\ \citenamefont {Marzari}}]{gennaro2022}%
  \BibitemOpen
  \bibfield  {author} {\bibinfo {author} {\bibfnamefont {R.}~\bibnamefont {De~Gennaro}}, \bibinfo {author} {\bibfnamefont {N.}~\bibnamefont {Colonna}}, \bibinfo {author} {\bibfnamefont {E.}~\bibnamefont {Linscott}},\ and\ \bibinfo {author} {\bibfnamefont {N.}~\bibnamefont {Marzari}},\ }\bibfield  {title} {\bibinfo {title} {{Bloch's theorem in orbital-density-dependent functionals: Band structures from Koopmans spectral functionals}},\ }\href {https://doi.org/10.1103/PhysRevB.106.035106} {\bibfield  {journal} {\bibinfo  {journal} {Phys. Rev. B}\ }\textbf {\bibinfo {volume} {106}},\ \bibinfo {pages} {035106} (\bibinfo {year} {2022})}\BibitemShut {NoStop}%
\bibitem [{\citenamefont {Linscott}\ \emph {et~al.}(2023)\citenamefont {Linscott}, \citenamefont {Colonna}, \citenamefont {De~Gennaro}, \citenamefont {Nguyen}, \citenamefont {Borghi}, \citenamefont {Ferretti}, \citenamefont {Dabo},\ and\ \citenamefont {Marzari}}]{linscott2023}%
  \BibitemOpen
  \bibfield  {author} {\bibinfo {author} {\bibfnamefont {E.~B.}\ \bibnamefont {Linscott}}, \bibinfo {author} {\bibfnamefont {N.}~\bibnamefont {Colonna}}, \bibinfo {author} {\bibfnamefont {R.}~\bibnamefont {De~Gennaro}}, \bibinfo {author} {\bibfnamefont {N.~L.}\ \bibnamefont {Nguyen}}, \bibinfo {author} {\bibfnamefont {G.}~\bibnamefont {Borghi}}, \bibinfo {author} {\bibfnamefont {A.}~\bibnamefont {Ferretti}}, \bibinfo {author} {\bibfnamefont {I.}~\bibnamefont {Dabo}},\ and\ \bibinfo {author} {\bibfnamefont {N.}~\bibnamefont {Marzari}},\ }\bibfield  {title} {\bibinfo {title} {{koopmans: An Open Source Package for Accurately and Efficiently Predicting Spectral Properties with Koopmans Functionals}},\ }\href {https://doi.org/10.1021/acs.jctc.3c00652} {\bibfield  {journal} {\bibinfo  {journal} {J. Chem. Theory Comput.}\ }\textbf {\bibinfo {volume} {19}},\ \bibinfo {pages} {7097} (\bibinfo {year} {2023})}\BibitemShut {NoStop}%
\bibitem [{\citenamefont {Dion}\ \emph {et~al.}(2005)\citenamefont {Dion}, \citenamefont {Rydberg}, \citenamefont {Schr{\"o}der}, \citenamefont {Langreth},\ and\ \citenamefont {Lundqvist}}]{dionerratum}%
  \BibitemOpen
  \bibfield  {author} {\bibinfo {author} {\bibfnamefont {M.}~\bibnamefont {Dion}}, \bibinfo {author} {\bibfnamefont {H.}~\bibnamefont {Rydberg}}, \bibinfo {author} {\bibfnamefont {E.}~\bibnamefont {Schr{\"o}der}}, \bibinfo {author} {\bibfnamefont {D.~C.}\ \bibnamefont {Langreth}},\ and\ \bibinfo {author} {\bibfnamefont {B.~I.}\ \bibnamefont {Lundqvist}},\ }\bibfield  {title} {\bibinfo {title} {{Erratum: Van der Waals Density Functional for General Geometries [Phys. Rev. Lett. {\bf 92}, 246401 (2004)]}},\ }\href@noop {} {\bibfield  {journal} {\bibinfo  {journal} {Phys. Rev. Lett.}\ }\textbf {\bibinfo {volume} {95}},\ \bibinfo {pages} {109902(E)} (\bibinfo {year} {2005})}\BibitemShut {NoStop}%
\bibitem [{\citenamefont {Ernzerhof}\ and\ \citenamefont {Perdew}(1998)}]{EP98}%
  \BibitemOpen
  \bibfield  {author} {\bibinfo {author} {\bibfnamefont {M.}~\bibnamefont {Ernzerhof}}\ and\ \bibinfo {author} {\bibfnamefont {J.~P.}\ \bibnamefont {Perdew}},\ }\bibfield  {title} {\bibinfo {title} {Generalized gradient approximation to the angle- and system-averaged exchange hole},\ }\href {https://doi.org/http://dx.doi.org/10.1063/1.476928} {\bibfield  {journal} {\bibinfo  {journal} {J. Chem. Phys.}\ }\textbf {\bibinfo {volume} {109}},\ \bibinfo {pages} {3313} (\bibinfo {year} {1998})}\BibitemShut {NoStop}%
\bibitem [{\citenamefont {Goerigk}\ \emph {et~al.}(2017)\citenamefont {Goerigk}, \citenamefont {Hansen}, \citenamefont {Bauer}, \citenamefont {Ehrlich}, \citenamefont {Najibi},\ and\ \citenamefont {Grimme}}]{gmtkn55}%
  \BibitemOpen
  \bibfield  {author} {\bibinfo {author} {\bibfnamefont {L.}~\bibnamefont {Goerigk}}, \bibinfo {author} {\bibfnamefont {A.}~\bibnamefont {Hansen}}, \bibinfo {author} {\bibfnamefont {C.}~\bibnamefont {Bauer}}, \bibinfo {author} {\bibfnamefont {S.}~\bibnamefont {Ehrlich}}, \bibinfo {author} {\bibfnamefont {A.}~\bibnamefont {Najibi}},\ and\ \bibinfo {author} {\bibfnamefont {S.}~\bibnamefont {Grimme}},\ }\bibfield  {title} {\bibinfo {title} {{A look at the density functional theory zoo with the advanced GMTKN55 database for general main group thermochemistry, kinetics and noncovalent interactions}},\ }\href@noop {} {\bibfield  {journal} {\bibinfo  {journal} {Phys. Chem. Chem. Phys.}\ }\textbf {\bibinfo {volume} {19}},\ \bibinfo {pages} {32184} (\bibinfo {year} {2017})}\BibitemShut {NoStop}%
\bibitem [{\citenamefont {Giannozzi}\ \emph {et~al.}(2009)\citenamefont {Giannozzi}, \citenamefont {Baroni}, \citenamefont {Bonini}, \citenamefont {Calandra}, \citenamefont {Car}, \citenamefont {Cavazzoni}, \citenamefont {Ceresoli}, \citenamefont {Chiarotti}, \citenamefont {Cococcioni}, \citenamefont {Dabo}, \citenamefont {Corso}, \citenamefont {de~Gironcoli}, \citenamefont {Fabris}, \citenamefont {Fratesi}, \citenamefont {Gebauer}, \citenamefont {Gerstmann}, \citenamefont {Gougoussis}, \citenamefont {Kokalj}, \citenamefont {Lazzeri}, \citenamefont {Martin-Samos}, \citenamefont {Marzari}, \citenamefont {Mauri}, \citenamefont {Mazzarello}, \citenamefont {Paolini}, \citenamefont {Pasquarello}, \citenamefont {Paulatto}, \citenamefont {Sbraccia}, \citenamefont {Scandolo}, \citenamefont {Sclauzero}, \citenamefont {Seitsonen}, \citenamefont {Smogunov}, \citenamefont {Umari},\ and\ \citenamefont {Wentzcovitch}}]{QE}%
  \BibitemOpen
  \bibfield  {author} {\bibinfo {author} {\bibfnamefont {P.}~\bibnamefont {Giannozzi}}, \bibinfo {author} {\bibfnamefont {S.}~\bibnamefont {Baroni}}, \bibinfo {author} {\bibfnamefont {N.}~\bibnamefont {Bonini}}, \bibinfo {author} {\bibfnamefont {M.}~\bibnamefont {Calandra}}, \bibinfo {author} {\bibfnamefont {R.}~\bibnamefont {Car}}, \bibinfo {author} {\bibfnamefont {C.}~\bibnamefont {Cavazzoni}}, \bibinfo {author} {\bibfnamefont {D.}~\bibnamefont {Ceresoli}}, \bibinfo {author} {\bibfnamefont {G.~L.}\ \bibnamefont {Chiarotti}}, \bibinfo {author} {\bibfnamefont {M.}~\bibnamefont {Cococcioni}}, \bibinfo {author} {\bibfnamefont {I.}~\bibnamefont {Dabo}}, \bibinfo {author} {\bibfnamefont {A.~D.}\ \bibnamefont {Corso}}, \bibinfo {author} {\bibfnamefont {S.}~\bibnamefont {de~Gironcoli}}, \bibinfo {author} {\bibfnamefont {S.}~\bibnamefont {Fabris}}, \bibinfo {author} {\bibfnamefont {G.}~\bibnamefont {Fratesi}}, \bibinfo {author} {\bibfnamefont {R.}~\bibnamefont {Gebauer}}, \bibinfo {author} {\bibfnamefont
  {U.}~\bibnamefont {Gerstmann}}, \bibinfo {author} {\bibfnamefont {C.}~\bibnamefont {Gougoussis}}, \bibinfo {author} {\bibfnamefont {A.}~\bibnamefont {Kokalj}}, \bibinfo {author} {\bibfnamefont {M.}~\bibnamefont {Lazzeri}}, \bibinfo {author} {\bibfnamefont {L.}~\bibnamefont {Martin-Samos}}, \bibinfo {author} {\bibfnamefont {N.}~\bibnamefont {Marzari}}, \bibinfo {author} {\bibfnamefont {F.}~\bibnamefont {Mauri}}, \bibinfo {author} {\bibfnamefont {R.}~\bibnamefont {Mazzarello}}, \bibinfo {author} {\bibfnamefont {S.}~\bibnamefont {Paolini}}, \bibinfo {author} {\bibfnamefont {A.}~\bibnamefont {Pasquarello}}, \bibinfo {author} {\bibfnamefont {L.}~\bibnamefont {Paulatto}}, \bibinfo {author} {\bibfnamefont {C.}~\bibnamefont {Sbraccia}}, \bibinfo {author} {\bibfnamefont {S.}~\bibnamefont {Scandolo}}, \bibinfo {author} {\bibfnamefont {G.}~\bibnamefont {Sclauzero}}, \bibinfo {author} {\bibfnamefont {A.~P.}\ \bibnamefont {Seitsonen}}, \bibinfo {author} {\bibfnamefont {A.}~\bibnamefont {Smogunov}}, \bibinfo {author}
  {\bibfnamefont {P.}~\bibnamefont {Umari}},\ and\ \bibinfo {author} {\bibfnamefont {R.~M.}\ \bibnamefont {Wentzcovitch}},\ }\bibfield  {title} {\bibinfo {title} {{\textsc{Quantum ESPRESSO}: a modular and open-source software project for quantum simulations of materials}},\ }\href@noop {} {\bibfield  {journal} {\bibinfo  {journal} {J. Phys.: Condens. Matter}\ }\textbf {\bibinfo {volume} {21}},\ \bibinfo {pages} {395502} (\bibinfo {year} {2009})}\BibitemShut {NoStop}%
\bibitem [{\citenamefont {Giannozzi}\ \emph {et~al.}(2017)\citenamefont {Giannozzi}, \citenamefont {Andreussi}, \citenamefont {Brumme}, \citenamefont {Bunau}, \citenamefont {Buongiorno~Nardelli}, \citenamefont {Calandra}, \citenamefont {Car}, \citenamefont {Cavazzoni}, \citenamefont {Ceresoli}, \citenamefont {Cococcioni}, \citenamefont {Collonna}, \citenamefont {Carnimeo}, \citenamefont {Dal~Corso}, \citenamefont {{de Gironcoli}}, \citenamefont {Delugas}, \citenamefont {{DiStasio Jr}}, \citenamefont {Feretti}, \citenamefont {Floris}, \citenamefont {Fratesi}, \citenamefont {Fugalio}, \citenamefont {Gebauer}, \citenamefont {Gerstmann}, \citenamefont {Giustino}, \citenamefont {Gorni}, \citenamefont {Jia}, \citenamefont {Kawamura}, \citenamefont {Ko}, \citenamefont {Kokalj}, \citenamefont {K{\"u}c{\"u}kbenli}, \citenamefont {Lazzeri}, \citenamefont {Marseli}, \citenamefont {Marzari}, \citenamefont {Mauri}, \citenamefont {Nguyen}, \citenamefont {Nguyen}, \citenamefont {{Otero-de-la-Roza}}, \citenamefont
  {Paulatto}, \citenamefont {Ponc{\'e}}, \citenamefont {Rocca}, \citenamefont {Sabatini}, \citenamefont {Santra}, \citenamefont {Schlipf}, \citenamefont {Seitsonen}, \citenamefont {Smogunov}, \citenamefont {Timrov}, \citenamefont {Thonhauser}, \citenamefont {Umari}, \citenamefont {Vast}, \citenamefont {Wu},\ and\ \citenamefont {Baroni}}]{Giannozzi17}%
  \BibitemOpen
  \bibfield  {author} {\bibinfo {author} {\bibfnamefont {P.}~\bibnamefont {Giannozzi}}, \bibinfo {author} {\bibfnamefont {O.}~\bibnamefont {Andreussi}}, \bibinfo {author} {\bibfnamefont {T.}~\bibnamefont {Brumme}}, \bibinfo {author} {\bibfnamefont {O.}~\bibnamefont {Bunau}}, \bibinfo {author} {\bibfnamefont {M.}~\bibnamefont {Buongiorno~Nardelli}}, \bibinfo {author} {\bibfnamefont {M.}~\bibnamefont {Calandra}}, \bibinfo {author} {\bibfnamefont {R.}~\bibnamefont {Car}}, \bibinfo {author} {\bibfnamefont {C.}~\bibnamefont {Cavazzoni}}, \bibinfo {author} {\bibfnamefont {D.}~\bibnamefont {Ceresoli}}, \bibinfo {author} {\bibfnamefont {M.}~\bibnamefont {Cococcioni}}, \bibinfo {author} {\bibfnamefont {N.}~\bibnamefont {Collonna}}, \bibinfo {author} {\bibfnamefont {I.}~\bibnamefont {Carnimeo}}, \bibinfo {author} {\bibfnamefont {A.}~\bibnamefont {Dal~Corso}}, \bibinfo {author} {\bibfnamefont {S.}~\bibnamefont {{de Gironcoli}}}, \bibinfo {author} {\bibfnamefont {P.}~\bibnamefont {Delugas}}, \bibinfo {author}
  {\bibfnamefont {R.~A.}\ \bibnamefont {{DiStasio Jr}}}, \bibinfo {author} {\bibfnamefont {A.}~\bibnamefont {Feretti}}, \bibinfo {author} {\bibfnamefont {A.}~\bibnamefont {Floris}}, \bibinfo {author} {\bibfnamefont {G.}~\bibnamefont {Fratesi}}, \bibinfo {author} {\bibfnamefont {G.}~\bibnamefont {Fugalio}}, \bibinfo {author} {\bibfnamefont {R.}~\bibnamefont {Gebauer}}, \bibinfo {author} {\bibfnamefont {U.}~\bibnamefont {Gerstmann}}, \bibinfo {author} {\bibfnamefont {F.}~\bibnamefont {Giustino}}, \bibinfo {author} {\bibfnamefont {T.}~\bibnamefont {Gorni}}, \bibinfo {author} {\bibfnamefont {J.}~\bibnamefont {Jia}}, \bibinfo {author} {\bibfnamefont {M.}~\bibnamefont {Kawamura}}, \bibinfo {author} {\bibfnamefont {H.-Y.}\ \bibnamefont {Ko}}, \bibinfo {author} {\bibfnamefont {A.}~\bibnamefont {Kokalj}}, \bibinfo {author} {\bibfnamefont {E.}~\bibnamefont {K{\"u}c{\"u}kbenli}}, \bibinfo {author} {\bibfnamefont {M.}~\bibnamefont {Lazzeri}}, \bibinfo {author} {\bibfnamefont {M.}~\bibnamefont {Marseli}}, \bibinfo
  {author} {\bibfnamefont {N.}~\bibnamefont {Marzari}}, \bibinfo {author} {\bibfnamefont {F.}~\bibnamefont {Mauri}}, \bibinfo {author} {\bibfnamefont {N.~L.}\ \bibnamefont {Nguyen}}, \bibinfo {author} {\bibfnamefont {H.-V.}\ \bibnamefont {Nguyen}}, \bibinfo {author} {\bibfnamefont {A.}~\bibnamefont {{Otero-de-la-Roza}}}, \bibinfo {author} {\bibfnamefont {L.}~\bibnamefont {Paulatto}}, \bibinfo {author} {\bibfnamefont {S.}~\bibnamefont {Ponc{\'e}}}, \bibinfo {author} {\bibfnamefont {D.}~\bibnamefont {Rocca}}, \bibinfo {author} {\bibfnamefont {R.}~\bibnamefont {Sabatini}}, \bibinfo {author} {\bibfnamefont {B.}~\bibnamefont {Santra}}, \bibinfo {author} {\bibfnamefont {M.}~\bibnamefont {Schlipf}}, \bibinfo {author} {\bibfnamefont {A.}~\bibnamefont {Seitsonen}}, \bibinfo {author} {\bibfnamefont {A.}~\bibnamefont {Smogunov}}, \bibinfo {author} {\bibfnamefont {I.}~\bibnamefont {Timrov}}, \bibinfo {author} {\bibfnamefont {T.}~\bibnamefont {Thonhauser}}, \bibinfo {author} {\bibfnamefont {P.}~\bibnamefont {Umari}},
  \bibinfo {author} {\bibfnamefont {N.}~\bibnamefont {Vast}}, \bibinfo {author} {\bibfnamefont {X.}~\bibnamefont {Wu}},\ and\ \bibinfo {author} {\bibfnamefont {S.}~\bibnamefont {Baroni}},\ }\bibfield  {title} {\bibinfo {title} {{Advanced capabilities for materials modelling with \textsc{Quantum ESPRESSO}}},\ }\href@noop {} {\bibfield  {journal} {\bibinfo  {journal} {J.Phys.: Condens. Matter}\ }\textbf {\bibinfo {volume} {29}},\ \bibinfo {pages} {465901} (\bibinfo {year} {2017})}\BibitemShut {NoStop}%
\bibitem [{\citenamefont {Lin}(2016)}]{linACE}%
  \BibitemOpen
  \bibfield  {author} {\bibinfo {author} {\bibfnamefont {L.}~\bibnamefont {Lin}},\ }\bibfield  {title} {\bibinfo {title} {{Adaptively Compressed Exchange Operator}},\ }\href@noop {} {\bibfield  {journal} {\bibinfo  {journal} {J. Chem. Theory Comput.}\ }\textbf {\bibinfo {volume} {12}},\ \bibinfo {pages} {2242} (\bibinfo {year} {2016})}\BibitemShut {NoStop}%
\bibitem [{\citenamefont {Carnimeo}\ \emph {et~al.}(2019)\citenamefont {Carnimeo}, \citenamefont {Baroni},\ and\ \citenamefont {Giannozzi}}]{PaoloElStruct1}%
  \BibitemOpen
  \bibfield  {author} {\bibinfo {author} {\bibfnamefont {I.}~\bibnamefont {Carnimeo}}, \bibinfo {author} {\bibfnamefont {S.}~\bibnamefont {Baroni}},\ and\ \bibinfo {author} {\bibfnamefont {P.}~\bibnamefont {Giannozzi}},\ }\bibfield  {title} {\bibinfo {title} {{Fast hybrid density-functional computations using plane-wave beasis sets}},\ }\href@noop {} {\bibfield  {journal} {\bibinfo  {journal} {Electron. Struct.}\ }\textbf {\bibinfo {volume} {1}},\ \bibinfo {pages} {015009} (\bibinfo {year} {2019})}\BibitemShut {NoStop}%
\bibitem [{\citenamefont {Hamann}(2013)}]{ONCV}%
  \BibitemOpen
  \bibfield  {author} {\bibinfo {author} {\bibfnamefont {D.~R.}\ \bibnamefont {Hamann}},\ }\bibfield  {title} {\bibinfo {title} {Optimized norm-conserving {Vanderbilt} pseudopotentials},\ }\href@noop {} {\bibfield  {journal} {\bibinfo  {journal} {Phys. Rev. B}\ }\textbf {\bibinfo {volume} {88}},\ \bibinfo {pages} {085117} (\bibinfo {year} {2013})}\BibitemShut {NoStop}%
\bibitem [{\citenamefont {Schlipf}\ and\ \citenamefont {Gygi}(2015)}]{sg15}%
  \BibitemOpen
  \bibfield  {author} {\bibinfo {author} {\bibfnamefont {M.}~\bibnamefont {Schlipf}}\ and\ \bibinfo {author} {\bibfnamefont {F.}~\bibnamefont {Gygi}},\ }\bibfield  {title} {\bibinfo {title} {Optimization algorithm for the generation of {ONCV} pseudopotentials},\ }\href@noop {} {\bibfield  {journal} {\bibinfo  {journal} {Comput. Phys. Commun.}\ }\textbf {\bibinfo {volume} {196}},\ \bibinfo {pages} {36} (\bibinfo {year} {2015})}\BibitemShut {NoStop}%
\bibitem [{\citenamefont {{Gharaee}}\ \emph {et~al.}(2017)\citenamefont {{Gharaee}}, \citenamefont {{Erhart}},\ and\ \citenamefont {{Hyldgaard}}}]{Gharaee2017}%
  \BibitemOpen
  \bibfield  {author} {\bibinfo {author} {\bibfnamefont {L.}~\bibnamefont {{Gharaee}}}, \bibinfo {author} {\bibfnamefont {P.}~\bibnamefont {{Erhart}}},\ and\ \bibinfo {author} {\bibfnamefont {P.}~\bibnamefont {{Hyldgaard}}},\ }\bibfield  {title} {\bibinfo {title} {{Finite-temperature properties of non-magnetic transition metals: {C}omparison of the performance of constraint-based semi and nonlocal functionals}},\ }\href@noop {} {\bibfield  {journal} {\bibinfo  {journal} {Phys. Rev. B}\ }\textbf {\bibinfo {volume} {95}},\ \bibinfo {pages} {085147} (\bibinfo {year} {2017})}\BibitemShut {NoStop}%
\bibitem [{\citenamefont {G\"orling}\ and\ \citenamefont {Levy}(1993)}]{Gorling93}%
  \BibitemOpen
  \bibfield  {author} {\bibinfo {author} {\bibfnamefont {A.}~\bibnamefont {G\"orling}}\ and\ \bibinfo {author} {\bibfnamefont {M.}~\bibnamefont {Levy}},\ }\bibfield  {title} {\bibinfo {title} {Correlation-energy functional and its high-density limit obtained from a coupling-constant perturbation expansion},\ }\href {https://doi.org/10.1103/PhysRevB.47.13105} {\bibfield  {journal} {\bibinfo  {journal} {Phys. Rev. B}\ }\textbf {\bibinfo {volume} {47}},\ \bibinfo {pages} {13105} (\bibinfo {year} {1993})}\BibitemShut {NoStop}%
\bibitem [{\citenamefont {Jiao}\ \emph {et~al.}(2018{\natexlab{b}})\citenamefont {Jiao}, \citenamefont {Schr{\"o}der},\ and\ \citenamefont {Hyldgaard}}]{JiScHy18b}%
  \BibitemOpen
  \bibfield  {author} {\bibinfo {author} {\bibfnamefont {Y.}~\bibnamefont {Jiao}}, \bibinfo {author} {\bibfnamefont {E.}~\bibnamefont {Schr{\"o}der}},\ and\ \bibinfo {author} {\bibfnamefont {P.}~\bibnamefont {Hyldgaard}},\ }\bibfield  {title} {\bibinfo {title} {{Extent of Fock-exchange mixing for a hybrid van der Waals density functional?}},\ }\href@noop {} {\bibfield  {journal} {\bibinfo  {journal} {J. Chem. Phys.}\ }\textbf {\bibinfo {volume} {148}},\ \bibinfo {pages} {194115} (\bibinfo {year} {2018}{\natexlab{b}})}\BibitemShut {NoStop}%
\bibitem [{\citenamefont {Perdew}\ and\ \citenamefont {Wang}(1992)}]{Perdew_1992:accurate_simple}%
  \BibitemOpen
  \bibfield  {author} {\bibinfo {author} {\bibfnamefont {J.~P.}\ \bibnamefont {Perdew}}\ and\ \bibinfo {author} {\bibfnamefont {Y.}~\bibnamefont {Wang}},\ }\bibfield  {title} {\bibinfo {title} {Accurate and simple analytic representation of the electron-gas correlation energy},\ }\href {https://doi.org/10.1103/PhysRevB.45.13244} {\bibfield  {journal} {\bibinfo  {journal} {Phys. Rev. B}\ }\textbf {\bibinfo {volume} {45}},\ \bibinfo {pages} {13244} (\bibinfo {year} {1992})}\BibitemShut {NoStop}%
\bibitem [{\citenamefont {Lee}\ \emph {et~al.}(2022)\citenamefont {Lee}, \citenamefont {Hyldgaard},\ and\ \citenamefont {Neaton}}]{MOFdobpdc}%
  \BibitemOpen
  \bibfield  {author} {\bibinfo {author} {\bibfnamefont {J.-H.}\ \bibnamefont {Lee}}, \bibinfo {author} {\bibfnamefont {P.}~\bibnamefont {Hyldgaard}},\ and\ \bibinfo {author} {\bibfnamefont {J.~B.}\ \bibnamefont {Neaton}},\ }\bibfield  {title} {\bibinfo {title} {{An assessment of density functionals for predicting CO$_2$ adsorption in diamine-functionalized metal-organic frameworks}},\ }\href@noop {} {\bibfield  {journal} {\bibinfo  {journal} {J. Chem. Phys.}\ }\textbf {\bibinfo {volume} {156}},\ \bibinfo {pages} {154113} (\bibinfo {year} {2022})}\BibitemShut {NoStop}%
\bibitem [{\citenamefont {Langreth}\ and\ \citenamefont {Perdew}(1975)}]{lape75}%
  \BibitemOpen
  \bibfield  {author} {\bibinfo {author} {\bibfnamefont {D.~C.}\ \bibnamefont {Langreth}}\ and\ \bibinfo {author} {\bibfnamefont {J.~P.}\ \bibnamefont {Perdew}},\ }\bibfield  {title} {\bibinfo {title} {The exchange-correlation energy of a metallic surface},\ }\href@noop {} {\bibfield  {journal} {\bibinfo  {journal} {Solid State Commun.}\ }\textbf {\bibinfo {volume} {17}},\ \bibinfo {pages} {1425} (\bibinfo {year} {1975})}\BibitemShut {NoStop}%
\bibitem [{\citenamefont {Langreth}\ and\ \citenamefont {Mehl}(1981)}]{lameprl1981}%
  \BibitemOpen
  \bibfield  {author} {\bibinfo {author} {\bibfnamefont {D.~C.}\ \bibnamefont {Langreth}}\ and\ \bibinfo {author} {\bibfnamefont {M.~J.}\ \bibnamefont {Mehl}},\ }\bibfield  {title} {\bibinfo {title} {{Easily Implementable Nonlocal Exchange-Correlation Energy Functional}},\ }\href@noop {} {\bibfield  {journal} {\bibinfo  {journal} {Phys. Rev. Lett.}\ }\textbf {\bibinfo {volume} {47}},\ \bibinfo {pages} {446} (\bibinfo {year} {1981})}\BibitemShut {NoStop}%
\bibitem [{\citenamefont {Rydberg}\ \emph {et~al.}(2003)\citenamefont {Rydberg}, \citenamefont {Dion}, \citenamefont {Jacobson}, \citenamefont {Schr{\"o}der}, \citenamefont {Hyldgaard}, \citenamefont {Simak}, \citenamefont {Langreth},\ and\ \citenamefont {Lundqvist}}]{rydberg03p126402}%
  \BibitemOpen
  \bibfield  {author} {\bibinfo {author} {\bibfnamefont {H.}~\bibnamefont {Rydberg}}, \bibinfo {author} {\bibfnamefont {M.}~\bibnamefont {Dion}}, \bibinfo {author} {\bibfnamefont {N.}~\bibnamefont {Jacobson}}, \bibinfo {author} {\bibfnamefont {E.}~\bibnamefont {Schr{\"o}der}}, \bibinfo {author} {\bibfnamefont {P.}~\bibnamefont {Hyldgaard}}, \bibinfo {author} {\bibfnamefont {S.~I.}\ \bibnamefont {Simak}}, \bibinfo {author} {\bibfnamefont {D.~C.}\ \bibnamefont {Langreth}},\ and\ \bibinfo {author} {\bibfnamefont {B.~I.}\ \bibnamefont {Lundqvist}},\ }\bibfield  {title} {\bibinfo {title} {{Van der Waals Density Functional for Layered Structures}},\ }\href@noop {} {\bibfield  {journal} {\bibinfo  {journal} {Phys. Rev. Lett.}\ }\textbf {\bibinfo {volume} {91}},\ \bibinfo {pages} {126402} (\bibinfo {year} {2003})}\BibitemShut {NoStop}%
\end{thebibliography}

%

\end{document}